\documentclass{aa}%[referee]
\usepackage[varg]{txfonts}
\usepackage{graphicx}
\usepackage[breaklinks, colorlinks, citecolor=blue, linkcolor=blue]{hyperref}%, draft
\usepackage{natbib}
\usepackage[caption = false]{subfig}
\bibpunct{(}{)}{;}{a}{}{,} % to follow the A&A style

\let\orgautoref\autoref
\renewcommand{\autoref}
        {\def\equationautorefname{Eq.}%
         \def\figureautorefname{Fig.}%
         \def\sectionautorefname{Sect.}%
         \def\subsectionautorefname{Sect.}%
         \def\subsubsectionautorefname{Sect.}%
         \orgautoref}

\newcommand*\samethanks[1][\value{footnote}]{\footnotemark[#1]}

\newcommand{\tx}[1]{\mathrm{#1}} %for text
\newcommand{\Steph}[1]{\textcolor{red}{\textbf{#1}}}%{#1}  %for commenting
\newcommand{\Stephb}[1]{\textcolor{red}{\textbf{#1}}}%{#1}  %for commenting
\newcommand{\Stephc}[1]{\textcolor{red}{\textbf{#1}}}%{#1}  %for commenting
\newcommand{\Stephd}[1]{\textcolor{red}{\textbf{#1}}}%{#1}  %for commenting
\newcommand{\Stephe}[1]{\textcolor{red}{\textbf{#1}}}%{#1}  %for commenting
\newcommand{\Stephf}[1]{\textcolor{red}{\textbf{#1}}}%{#1}  %for commenting
\renewcommand{\Steph}[1]{#1}  %for commenting
\renewcommand{\Stephb}[1]{#1}  %for commenting
\renewcommand{\Stephc}[1]{#1}  %for commenting
\renewcommand{\Stephd}[1]{#1}  %for commenting
\renewcommand{\Stephe}[1]{#1}  %for commenting
\renewcommand{\Stephf}[1]{#1}  %for commenting
\makeatletter
\def\instrefs#1{{\def\scsep{\def\scsep{,}}\@for\w:=#1\do{\scsep\ref{inst:\w}}}}
\renewcommand{\inst}[1]{\unskip$^{\instrefs{#1}}$}

\begin{document}

\title{The CARMENES search for exoplanets around M dwarfs}

\subtitle{Three temperate-to-warm super-Earths}

\author{S.~Stock\inst{lsw}\thanks{Fellow of the International Max Planck Research School for Astronomy and Cosmic Physics at the University of Heidelberg (IMPRS-HD).}
\and E.~Nagel\inst{hs,tls}
\and J.~Kemmer\inst{lsw}\samethanks 
\and V.\,M.~Passegger\inst{hs,ou}
\and S.~Reffert\inst{lsw}
\and A.~Quirrenbach\inst{lsw}
\and J.\,A.~Caballero\inst{cabesac}
\and S.~Czesla\inst{hs}
%alphabetetic list 1 other important contributers
\and V.\,J.\,S.~B\'ejar\inst{iac,ull}
\and C.~Cardona\inst{iac}
\and E.~D\'iez-Alonso\inst{ucm,uovi}
\and E. Herrero\inst{ice}
\and S.~Lalitha\inst{iag}
\and M.~Schlecker\inst{mpia}\samethanks
\and L.~Tal-Or \inst{dpi,iag}
\and E.~Rodr\'\i guez\inst{iaa}
\and C.~Rodr\'iguez-L\'opez\inst{iaa}
% keynames
\and I.~Ribas\inst{ice,iceb}
\and A.~Reiners\inst{iag}
\and P.~J.~Amado\inst{iaa}
%alphabetetic list 2 (rest of contributors)
\and F.\,F.~Bauer\inst{iaa}
\and P.~Bluhm\inst{lsw}\samethanks
\and M.~Cort\'es-Contreras\inst{cabesac}
\and L.~Gonz\'alez-Cuesta\inst{iac,ull}
\and S.~Dreizler \inst{iag} 
\and A.~P.~Hatzes\inst{tls}
\and Th.~Henning\inst{mpia}
\and S.\,V.~Jeffers\inst{iag}
\and A.~Kaminski\inst{lsw}  
\and M.~K\"urster\inst{mpia}
\and M.~Lafarga\inst{ice,iceb}
\and M.~J.~López-González\inst{iaa}
\and D.~Montes\inst{ucm}
\and J.~C.~Morales\inst{ice,iceb}
\and S.~Pedraz\inst{caha}
\and P.~Schöfer\inst{iag}
\and A.~Schweitzer \inst{hs}
\and T.~Trifonov\inst{mpia}  
\and M.~R.~Zapatero~Osorio\inst{cabesac}
\and M.~Zechmeister\inst{iag}
}

\institute{
\label{inst:lsw}Landessternwarte, Zentrum für Astronomie der Universität Heidelberg, Königstuhl 12, 69117 Heidelberg, Germany\\ \email{sstock@lsw.uni-heidelberg.de}
 \and
 \label{inst:hs}Hamburger Sternwarte, Gojenbergsweg 112, 21029 Hamburg, Germany
 \and 
 \label{inst:tls}Th\"uringer Landessternwarte Tautenburg, Sternwarte 5, 07778 Tautenburg, Germany
 \and
 \label{inst:ou}Homer L. Dodge Department of Physics and Astronomy, University
 of Oklahoma, 440 West Brooks Street, Norman, OK 73019, United
 States of America
  \and 
 \label{inst:mpia}Max-Planck-Institut f\"ur Astronomie, K\"onigstuhl 17, 69117 Heidelberg, Germany
 \and
 \label{inst:cabesac}Centro de Astrobiolog\'ia (CSIC-INTA), ESAC, Camino bajo del castillo s/n, 28692 Villanueva de la Ca\~nada, Madrid, Spain
 \and 
 \label{inst:iag}Institut f\"ur Astrophysik, Georg-August-Universit\"at, Friedrich-Hund-Platz 1, 37077 G\"ottingen, Germany
 \and 
 \label{inst:ice}Institut de Ci\`encies de l’Espai (ICE, CSIC), Campus UAB, C/Can Magrans s/n, E-08193 Bellaterra, Spain
 \and 
 \label{inst:iceb}Institut d’Estudis Espacials de Catalunya (IEEC), E-08034 Barcelona, Spain
 \and 
 \label{inst:iaa}Instituto de Astrof\'isica de Andaluc\'ia (IAA-CSIC), Glorieta de la Astronom\'ia s/n, 18008 Granada, Spain
 \and 
 \label{inst:iac}Instituto de Astrof\'isica de Canarias (IAC), 38205 La Laguna, Tenerife, Spain
 \and 
 \label{inst:ull}Departamento de Astrof\'isica, Universidad de La Laguna (ULL), 38206, La Laguna, Tenerife, Spain
 \and
 \label{inst:ucm}Departamento de F\'isica de la Tierra y Astrof\'isica \& IPARCOS-UCM (Instituto de F\'isica de Part\'iculas y del Cosmos de la UCM), Facultad de Ciencias F\'isicas, Universidad Complutense de Madrid, 28040 Madrid, Spain
 \and 
 \label{inst:uovi}Department of Exploitation and Exploration of Mines, University of Oviedo, Oviedo, Spain
 \and
 \label{inst:dpi}Department of Physics, Ariel University, Ariel 40700, Israel
 \and
\label{inst:caha}Observatorio de Calar Alto, Sierra de los Filabres, 04550 G\'ergal, Almer\'ia, Spain
 }

\date{Received dd June 2020 / Accepted dd Month 2020}

\abstract
{We announce the discovery of two planets orbiting the M dwarfs \object{GJ~251} \Steph{($0.360\pm0.015$\,M$_\odot$)} and \object{HD~238090} \Steph{($0.578\pm0.021$\,M$_\odot$)} based on CARMENES radial velocity (RV) data. In addition, we independently confirm with CARMENES data the existence of \object{Lalande~21185}~b, a planet that has recently been discovered \Steph{with} the SOPHIE spectrograph. All three planets \Stephb{belong to} the class of warm or temperate super-Earths and share similar properties. 
The orbital periods are 14.24\,d, 13.67\,d, and 12.95\,d and the minimum masses are $4.0\pm0.4\,M_\oplus$, $6.9\pm0.9\,M_\oplus$, and $2.7\pm0.3\,M_\oplus$ for GJ~251~b, HD~238090~b, and Lalande~21185~b\Steph{, respectively}. \Steph{Based on the orbital and stellar properties, we estimate equilibrium temperatures of $351.0\pm1.4$\,K for GJ~251~b, $469.6\pm2.6$\,K for HD~238090~b, and $370.1\pm6.8$\,K for Lalande~21185~b}. 
For the latter
we \Steph{resolve} the daily aliases that were \Stephb{present} in the SOPHIE data \Stephb{and that} hindered an unambiguous determination of the orbital period. We find no \Steph{significant signals} in any of our spectral activity indicators at the planetary periods. \Stephb{The} RV observations were accompanied by \Steph{contemporaneous} photometric observations. We derive stellar rotation periods of $122.1\pm2.2$\,d and $96.7\pm3.7$\,d for GJ~251 and HD~238090\Steph{, respectively}. The RV data of all three stars exhibit significant signals at the rotational period or its first harmonic. For GJ~251 and Lalande~21185, we also find long-period signals around 600\,d, and 2900\,d, respectively, which we tentatively attribute to long-term magnetic cycles. \Stephc{We apply a Bayesian approach to carefully model the Keplerian signals simultaneously with the stellar activity using Gaussian process regression models and extensively search for additional significant planetary signals hidden behind the stellar activity.} \Steph{Current planet formation theories suggest that the three systems represent a common architecture, consistent with formation following the core accretion paradigm.}  
   }

\maketitle

\titlerunning{GJ 251~b, HD 238090~b, and Lalande 21185~b}

\authorrunning{S. Stock et~al.}

\section{Introduction}
\Steph{More than 4200 exoplanets have been confirmed so far\footnote{\url{http://exoplanet.eu/catalog/} (25 May 2020)}. A significant fraction have been discovered with the radial velocity (RV) method.} \Steph{In the past decades, the development of new high-precision spectrographs allowed probing a large variety of planets with minimum masses of several Jupiter masses down to only $0.7$\,M$_\oplus$ for \object{YZ~Cet}~b, which is the least massive planet detected so far with the RV technique \citep{Astudillo, Stock2020b}.  One such high-precision spectrograph is the CARMENES instrument \citep{Quirrenbach2014, Quirrenbach2018}, which is used to conduct a survey for detecting exoplanets around M dwarfs, which are the most abundant stars of our Galaxy \citep{Kroupa2001, Chabrier2003, Henry2006}. The detection of the large number of exoplanets has resulted in the discovery of exotic new types of planets that have no counterpart in our own Solar System, such as super-Earths \citep[$M$ = 1.9--10\,$M_\oplus$;][]{Rivera2005, Valencia2007, Charbonneau2009}. These super-Earths are abundant around M dwarfs \citep{Dressing2015}}.

\Steph{The detection of planets close to or inside the habitable zones \citep[HZ, see][]{Kasting1993, Kopparapu2013} of their parent stars is of particular interest. With the current technology, M dwarfs are ideal targets for detecting such temperate planets because the HZ of these stars corresponds to a relatively small orbital radius. \Stephb{The lower host star masses} result in a higher Doppler amplitude ($\text{higher by a}$ few m\,s$^{-1}$) than those of more massive stars, which can be measured by current techniques}.
However, M dwarfs tend to be very active \citep{Johns-Krull1996, Delfosse1998, Mohanty2003, Reiners2012}.
\Steph{The activity can make the detection of small planets difficult by inducing distortions in the shape of the spectral line profiles; this mimicks a planetary signal \citep{Queloz2001, Desort2007, Barnes2011, Robertson2014, Robertson2015}.}

\Stephc{Various methods can be used to distinguish stellar astrophysical signals from planet-induced signals.}
Photometric observations, ideally contemporaneous with the RV observations, as well as different spectral activity indicators can be used to derive more information on the stellar rotation period and activity-induced RV variations. \Stephc{In addition}, many novel techniques have been developed to analyze the coherence of a signal, for example, Bayesian-stacked periodograms \citep{Mortier2015, Mortier2017}, growth of the Lomb-Scargle power, or the evolution of the significance \citep{Hatzes2013, Ribas2018, Reichert2018}. These tools can provide strong indications that a signal has a nonplanetary origin because an RV signal caused by Keplerian motion should \Steph{be coherent and long-lived}. \Stephb{When} both planetary signals and activity contribute significantly to the RV variations, it can be necessary to simultaneously fit for these signals using Gaussian process (GP) regression \citep{Rajpaul2015} or similar models, such as sinusoids \citep{Boisse2011, Dumusque2012}. Modeling the stellar activity simultaneously with the Keplerian fit is essential because this contamination can have a significant effect on the derived planetary parameters \citep[see, e.g, ][]{Stock2020b}. 

\Stephb{In the following, we present a detailed analysis of photometric and spectroscopic data of GJ 251, HD 238090, and Lalande 21185.} 
For GJ~251, \cite{Butler2017} reported a possible planet candidate at a period of 1.74\,d, \Stephd{but with the more precise CARMENES data, we cannot confirm this claim.} 
Lalande~21185 is the brightest M dwarf in the northern hemisphere, the fourth closest main-sequence star system after $\alpha$~Centauri, Barnard's star, and CN Leo, and the third closest planetary system.
Lalande~21185 has a remarkable history regarding former planet claims. 
\Steph{Those by \cite{Lalande1951} and \cite{Lalande1996} were based on astrometric data, but have never been confirmed independently. Later, \cite{Butler2017} reported that Lalande~21185 has a planet candidate with an orbital period of 9.87\,d.} 
A recent study by \cite{Lalande2019} was unable to provide evidence for these \Steph{previous planet claims.} However, \cite{Lalande2019} announced the discovery of a super-Earth planet orbiting Lalande~21185 \Stephc{with a period of} 12.93\,d. \Stephc{The analysis of our CARMENES data agrees with the findings from \citet{Lalande2019} and confirms a single planet orbiting Lalande~21185.} 
HD~238090 has no reported planet to date.

In Sect.~\ref{Sect: Data} and Sect.~\ref{Sect: Methods} we describe the data, instruments, and methods we used within this study, while in Sect.~\ref{Sect: Stellar_props} we compile the basic stellar properties of GJ~251, HD~238090, and Lalande~21185. \Steph{We then analyze our photometric and RV data for the three stars in Sect.~\ref{Sect: GJ251}, Sect.~\ref{Sect: HD238090}, and Sect.~\ref{Sect: Lalande}}, and provide a star-by-star discussion in Sect.~\ref{Sect: Discussion} and a general summary in Sect.~\ref{Sect: Summary}.

\section{Data}
\label{Sect: Data}

\subsection{High-resolution spectroscopy}

\begin{table}
\caption{Number and quality of the RV observations}
\label{Tab: datasets}
\centering
\begin{tabular}{l c c c c c}
\hline\hline
Instr. & \Stephc{Obs$_{\mathrm{start}}$}& \Stephc{Obs$_{\mathrm{end}}$} & $N_{\mathrm{obs.}}$ & $\sigma_{\tx{RV}}$ & $\mathrm{rms}$  \\
  & mm/yyyy & mm/yyyy &  & [$\mathrm{m\,s^{-1}}$] & [$\mathrm{m\,s^{-1}}$] \\
\hline
\noalign{\smallskip}
\multicolumn{6}{c}{GJ~251}\\
\noalign{\smallskip}
CARM. & \Stephc{01/2016} & \Stephc{01/2020} & 212 & 1.27 & 3.69\\
HIRES & \Stephc{10/1997} & \Stephc{11/2013} &75 & 2.13 & 4.63 \\
\noalign{\smallskip}
\multicolumn{6}{c}{HD~238090}\\
\noalign{\smallskip}
CARM. & \Stephc{01/2016} & \Stephc{04/2019} &\Stephc{108} & 1.67 & 3.28\\
\noalign{\smallskip}
\multicolumn{6}{c}{Lalande~21185}\\
\noalign{\smallskip}
CARM. & \Stephc{01/2016} & \Stephc{01/2020} &321 & 1.40 & 4.38\\
%C. NIR &  & &271 & 7.14 & 9.49\\
HIRES & \Stephc{06/1997} & \Stephc{07/2014} &261 & 1.38 & 4.63 \\
SOPHIE & \Stephc{10/2011} & \Stephc{06/2018} &155 & 1.32 & 2.54 \\
\hline
\end{tabular}
\end{table}

\paragraph{CARMENES.}
GJ~251, HD~238090, and Lalande~21185 were observed as part of our CARMENES\footnote{\url{http://carmenes.caha.es}} guaranteed-time observation program (GTO) to search for exoplanets around M dwarfs \citep{Reiners2018}. CARMENES is a double-channel \'echelle spectrograph installed at the 3.5\,m telescope of the Calar Alto Observatory in Almer\'ia, Spain. Details regarding the instrument and its performance are given in \cite{Quirrenbach2014, Quirrenbach2018}, \cite{Reiners2018}, and \cite{Trifonov2018}.
The data were processed with the standard pipelines and were reduced with {\tt caracal} \citep{Cab16a}.
The RVs obtained with {\tt serval} \citep{Zechmeister2018} were corrected for barycentric motion, secular perspective acceleration, instrumental drift, and nightly zero-point variations \citep[][]{Trifonov2018, Tal-Or2019, Trifonov2020}. 
Table \ref{Tab: datasets} shows a summary of the CARMENES visual arm (VIS) 
RVs and their overall quality. \Stephc{The median exposure times in the VIS channel were 509\,s, 1000\,s, and 95\,s, resulting in a median signal-to-noise (S/N) of 116, 154, and 132 for GJ~251, HD~238090, and Lalande~21185, respectively.} 

\Stephc{In the CARMENES near-infrared (NIR) data, the scatter was not sufficiently small for the RV analysis of the planetary signals in this work; it was on the order of a few m\,s$^{-1}$ \citep[see][for a detailed analysis of the performance of CARMENES]{Bauer2020}.
The RV time series and their uncertainties for the CARMENES VIS data of GJ~251, HD~238090, and Lalande~21185 are listed in the appendix %Tables~\ref{Tab: RV_Data1_GJ251},~\ref{Tab: RV Data HD238090}, and~\ref{Tab: RV Data Lalande21185} 
together with some activity indicators.}

\paragraph{HIRES.}
The High Resolution Echelle Spectrometer \citep[HIRES;][]{Vogt1994} is installed at the Keck I telescope in Hawai'i, USA. HIRES uses the iodine cell technique \citep{Butler1996} to obtain RV measurements with a typical precision of a few $\mathrm{m\,s^{-1}}$. We \Stephb{used archival} HIRES data for GJ~251 and Lalande~21185 to confirm the planetary signals and to extend the time baseline, and to search for long-period signals. For our analysis, we used the HIRES data corrected by \cite{Tal-Or2019}, which account for nightly zero-point offsets and an instrumental jump in 2004, which is an improvement over the original data reduction by \citet{Butler2017}. 
\Stephb{Details on the quality of the data are given in Table~\ref{Tab: datasets}.} \Stephe{The median exposure times for GJ~251 and Lalande~21185 were 500\,s and 135\,s, respectively.}

\paragraph{SOPHIE.}
We also used RV data for Lalande~21185 obtained with the SOPHIE instrument \citep{Perruchot2008}. \Stephb{These data were made public by \cite{Lalande2019}, and further information on the acquisition and properties of these data is provided in their study. \Stephb{We show a summary of the quality of the RV data in Table~\ref{Tab: datasets}.} }

\subsection{Photometry}

\Stephb{We carried out a contemporaneous photometric follow-up of GJ~251 and HD~238090 during 2018 and 2019. We also compiled photometric data publicly available as described below.}

\paragraph{T90.}  
We monitored GJ~251 and HD~238090 in the Johnson $V$ and $R$ bands with the T90 telescope at the Observatorio de Sierra Nevada (OSN) in Granada, Spain.
The T90 telescope is a 90\,cm Ritchie-Chr\'etien telescope equipped with a 2k\,$\times$\,2k pixel VersArray CCD camera, with a field of view of $13.2\times13.2$\,arcmin$^2$ \citep{Rodriguez2010}.
The observations of GJ~251 and HD~238090 were carried out on 42 nights from October 2018 to February 2019 and on 53 nights from February 2019 to July 2019, respectively.
The typical number of exposures per night and target was around 35.
We did not apply any binning, corrected each CCD frame in a standard way for bias and flat-fielding \Stephc{with IRAF}, and selected the best aperture sizes and reference stars for the synthetic aperture photometry.
\Stephc{In particular, we used the same aperture size as in \cite{Perger2019}}.

\paragraph{TJO.}
Observations of GJ~251 and HD~238090 with the 80\,cm Telescopi Joan Oró (TJO) at Observatori Astron\`omic del Montsec in Lleida, Spain, were conducted using a Johnson $R$ filter and its main imaging camera LAIA, a 4k\,$\times$\,4k back-illuminated CCD with a pixel scale of 0.4\,arcsec and a field of view of $30\times30$\,arcmin$^2$.
The TJO data for GJ~251 were collected between February and November 2019 during 157 nights and for HD~238090 between February and November 2019 during 149 nights. 
We obtained several \Steph{batches} of five images per night. The images were calibrated with darks, bias, and flat fields with the {\tt icat} pipeline \citep{Colome2006}. Differential photometry was extracted with {\tt AstroImageJ} \citep{Collins2017} using the aperture size and the set of comparison stars that minimized the root mean square (rms) of the photometry. Data with low S/N due to bad weather conditions or high airmass were removed. \Stephc{For GJ~251, we removed 392 low S/N measurements from the initial 2746 data points and for GJ458A, we removed 477 from the initial dataset of 6207 measurements. These correspond to the measurements for which the S/N of the target is below 30\% of the best measurement.} The resulting light curves were binned to one data point per hour.

\paragraph{LCO.}
We observed GJ\,251 on 44 epochs using the 40\,cm telescopes of Las Cumbres Observatory (LCO) in the $V$ band at the Teide, Haleakal{\=a}, and McDonald observatories between 13 January and 3 March 2019. 
The telescopes are equipped with a 3k\,$\times$\,2k SBIG CCD camera 
with a pixel scale of 0.571\,arcsec, providing a field of view of 29.2\,$\times$\,19.5\,arcmin$^2$. 
We acquired 50 individual exposures of 30\,s per epoch. 
Weather conditions were mostly clear, and the average seeing varied from 1.5\,arcsec to 3.0\,arcsec.
Raw data were processed using the {\tt banzai} pipeline \citep{McCully2018} \footnote{\url{https://banzai.readthedocs.io/en/latest/}}, which includes bad pixel, bias, dark, and flat-field corrections for each individual night. 
We performed aperture photometry for GJ\,251 and three reference stars in the field and obtained the relative differential photometry. We adopted an aperture of 13 pixels (7.4\,arcsec), which minimized the dispersion of the differential light curve. 

\paragraph{{\em TESS}.}
\Stephb{The Transiting Exoplanet Survey Satellite ({\em TESS}) is a space-borne instrument that searches for transiting planets around nearby stars \citep{Ricker2015}. The primary mission goal consists of observations of 26 sectors with 24$\times$96 deg$^2$ in the northern and southern hemisphere, which are still ongoing. Each sector is observed for about 28\,d. We obtained for all three targets of this work the pre-search data conditioning simple aperture photometry (PDCSAP) light curves. These are provided by the Science Processing Operations Center (SPOC; \citealt{Jenkins2016}) at the Mikulski Archive for Space Telescopes (MAST)\footnote{\url{https://mast.stsci.edu/portal/Mashup/Clients/Mast/Portal.html}}.}

\paragraph{MEarth.}
We used data of HD~238090 from the seventh data release (DR7\footnote{\url{https://www.cfa.harvard.edu/MEarth/DataDR7.html}}) of the MEarth project \citep{Berta2012}. 
The MEarth project is an all-sky transit survey that has been conducted since 2008.
It consists of 16 robotic 40\,cm telescopes, 8 located in the Northern Hemisphere at the Fred Lawrence Whipple Observatory in Arizona, USA, and the other 8 in the Southern Hemisphere located at Cerro Tololo Inter-American Observatory, Chile. The project monitors several thousand nearby mid- and late-M dwarfs over the whole sky. Each telescope is equipped with a 2k$\times$2k CCD that provides a field of view of 26$\times$26\,arcmin$^2$. 
MEarth uses an $RG715$\footnote{\url{https://www.pgo-online.com/intl/curves/optical_glassfilters/RG715_RG9_RG780_RG830_850.html}} long-pass filter, except for the 2010--2011 season, when an $I_{715-895}$ interference filter was chosen. 

\paragraph{NSVS.}
The Northern Sky Variability Survey \citep[NSVS]{NSVS2004} was a robotic survey that primarily targeted the northern sky with telephoto lenses located at the Los \'Alamos National Laboratory in New Mexico, USA. The survey provided data for 14 million objects in the magnitude range between 8\,mag to 15.5\,mag. For details on the instrumental setup and the conducted observations, we refer to the survey paper \citep{NSVS2004}. We used public NSVS data for HD~238090, which we obtained from their public webpage\footnote{\url{https://skydot.lanl.gov/nsvs/nsvs.php}}.

\paragraph{SuperWASP.}
For GJ~251 we used public data \Stephc{processed and} collected by the Wide Angle Search for Planets (WASP) survey \citep{superWASP}\Stephc{\footnote{\url{https://wasp.cerit-sc.cz}}}, in particular SuperWASP-North at the Observatorio del Roque de los Muchachos in La Palma, Spain. \Steph{SuperWASP-North consisted of one wide-field array of eight cameras, each with a 200\,mm, f/1.8 lens, a broadband filter spanning the wavelength range between 400\,nm and 700\,nm, and a 2k\,$\times$\,2k CCD. The resulting plate scale was 13.7\,arcsec\,pixel$^{-1}$.}

\section{Methods}
\label{Sect: Methods}

\subsection{Periodograms}

We used generalized Lomb-Scargle (GLS) periodograms \citep{Zechmeister2009} to assess significant periodicities in the photometric and spectroscopic data. We applied the normalization as given in  \cite{Zechmeister2009}, which is abbreviated as $P_{\tx{ZK}}$ throughout this work. For each periodogram, we computed false-alarm probabilities (FAPs) by applying bootstrapping with $n=10\,000$ iterations. Our detection threshold for a signal deemed to be significant was at an $\mathrm{FAP} < 0.001$. The uncertainties on the periods of significant GLS signals were estimated from the local $\chi^2$ curvature by the GLS routine.

To assess the coherence of a periodic signal over the observation time, we used the stacked-Bayesian GLS periodogram \citep[s-BGLS;][]{Mortier2015, Mortier2017}. The Bayesian GLS periodogram allows the comparison of probabilities of periodic signals in the data, while the stacking examines the coherence of the signal with an increasing number of observations. As in \cite{Mortier2017}, we normalized all s-BGLS periodograms to their respective minimum values, which means that the probability of each signal and its growth or decrease over time is a relative measure compared to the lowest probability obtained within one calculated s-BGLS over a specific period range.

\begin{table*}
\caption{Stellar parameters of HD~238090, GJ~251, and Lalande~21185.}
\label{Tab: stellar_parameters}
\centering
\begin{tabular}{l c c c r}
\hline\hline
Parameter & GJ~251& HD~238090&  Lalande~21185 & Ref. \Stephc{(GJ~251/HD~238090/Lalande~21185)} \\
\hline
\noalign{\smallskip}
\multicolumn{5}{c}{\em Identifiers}\\
\noalign{\smallskip}
Gliese-Jahreiß & GJ 251 & \object{GJ 458~A} &  \object{GJ 411} & Gli79 \\
Karmn &J06548+332 &J12123+544S &  J11033+359 & Cab16\\
\noalign{\smallskip}
\multicolumn{5}{c}{ \em Coordinates and spectral type}\\
\noalign{\smallskip}
Epoch& J2015.5& J2015.5& J2000.0 & \emph{Gaia} DR2\,/\,\emph{Gaia} DR2\,/\,vLe07\\
$\alpha$&$06\,54\, 48.06$ & $12\, 12\, 21.27$ &  $      11\, 03\, 20.19$ & \emph{Gaia} DR2\,/\,\emph{Gaia} DR2\,/\,vLe07\\
$\delta$& $+33\, 15\, 59.3$& $+54\, 29\, 10.2$ &  $+35\, 58\, 11.6$ & \emph{Gaia} DR2\,/\,\emph{Gaia} DR2\,/\,vLe07\\
Sp. type & M3.0\,V & M0.0\,V &  M1.5\,V &Alo15\,/\,PMSU\,/\,Alo15 \\
$G$\,[mag]&  $8.8552\pm0.0011$&$9.0379\pm0.0005$& \ldots &\emph{Gaia} DR2\\
$J$\,[mag]&  $6.10\pm0.02$& $6.88\pm0.02$ & $4.20\pm0.24$ & 2MASS\\
\noalign{\smallskip}
\multicolumn{5}{c}{ \em Parallax and kinematics}\\
\noalign{\smallskip}
$\mu_{\alpha}\cos\delta$\,[mas/yr]& $-726.39\pm0.13$& $+232.38\pm0.04$ &  $-580.27\pm0.62$ &\emph{Gaia} DR2\,/\,\emph{Gaia} DR2\,/\,vLe07 \\
$\mu_\delta$\,[mas/yr]& $-398.13\pm0.12$& $+92.09\pm0.04$ &  $-4765.85\pm0.64$ &\emph{Gaia} DR2\,/\,\emph{Gaia} DR2\,/\,vLe07\\
$\pi$\,[mas]& $179.16\pm0.06$& $65.61\pm0.03$ & $392.64\pm0.67$ &\emph{Gaia} DR2\,/\,\emph{Gaia} DR2\,/\,vLe07 \\
$d$\,[pc]& $5.581\pm0.002$ & $15.24\pm0.01$ &  $2.547\pm0.004$ &\emph{Gaia} DR2\,/\,\emph{Gaia} DR2\,/\,vLe07 \\
$\gamma$\,[km/s] & \Stephc{$22.654\pm0.025$} & \Stephc{$-17.668\pm0.018$} & \Stephc{$-85.016\pm0.023$} & \Stephc{Laf19}\\ 
$U$\,[km/s]& \Stephc{$-27.41\pm0.02$ }& \Stephc{$18.08\pm0.01$}&  \Stephc{$46.29\pm0.03$}  & This work \\
$V$\,[km/s]& \Stephc{$-3.67\pm0.01$} & \Stephc{$7.41\pm0.01$} &  \Stephc{$-53.68\pm0.09$ } & This work \\
$W$\,[km/s]& \Stephc{$-15.13\pm0.01$}&\Stephc{$-16.01\pm0.02$} & \Stephc{$-74.59\pm0.02$  }& This work \\
\noalign{\smallskip}
\multicolumn{5}{c}{\em Photospheric parameters}\\
\noalign{\smallskip}
$T_\tx{eff}$\,[K]&$3451\pm51$ & $3933\pm51$ &  $3601\pm51$ &Sch19 \\
$\log{g}$\,[dex]&$4.96\pm0.07$ &  $4.70\pm0.07$ &$4.87\pm0.07$ &Sch19 \\
$[\tx{Fe/H}]$\,[dex]& $-0.03\pm0.16$ & $-0.03\pm0.16$ & $-0.09\pm0.16$ & Sch19 \\
\noalign{\smallskip}
\multicolumn{5}{c}{\em Physical parameters}\\
\noalign{\smallskip}
$L$\,[$L_\odot$]& $0.0169\pm0.0003$ & $0.0702\pm0.0015$ & $0.0195\pm0.0013$ & Sch19 \\
$R$\,[$R_\odot$]& $0.364\pm0.011$  &  $0.570\pm0.016$& $0.392\pm0.004$ &Sch19\,/Sch19\,/Boy12 \\
$M$\,[$M_\odot$]& $0.360\pm0.015$ & $0.578\pm0.021$ & $0.390\pm0.011$ &Sch19\,/Sch19\,/This work \\
\multicolumn{5}{c}{\Steph{ \em Activity parameters}}\\
pEW (H$\alpha$) [\AA] & $0.00\pm0.01$ & $+0.04\pm0.01$ & $-0.04\pm0.01$ & Schf19 \\
$v\sin{i}$\,[km/s]& $< 2$ & $< 2$ & $< 2$ & Rei18 \\
$P_\mathrm{rot}$ [d] & $122.1^{+1.9}_{-2.2}$  & $96.7^{+3.7}_{-3.2}$ & $56.15\pm0.27$ & This work\,/This work\,/Dia19 \\
\noalign{\smallskip}
\hline
\end{tabular}
\tablebib{
    2MASS: \citet{2MASS};
    Alo15: \citet{Alo15};
    Cab16: \citet{Cab16};
    Gli79: \citet{Gliese};
    \emph{Gaia} DR2: \citet{Gaia};
    PMSU: \citet{PMSU96};
    Sch19: \citet{Schweitzer2019};
    vLe07: \citet{vanLeeuwen2007};
    Boy12: \citet{Boyes2012};
    Dia19: \citet{Lalande2019};
    Schf19: \citet{Schoefer2019};
    Laf19: \citet{Lafarga2020}.
}
\end{table*}

\subsection{Modeling of RV and photometric data}

For the modeling, we used \texttt{juliet} \citep{Espinoza2018}, which allows the fitting of photometric and RV data \Steph{by searching for the global posterior maximum based on the evaluation of the Bayesian log-evidence} \Stephc{($\ln{\mathcal{Z}}$)} within a provided prior volume of the fitting parameters. \texttt{juliet} allows us to statistically compare models with different numbers of parameters within a Bayesian framework through the log-evidence, which includes the model complexity and the number of degrees of freedom within its assessment. 
Following \citet{Trotta2008}, a model is considered as a significant improvement if \Stephc{$\Delta \ln{\mathcal{Z}}>5$}.
The \texttt{juliet} calculation of the log-evidence is conducted with nested sampling algorithms. In particular, we used the dynamic nested sampling algorithm \texttt{dynesty} \citep{Speagle2019}. 

We used \texttt{radvel} \citep{radvel} to model Keplerian RV signals, and \texttt{george} \citep{george} for GP modeling of both photometric and RV data.
In all cases, we used an exp-sin-squared kernel multiplied with a squared-exponential kernel, which is included as a default kernel within \texttt{juliet}. This kernel, also known as the quasi-periodic (QP) kernel, has the form
\begin{equation}
\label{Eq: kernel}
k(\tau)=\sigma^2_{\textnormal{GP}}\exp{(-\alpha_\textnormal{GP}\tau^2-\Gamma\sin^2{(\pi\tau/P_{\tx{rot}}})),}    
\end{equation}
where $\sigma_{\tx{GP}}$ is the amplitude of the GP component given in parts per million (ppm) for photometric data or $\mathrm{m\,s^{-1}}$ for RV data, $\Gamma$ is the amplitude of the GP sine-squared component \Stephb{and is dimensionless}, $\alpha$ is the inverse length-scale of the GP exponential component given in d$^{-2}$, $P_{\tx{rot}}$ the period of the GP QP component given in d, and $\tau$ %= |t_{i} - t_{j}|$ 
is the time lag. 
This choice of kernel represents one part of our prior knowledge, as it provides the framework of how an effective model of stellar activity should fit the data. 
The timescale \Stephe{$P_{\tx{dec}}$} in days of the exponential decay can be approximated with 
\begin{equation}
P_{\tx{dec}}=(2\alpha_{\rm GP})^{-1/2}.
\end{equation}

The $\alpha_{\tx{GP}}$ parameter is of particular interest with regard to the stability of a QP signal. 
A smaller $\alpha$ describes a more stable periodic signal in which data points are more strongly correlated with each other. For a review and a detailed description of each kernel hyperparameter and a possible physical interpretation, we refer to \cite{Angus2018}.

\Stephb{The evaluation of the GP likelihood with \texttt{george} is computationally expensive and scales as $N \ln{N}$, where $N$ is the number of data points \citep{george}. 
For the derivation of the stellar rotation, we searched for periods on timescales of days. To do this, it is reasonable to create nightly bins of the photometric data. 
This reduces the computation time of the GP log-likelihood evaluation and short-term variations based on the jitter of the star.}

\Stephb{
For the photometric analysis, we applied distinct GP hyperparameters for the amplitudes $\sigma_{{\rm GP}}$ and $\Gamma$, to account for the effect that stellar activity depends on wavelength, but we used global GP hyperparameters for the timescale of the amplitude modulation and the rotation period.
In addition, we fit an offset and a jitter term (in quadrature to the diagonal of the resulting covariance matrix of the GP) for each data set. Table~\ref{Tab: Priors_photometry} shows the priors of the photometric GP analysis.}

\Stephc{
For the final RV analysis, we applied global GP hyperparameters. A statistical comparison with models using distinct GP hyperparameters for each RV instrument did not show any significant improvement in log-evidence.
For each data set, we fit an offset and a jitter term. Our priors for the RV GP analysis are provided in Table~\ref{Tab: Priors_GP_RV}}.

\subsection{De-aliasing}

\Stephc{Aliases are spurious signals caused by the sampling of the data, which are often indistinguishable from the true signal.}
The significance of a signal or the goodness of a fit is not a sufficient criterion to differentiate between true signals and alias signals, especially in cases of non-optimal sampling, where the results of these metrics can be similar. \Stephc{For example, it is a common misconception that peaks close to one day are always alias frequencies, but a priori, it is not clear which of the peaks represents the true frequency of the signal and which represents the alias \citep[][]{Dawson2010}. 
Alias frequencies can be calculated by $f_a = f_t \pm m f_s$, 
where $f_t$ is the assumed true frequency, $f_s$ the sampling frequency, and $f_a$ the alias frequency. Because RV measurements are usually taken with a rather irregular sampling \citep{Garcia2017}, more than one sampling frequency is often apparent in the window function of the data. This results in several peaks at alias frequencies related to the different sampling frequencies, and can make it even harder to distinguish the true underlying signal}.

We used the \texttt{AliasFinder} \citep{Stock2020a}\footnote{\url{https://github.com/JonasKemmer/AliasFinder}} to confirm that the assumed planetary signal is the true signal and not an alias. The method on which the \texttt{AliasFinder} is based is described in \cite{Dawson2010}, \cite{Stock2020a}, and \cite{Stock2020b}. \Steph{For each frequency under consideration, \texttt{AliasFinder}  simulates} 1000 data sets based on the true sampling of the observed data and inserts one sinusoidal signal with one of the frequencies. \Stephb{\texttt{AliasFinder}  also includes} a noise contribution based on the RV jitter of the star, \Stephc{which we made use of for our analyses in this paper.}  We compared the resulting ensemble periodograms for each simulated frequency to the periodogram obtained from the observed data. Peak position, power, and phase are the parameters compared in this test. If the ensemble of the periodograms of one simulated frequency reproduces the data periodogram significantly better than the periodograms of the other simulated frequencies, then the most probable planetary period has successfully been identified.

\section{Stellar properties}
\label{Sect: Stellar_props}

Photospheric parameters, \Stephb{such as effective temperature, surface gravity, and metallicity,} were determined by \cite{Schweitzer2019} by fitting an updated set of PHOENIX-ACES atmosphere models \citep{Husser2013} to high-resolution CARMENES spectra. These updated PHOENIX models incorporated the latest solar abundances, molecular and atomic line lists, and a new equation of state \citep{Meyer2017}, which were especially designed to treat low-temperature stellar atmospheres. 
The parameters were determined assuming a rotational velocity of $v \sin{i}$ = 2\,km\,s$^{-1}$ \citep{Reiners2018b}. \Stephc{To reduce degeneracies between the parameters, \cite{Schweitzer2019} constrained the surface gravity $log{g}$ with the help of evolutionary models  \citep[PARSEC,][]{Bressan2012,Chen2014,Chen2015,Tang2014} and stellar ages estimated by \cite{Passegger2019}.
The actual ages of the three investigated stars are probably older than tabulated, as derived from a new kinematics analysis (Cort\'es-Contreras et al., in prep.).
The galactocentric space velocities in Table~\ref{Tab: stellar_parameters} were calculated from the latest {\em Hipparcos} and {\em Gaia} DR2 proper motions and parallaxes \citep{vanLeeuwen2007, Gaia} and absolute RVs of \cite{Lafarga2020}, following the approach of \cite{Montes2001} and \cite{cortes2016}.}

Physical parameters, \Stephb{such as luminosity, radius, and mass,} were derived by \cite{Schweitzer2019}.
\Stephc{\cite{Cifuentes2020} exhaustively described the luminosity determination in M dwarfs}.
The radius (and hence mass) for Lalande~21185 was an outlier in \cite{Schweitzer2019} because the photometry for this star was of low quality, suggesting an uncertain and too low luminosity. \Steph{They derived a stellar radius and mass of} $0.3587\pm0.0157\,R_\odot$ and $0.355\pm0.019\,M_\odot$, respectively.
We therefore used a slightly different approach to derive the mass and radius. \Stephb{We used its radius $0.3921\pm0.0037\,R_\odot$, which was derived by \cite{Boyes2012} using the interferometric angular diameter.} \Stephb{Applying the} same empirical mass-radius relationship as was used for the other two targets, we derived a stellar mass of $0.390\pm0.011\,M_\odot$, \Steph{which agrees better with the typical parameters of the ensemble.} The detailed stellar parameters of all three stars and their references are given in Table~\ref{Tab: stellar_parameters}. 

\Steph{The M0.0\,V star HD~238090 is the primary component of a wide binary. The secondary component is the M3.0\,V star \object{GJ\,458~B}, with a stellar mass of $0.230\pm0.005$\,$M_\odot$. The angular separation of the two components of $14.68\pm0.44$\,arcsec \citep{cortes2016} results in a projected separation of approximately $224$\,au.}
We computed the stellar mass of the secondary component using the mass-luminosity-metallicity relation of \cite{Mann2019}. Based on the masses and projected minimum separation, we estimated the minimum orbital period of this binary to be longer than 3700\,yr. This long binary period agrees with the 34 observations between 1955 and 2015 tabulated in the Washington Double Star Catalog \citep{Mason2001}, which do not indicate any change in the position angle.

\begin{figure}
\centering
\includegraphics[width=0.49\textwidth]{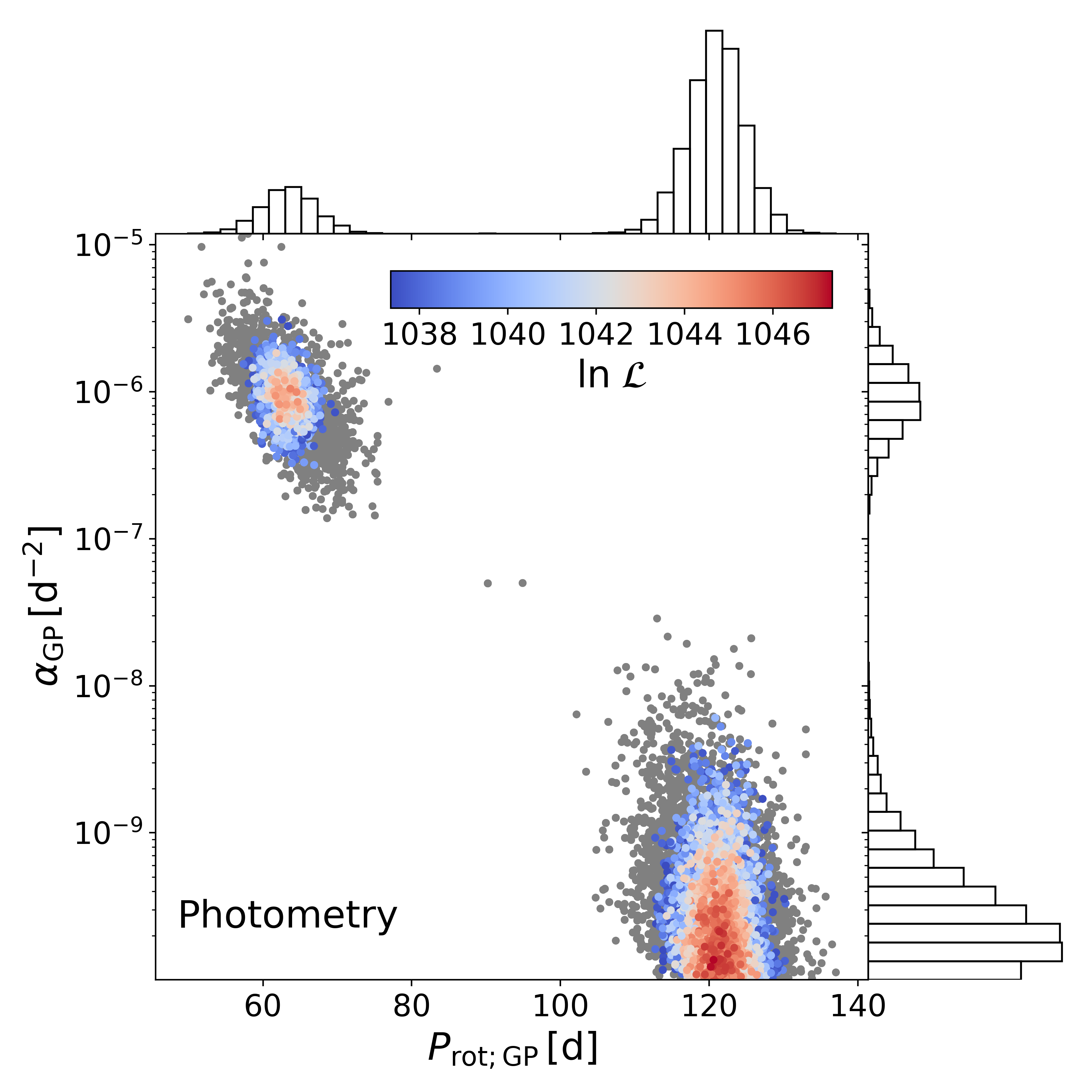}
\caption{Posterior distribution in the $\alpha_{\rm GP}$ vs. $P_{\rm rot}$ plane of the GP fit to the combined photometric data of T90, TJO, LCO, and SuperWASP for GJ~251. The color-coding shows the log-likelihood normalized to the highest \Stephb{achieved log-likelihood} value within the posterior sample. Gray samples indicate solutions with \Steph{$\Delta \ln{L}$ lower than 10.} }
\label{Fig: GP_GJ251}
\end{figure}

\section{GJ 251}
\label{Sect: GJ251}
\subsection{Photometric monitoring}
\label{SubSect: GJ251_photometry}
For a significant fraction of the CARMENES RV observations of GJ~251, we obtained quasi-simultaneous photometry with the T90, TJO, and LCO telescopes. We combined these data with public data from SuperWASP. 
A joint GLS periodogram analysis, where we fit for offsets and jitter of each data set, indicated significant signals at periods of 30\,d, 70\,d, and 120\,d. However, a sinusoidal model, as used in the GLS analysis, is an imperfect description of stellar activity, which is often better represented by a QP signal. Therefore we fit a more sophisticated model to the photometric data in the form of a QP GP to derive the stellar rotation period.

\begin{figure}
\centering
\includegraphics[width=9cm]{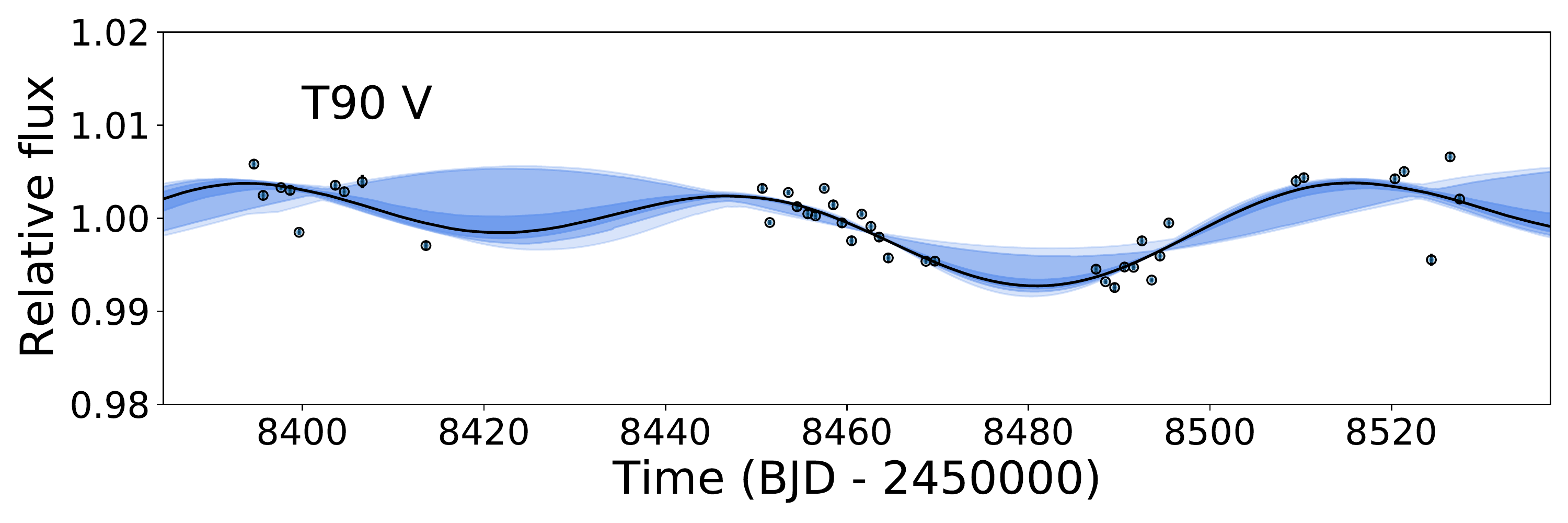}

\includegraphics[width=9cm]{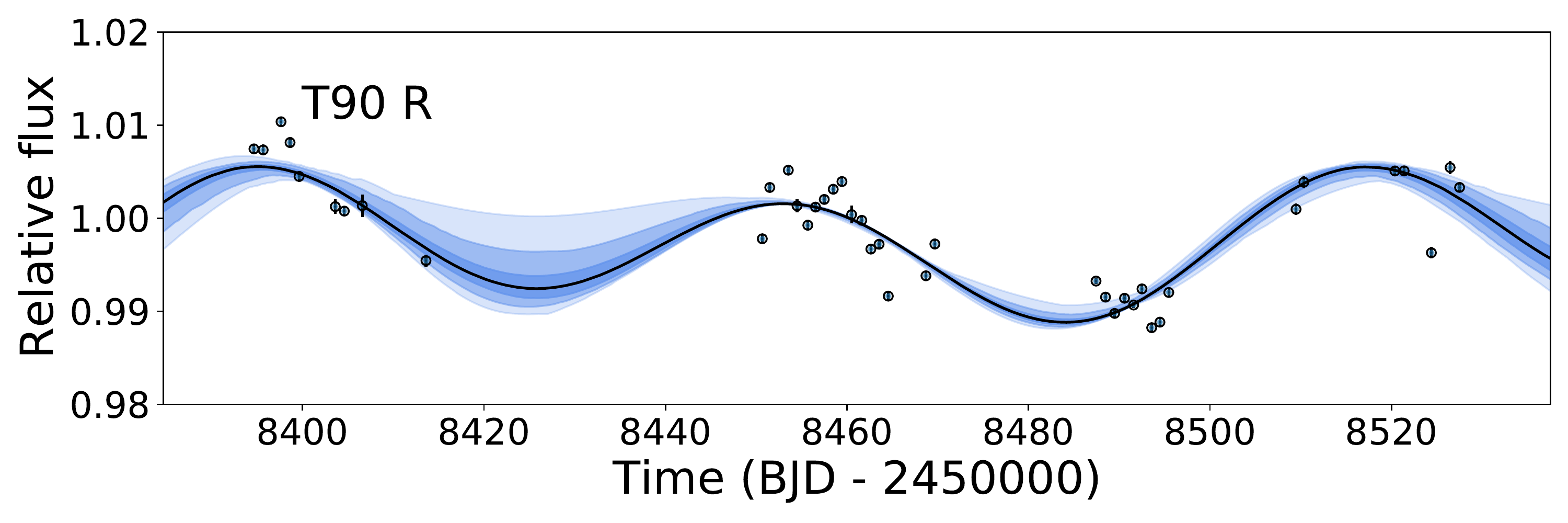}

\includegraphics[width=9cm]{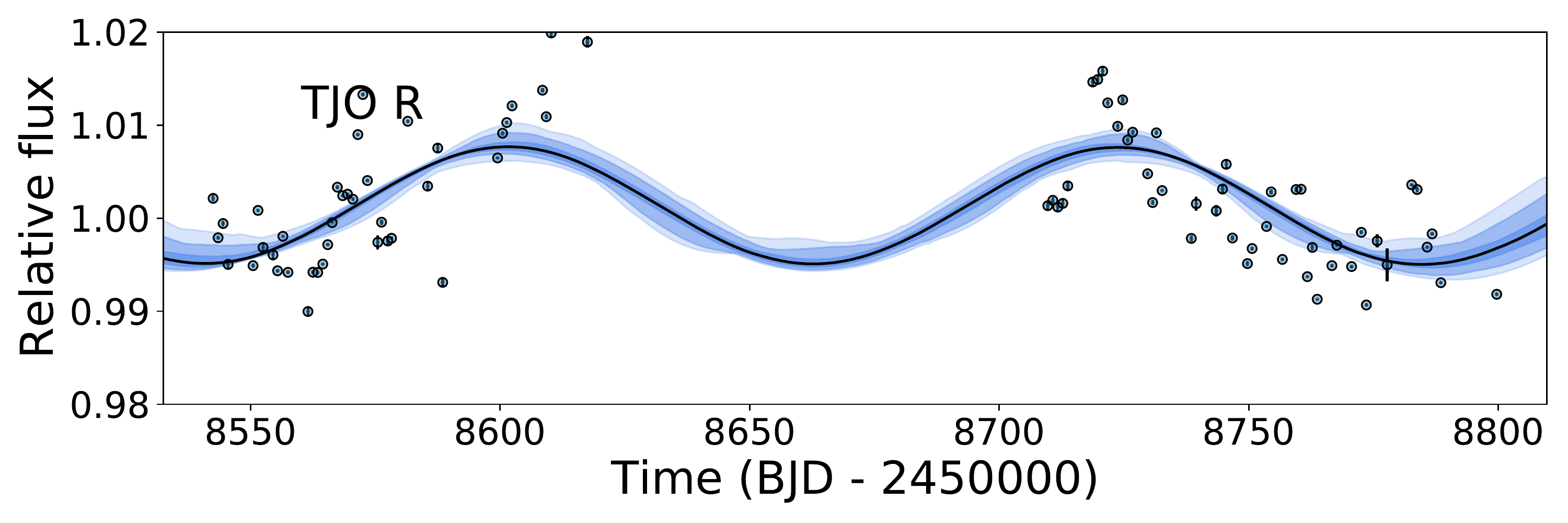}

\includegraphics[width=9cm]{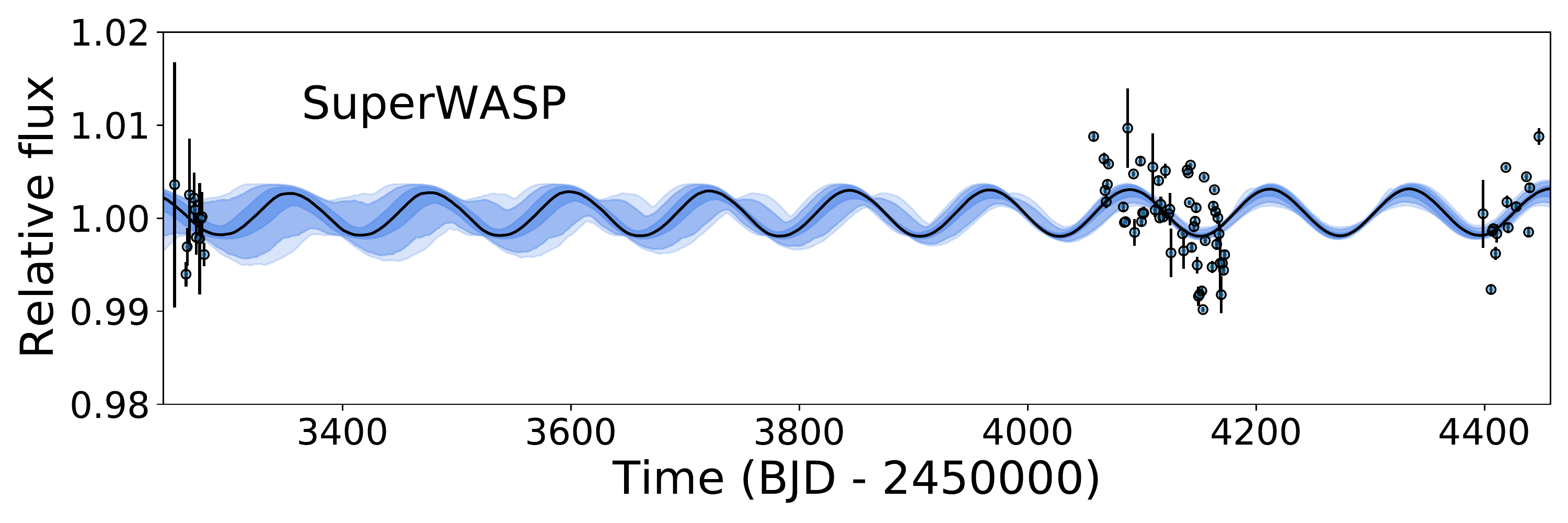}
\caption{Joint GP model of the nightly binned photometric data of GJ~251. From top to bottom: T90 $V$, T90 $R$, TJO $R$, and SuperWASP.} 
\label{Fig: GJ251_photometric_GP}
\end{figure}

\begin{figure*}[!ht]
\centering
\includegraphics[width=18cm]{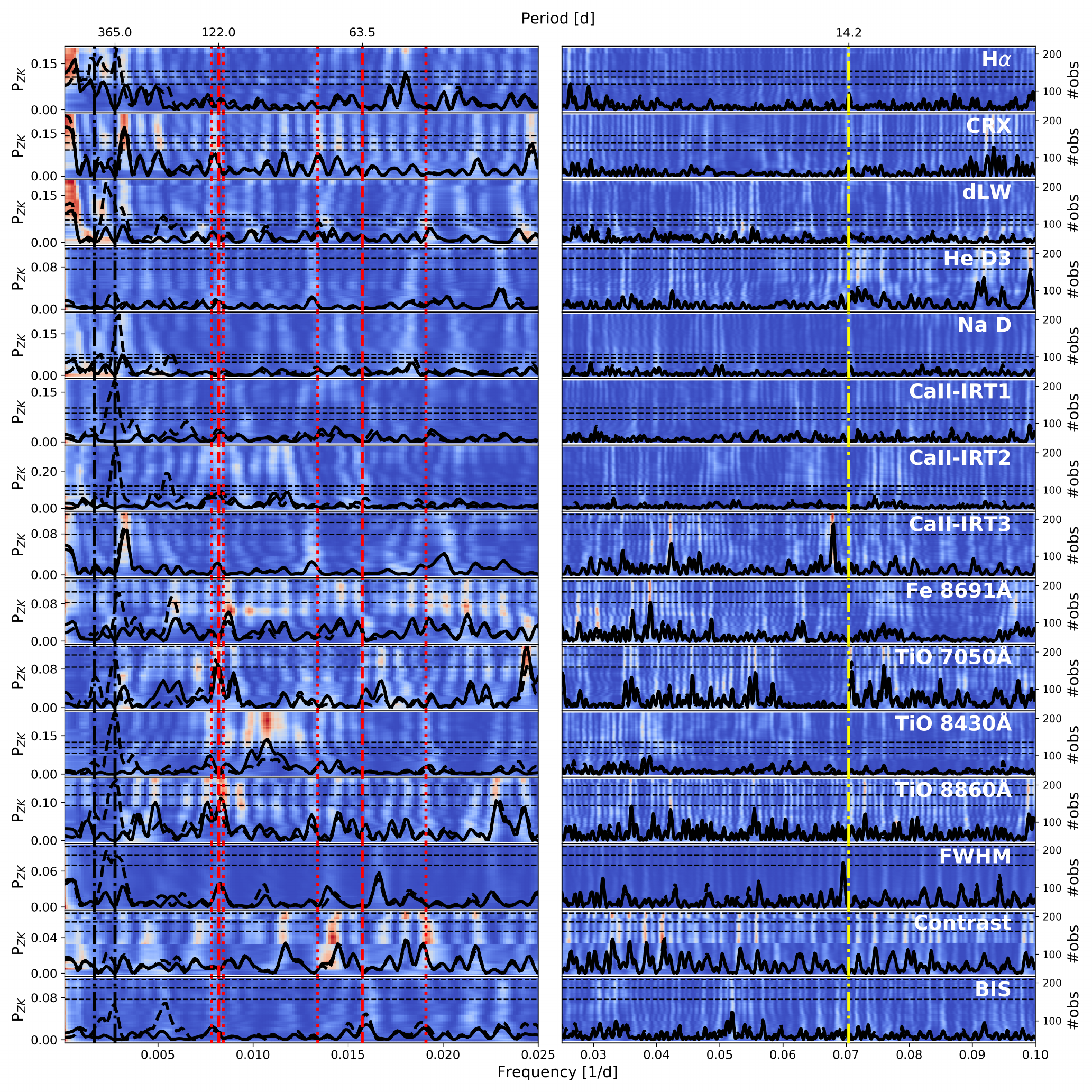}
\caption{Generalized Lomb-Scargle periodograms of several activity indicators of GJ~251 from CARMENES spectroscopic data. The dashed black periodograms represent the GLS of the activity indicators, and the solid GLS periodogram represents the residuals after subtracting a 365\,d signal. For the residuals from which the 365\,d signal was subtracted, we also overplot the s-BGLS periodogram, \Stephc{where the probability increases from blue to white to red}. The red dashed lines mark the rotation period and the first harmonic estimated from photometric data, while the dotted red lines show the 3$\sigma$ uncertainties. The dashed black line marks a significant HIRES signal around 600\,d, and the yearly period of 365\,d. The dashed yellow line marks the period of the planetary signal published in this work.}
\label{Fig: GJ251_GLS_act}
\end{figure*}

\begin{figure}
    \centering
    \includegraphics[width=9cm]{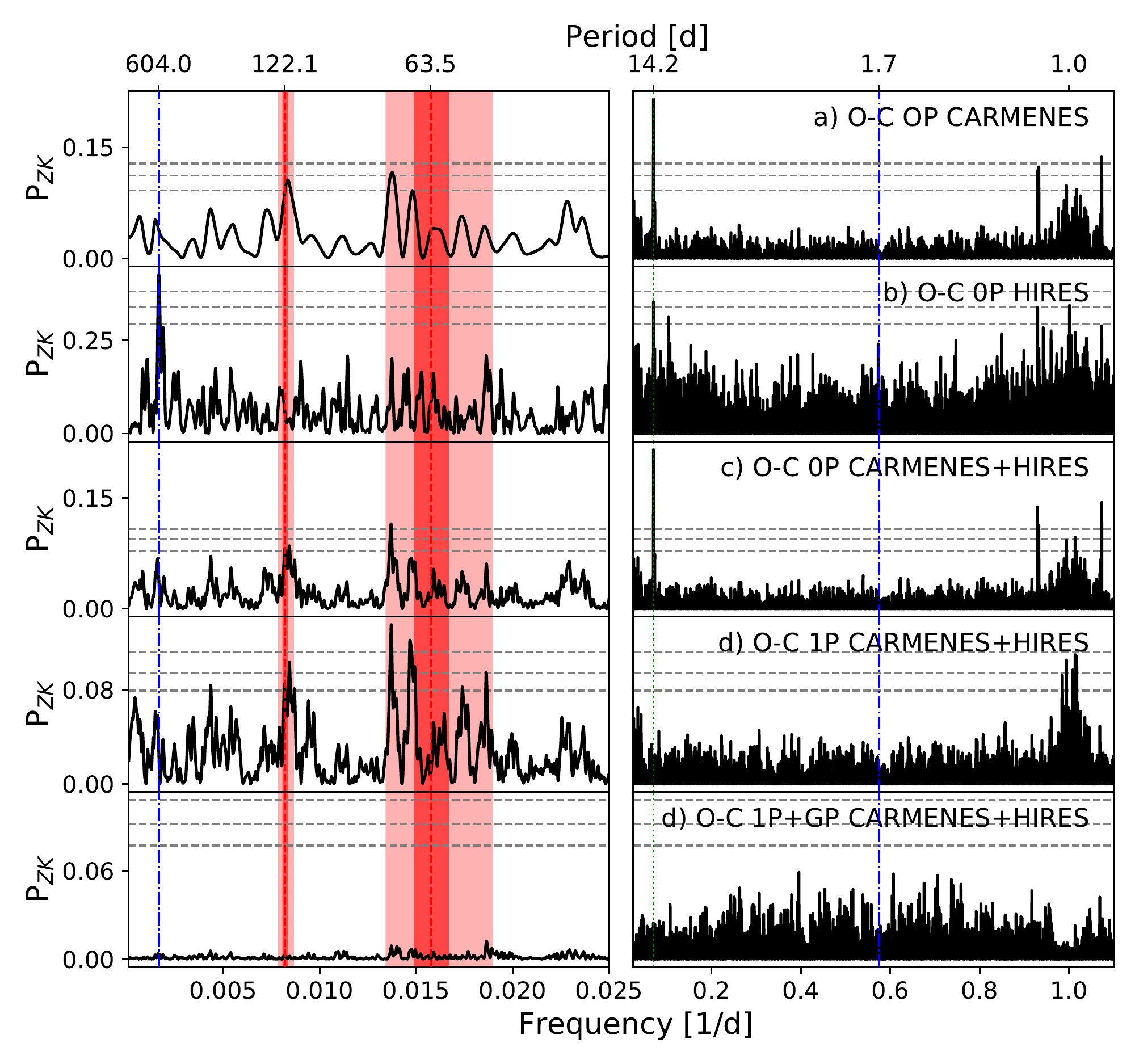}
    \caption{Generalized Lomb-Scargle periodograms \Stephc{of RV data} for GJ~251 for CARMENES, HIRES, and a combination of both. The stellar rotational period derived by photometry is plotted as the dashed red line, and the $1\sigma$ and $3\sigma$ uncertainties are highlighted in red. We also indicate the harmonic of the rotational period and its uncertainty. \Steph{The green line marks the suspected planetary signal.} The blue lines mark the periods of the published planetary candidates by \cite{Butler2017} at 1.7\,d, and the significant HIRES signal around 604\,d.}
    \label{fig: GJ251_GLS_rv}
\end{figure}

\begin{figure*}
    \centering
    \includegraphics[width=4.8cm]{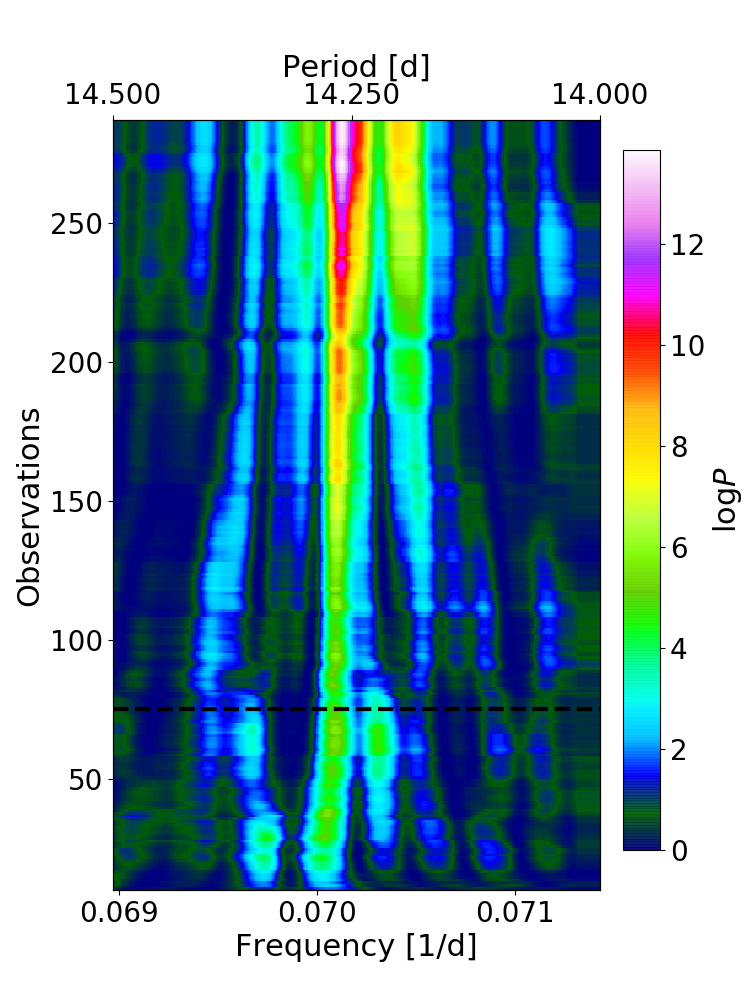}
    \includegraphics[width=13.2 cm]{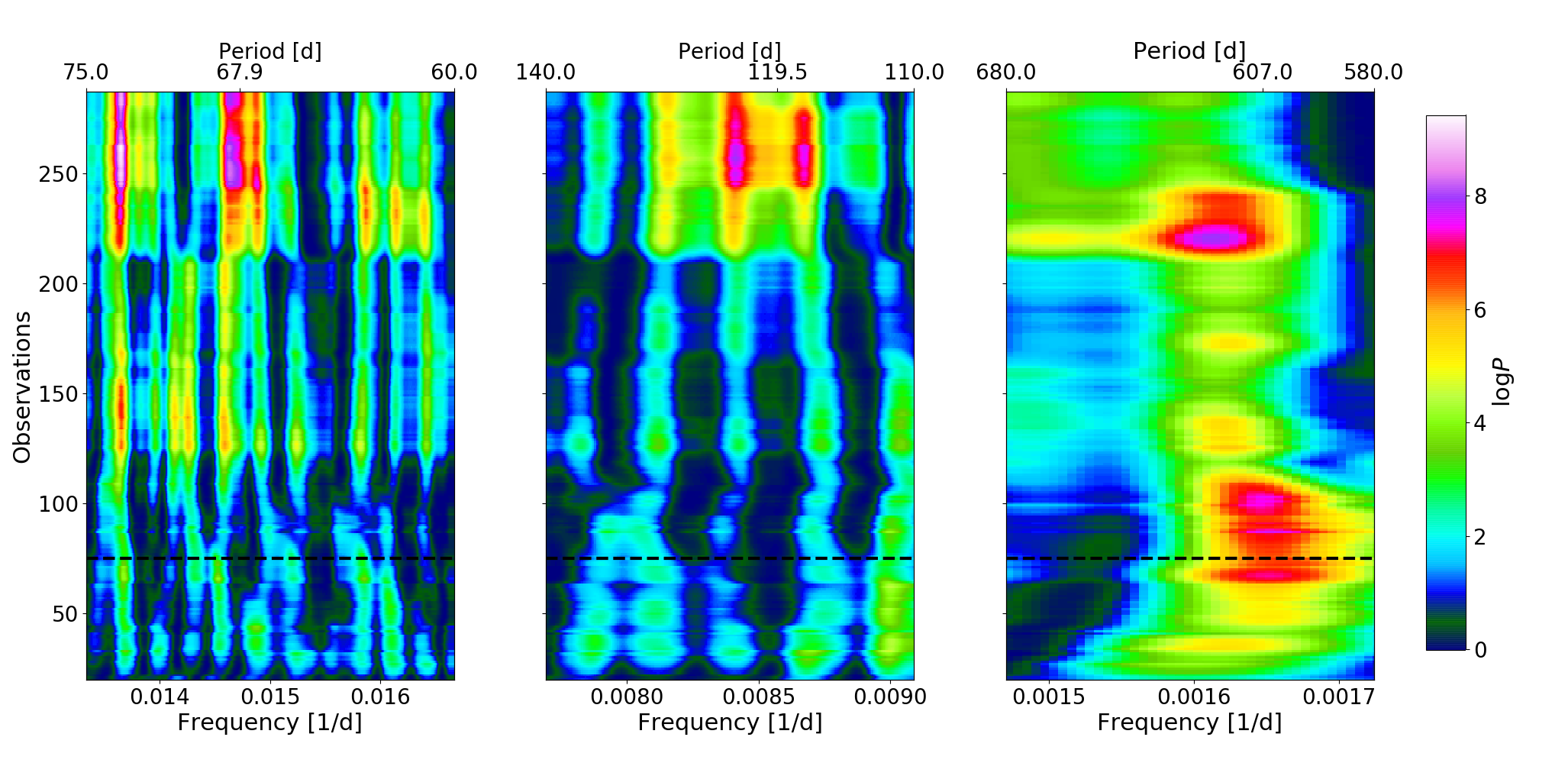}
    \caption{Stacked-Bayesian GLS periodogram of the planetary signal at 14.22\,d and on the zero-planet residuals (left diagram with its own normalization). The three s-BGLS on the right were calculated using the one-planet Keplerian RV residuals and show signals that we attributed to stellar activity: the forest of signals between 60\,d and 75\,d corresponding to roughly half of the rotation period (left), the RV signals around the photometrically derived rotation period at 119.5\,d (middle), and the long period signal around 600\,d. In all four plots, the dashed black line indicates the boundary between HIRES and CARMENES data, which were taken successively.}
    \label{fig: GJ251_BGLS}
\end{figure*}

Our GP analysis of the photometry of GJ~251 based on our T90, TJO, LCO, and SuperWASP data resulted in a bimodal distribution for the rotational period with \Stephc{posterior solutions} around 120\,d and 60\,d. 
%,where the latter peak probably represents the first harmonic of the former. 
We plot the informative GP $\alpha$-period diagram ($\alpha_{\textnormal{GP}}$ versus\ $P_{\textnormal{GP}}$) in Fig.~\ref{Fig: GP_GJ251}. This plane of parameters \Stephc{shows the decay-timescale over the rotation period, and it} is useful for identifying whether \Stephc{stronger} correlated noise (small $\alpha$) favors a certain periodicity (see also
\citealt{Stock2020b} for a more detailed explanation). 
\Stephc{Within this plane, we identified that the likelihood and number of posterior samples at 120\,d is higher than that of the posterior samples around 60\,d. 
Furthermore, the $\alpha$ values of the 120\,d signal converge toward our prior boundary of $10^{-10}$\,d$^{-2}$, representing a decay timescale longer than 70\,000\,d, which indicates a stable periodic signal over the entire time of observations. 
We determined the rotational period for each posterior solution and derived $63.5^{+3.7}_{-3.6}$\,d and $P_{\text{rot,phot.}}=122.1^{+1.9}_{-2.2}$\,d}. 

The latter we regard formally as the derived rotational period of GJ~251 because\Stephc{ on average, its likelihood of posterior samples is higher than the former solution, because of the stronger coherence of the signal, and because the 60\,d signal can be explained as the first harmonic of a signal with a fundamental period of about 120\,d. A stronger coherence of signals related to stellar activity would be expected for M dwarfs because the spot lifetime increases with decreasing effective temperature \citep{Giles2017, Shapiro2020}. Additionally, if GJ~251 is a slowly rotating star, which means that it is
relatively inactive,
%less active, % "less active than what?"
there is evidence that faculae, which are in general longer-lived then starspots, are dominant surface features \citep[see][and references there]{Shapiro2020}. These long-lived faculae, in particular, affect the photometric variability of the star \citep{Rheinhold2019} and less so the RVs, which are typically spot-dominated. For this reason, among others, the same decay timescales $\alpha_{\textnormal{GP}}$ of the rotational signal between RV and photometric data should not be assumed.}
We show the binned photometric data overplotted with the median GP model and its uncertainties in Fig.~\ref{Fig: GJ251_photometric_GP}.

Based on our estimate of the stellar rotation period, we used Eqs. 1 and 2 by \cite{Suarez2018} to estimate $\tx{log}(R'_{HK})$ and based on this, the expected RV semiamplitude of the stellar rotational signal. By propagating all uncertainties of the parameters given by \cite{Suarez2018} for M0-M3 stars and our measurement uncertainty of the rotation period (in the form of their actual distributions), we derived for the medians and $1\sigma$ uncertainties $\tx{log}(R'_{HK})=-5.79^{+0.53}_{-0.61}$ (mean at $-5.83$) and $K_{\tx{exp.}}=0.68^{+3.71}_{-0.58}\,\mathrm{m\,s^{-1}}$ (mean at 3.59\,$\mathrm{m\,s^{-1}}$).

\subsection{Spectroscopic activity indicators} 
\label{SubSect: GJ251_activity}

We analyzed a number of spectral activity indicators for GJ~251 obtained from the CARMENES spectra using the indicators provided by {\tt serval}, which includes the chromatic index and the differential line width \citep[CRX and dLW, see ][]{Zechmeister2018}. We also investigated the cross-correlation function \citep[CCF, see ][]{Lafarga2020, Reiners2018A} \Stephb{to derive} the full width at half maximum (FWHM), contrast (CON), and bisector span (BIS), \Stephc{and we derived a large number of additional indicators \citep[see][]{Schoefer2019}.} We searched for periodicities of all these indicators using the GLS periodogram. Many indicators show significant long-term signals around 365\,d, and its 1\,d aliases. \Stephc{ The occurrence of this period in activity indicators of several other stars of our survey, in particular, of the other two targets discussed in this work, and the fact that it is compatible with one yearly cycle, makes it unlikely that stellar activity is the origin. This periodicity might be caused by small yearly environmental changes on the instrument or micro-tellurics that might affect the spectral line shapes to which the CCF and the measured pEW's are more sensitive than the actual RV measurements.}

\Stephc{Because this yearly signal is not believed to be of stellar activity, and most importantly, because is far away from the planetary periods and derived stellar rotational period, we subtracted it so that we would be more sensitive to periods in the high-frequency regime}. We show the residual GLS periodogram and its s-BGLS periodogram in Fig.~\ref{Fig: GJ251_GLS_act}. 
We found a signal \Stephc{at 121.0\,d that is} within the 1$\sigma$ uncertainty of the photometric rotation period in TiO at 8430\,\AA\,with an $\mathrm{FAP}<10^{-2}$. Within the uncertainty of the first harmonic of the rotation period, we observed a peak in H$\alpha$ with an FAP reaching almost $10^{-3}$. From the s-BGLS, the star showed the strongest activity in most indicators at periods attributed to the stellar rotation, whether at 120\,d or 60\,d, between \Stephb{January 2019 and October 2019 (CARMENES observation numbers 130 to 180).

\Stephc{
Recently, 
signals at approximately 90\,d in TiO 8430\,\r{A} and around 45\,d in TiO 7050\,\r{A} have become significant. It is not clear where these signals originate. Recent works, for example, \Stephf{\cite{Shapiro2020} and \citet{Nava2020}}, have shown that the interplay of activity signals that is due to the distribution and different lifetimes of starspots and faculae on the stellar surface, may result in signals that cannot be directly attributed to the stellar rotation. However, we find a good agreement between our photometric results and spectroscopic results (see further down).}
The measured median pEW of the H$\alpha$ line is $+0.00\pm0.01$\,\r{A}, and indicates that GJ~215 is not a H$\alpha$ active star \citep{Jeffers.2018, Schoefer2019},} \Stephc{which is in line with the long rotational period derived for this star, as is the upper limit of $v \sin i < 2\,\mathrm{km\,s^{-1}}$ measured by \cite{Reiners2018}.}

\subsection{Periodogram analysis and RV modeling}
\label{SubSect: GJ251_periodograms}
We show the results of the periodogram analysis of the CARMENES RV data in Fig.~\ref{fig: GJ251_GLS_rv}. A significant peak with an $\mathrm{FAP}<10^{-7}$ is visible at $14.22\pm0.01$\,d with an amplitude of $2.13\pm0.23$\,m\,s$^{-1}$, as well as two additional peaks close to one day that we attributed to daily aliases \Steph{of the 14\,d period}. Although the absolute GLS power and FAPs of the suspected aliases were smaller than the frequency of the 14.22\,d signal, we used \texttt{AliasFinder} to verify that the 14.22\,d signal represents the most probable true signal, which we confirmed. Additional strong secondary signals were identified at 73\,d, and 119.5\,d, each with an FAP of about $10^{-2}$. 

We also performed an independent periodogram analysis of the HIRES data.
The strongest signal is at 604\,d with an FAP of almost $10^{-5}$, followed by one at 14.2\,d with $\mathrm{FAP}<10^{-2}$ (see Fig.~\ref{fig: GJ251_GLS_rv}). The latter is consistent with our strongest signal in the CARMENES data.

\begin{table}
\caption{Bayesian log-evidence for GJ~251 for different models$^{a}$.}
\label{Tab: GJ251_fit_evidence}
\centering
\begin{tabular}{l c c c}
\hline\hline
Model & $P$ [d] &$\ln{\mathcal{Z}}$ & $\Delta\ln{\mathcal{Z}}$ \\
\hline
\noalign{\smallskip}
\multicolumn{4}{c}{CARMENES}\\
\noalign{\smallskip}
0p & \ldots & $-545.4\pm0.1$ & 0\\
1p &  14.2 & $-525.8\pm0.2$ & 19.6\\
2p & 14.2, 1.7 & $-528.7\pm0.2$ & 16.7\\
2p &  14.2, 656.0 & $-524.5\pm0.2$ & 20.9\\
1p+GP &  14.2  & $-492.2\pm0.1$ & 53.2\\
\noalign{\smallskip}
\multicolumn{4}{c}{HIRES}\\
\noalign{\smallskip}
0p & \ldots & $-227.3\pm0.1$ & 0\\
1p &  14.2 & $-222.1\pm0.1$ & 5.2\\
2p & 14.2, 1.7  & $-222.2\pm0.2$ & 5.1\\
2p & 14.2, 601.9 & $-211.4\pm0.2$ & 15.9\\
1p+uGP & 14.2  & $-213.3\pm0.1$ & 14.0\\
\noalign{\smallskip}
\multicolumn{4}{c}{CARMENES + HIRES}\\
\noalign{\smallskip}
0p & \ldots & $-772.0\pm0.2$ & 0\\
1p & \Stephc{14.2} & $-747.2\pm0.2$ & 24.8\\
1cp& \Stephc{14.2}  & $-744.5\pm0.2$ & 27.5\\
2p & 14.2, 1.7  & $-748.9\pm0.2$ & 23.1\\
2p & 14.2, 629.2 & $-743.5\pm0.2$ & 28.5\\
\Stephc{GP} & \Stephc{\ldots} & \Stephc{$-743.3\pm0.2$} & \Stephc{28.9}\\
\Stephc{2p+GP} & \Stephc{1.4, 14.2} & \Stephc{$-708.9\pm0.4$} & \Stephc{63.1}\\
\Stephc{2p+GP} & \Stephc{14.2, 667.1} & \Stephc{$-707.1\pm0.3$} & \Stephc{64.9}\\
1p+uGP &  14.2  & $-706.2\pm0.1$ & 65.8\\
1p+GP &  14.2  & $-704.8\pm0.2$ & 67.2\\
1cp+GP &  14.2  & $-703.8\pm0.3$ & 68.2\\
\hline
\end{tabular}
 \tablefoot{
        \tablefoottext{a}{Planetary models based on CARMENES, HIRES, and combined CARMENES+HIRES RV data.
        0p: 0 planets, 1p: 1 planet, 1cp: 1 planet on a circular orbit ($e = 0$), 2p: 2 planets.
        GP and uGP: additional constrained and unconstrained Gaussian processes, respectively.
        Orbital periods rounded to one decimal.}
    }
\end{table}

We combined the CARMENES and HIRES spectroscopic RV data by fitting an offset and jitter term for each instrument. The combined periodogram showed the highest peak at 14.24\,d with an $\mathrm{FAP}<10^{-11}$, and the daily aliases of this signal were the second and third highest signals.
Within our activity indicators, we did not identify any significant GLS periodogram peak with an $\mathrm{FAP}<10^{-1}$ at the frequency of the 14.2\,d RV signal. A signal at $14.731\pm0.026$\,d in the Ca~{\sc ii} IRT$_3$ line reaches almost 1\,\% FAP, \Steph{but is still larger} than and can be well separated from the 14.24\,d signal within the resolution of the GLS periodogram over the observed time baseline.
We fit a Keplerian model to the signal at 14.24\,d. The log-evidence of the different model fits applied to the data sets is given in Table \ref{Tab: GJ251_fit_evidence}. 

The residual periodogram of the one-planet Keplerian fit on the HIRES and CARMENES combined data shows several remaining significant peaks at periods of $73.02$\,d ($\mathrm{FAP}<10^{-5}$), $68.15$\,d ($\mathrm{FAP}<10^{-4}$), and $67.86$\,d  ($\mathrm{FAP}<10^{-2}$). These periods are close to half of the derived rotational period of the star. 
We also observed a peak at 118.78\,d with $\mathrm{FAP}<10^{-2}$, which is very close to the rotation period derived from photometry. As a simple test, we fit a sinusoid to the 118.78\,d period. The signal at $67.86$\,d was then the most significant. 
It was necessary to fit an additional sinusoid for the $67.86$\,d signal to obtain a periodogram that did not show any signal with an $\mathrm{FAP}<0.01$, which showed that the other signals were connected through aliasing.
The necessity of modeling two sinusoidal functions with periods close to the rotational period and its half suggests that these signals are caused by stellar activity, for instance, a multi-spot pattern, or amplitude variations caused by decreasing spot areas. To rule out the possibility of independent planet signals, we analyzed the coherence of these signals.

We used the s-BGLS periodogram to assess the coherence of the significant RV signals with increasing numbers of observations. We show the resulting s-BGLS diagrams in Fig.~\ref{fig: GJ251_BGLS}. We identified that neither the signals around 120\,d nor the forest of signals between 50\,d, and 73\,d were stable over the observational time baseline. The 73\,d signal lost about three orders of magnitude in signal probability after roughly observation 160 (May 2018), but reappeared in observation 210 (January 2019). 
All these mentioned signals showed a lack of coherence in the latest observations between observation 200 (January 2019) and 250 (October 2019). In contrast to these signals, the suspected planetary signal at 14.2\,d never showed strong dips in its probability \Stephb{during the time of observations. These results together with the analysis of the activity indicators mean that this signal probably is of planetary origin. We refer to it as GJ~251~b.}

\subsection{A second planet in the system?}

\begin{figure*}
    \centering
    \includegraphics[width=9cm]{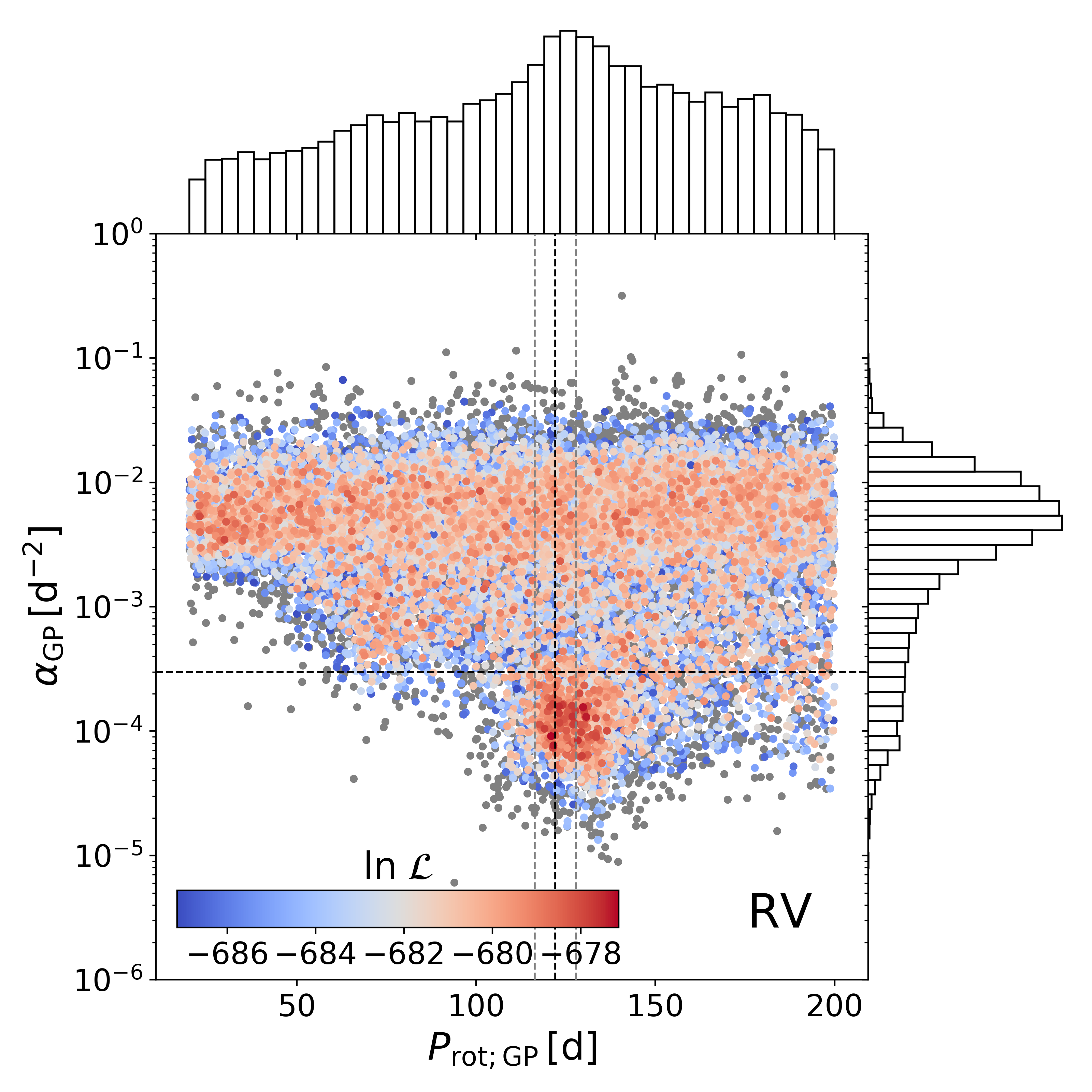}
    \includegraphics[width=9cm]{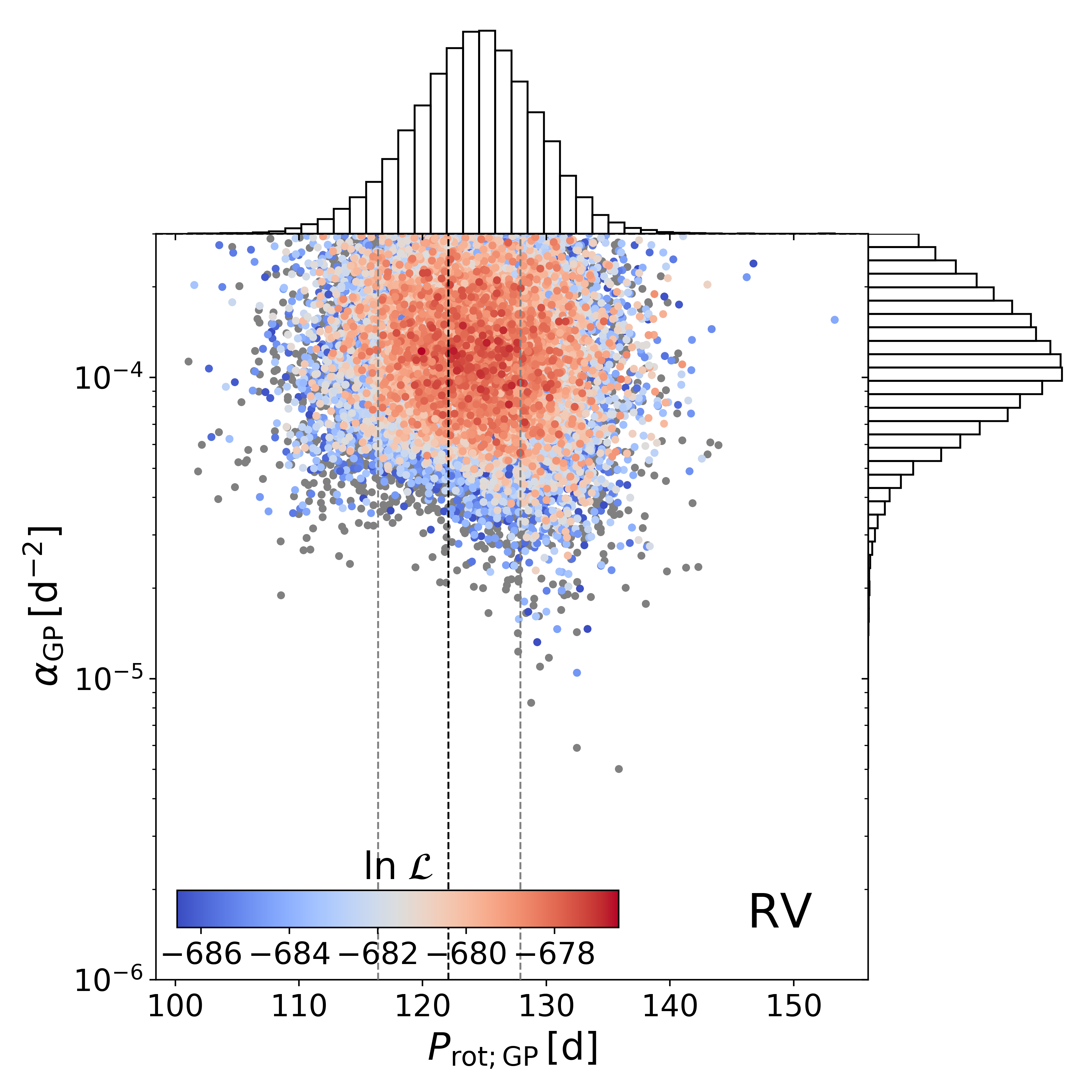}
    \caption{Posterior distribution of the GP fit to the RV data in the $\alpha_{\tx{GP}}$ vs. $P_{\tx{rot}}$ plane for GJ~251. The color-coding shows the log-likelihood normalized to the highest value in the posterior sample. Gray samples indicate solutions with a $\Delta \ln{L}>10$ compared to the best solution. \textit{Top:} GP fit to the RV data with a wide uniform prior for the rotational period. \textit{Bottom:} GP fit to the RV data with an informative normal prior based on the photometric GP results and additional constraints on the other hyperparameters. We overplot the derived rotational period of the photometric GP and its 3$\sigma$ uncertainties with vertical lines. The horizontal line marks the cut in $\alpha_{\tx{GP}}$ used to constrain the GP fit shown in the lower plot. }
    \label{fig: GJ251_GP_RV}
\end{figure*}{}

\begin{figure}
\centering
\includegraphics[trim=0cm 0.7cm 1cm 1.6cm, clip, width=9cm]{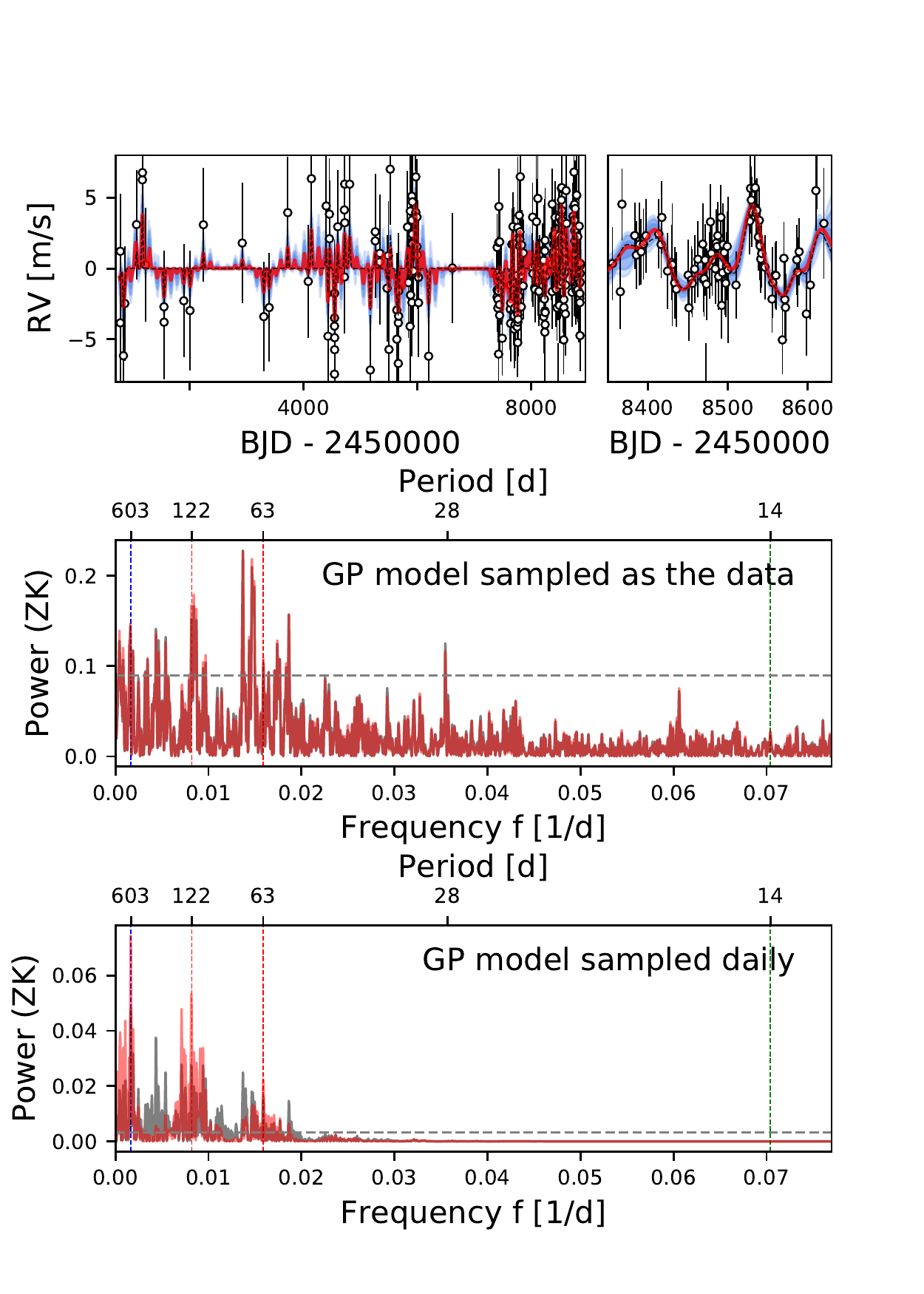}
\caption{Gaussian process model for the RV data of GJ~251 \Stephb{without the planetary model, which is subtracted from the RV data}. The constrained GP model is shown in red. The blue regions show 1$\sigma$, 2$\sigma$, and 3$\sigma$ uncertainties. We show a zoom to some CARMENES observations (top right). The GLS is evaluated on the GP model at each observed data point (top GLS) and daily (bottom GLS). The dashed line in the GLS periodograms indicates an FAP \Stephb{level of 0.1\,\%}. In addition, we show the unconstrained GP model as the dashed black line in the upper plots and as the gray periodograms in the lower plots.}
\label{Fig: GJ251_GPmodel}
\end{figure}

\begin{figure*}
\centering
\includegraphics[scale=0.5]{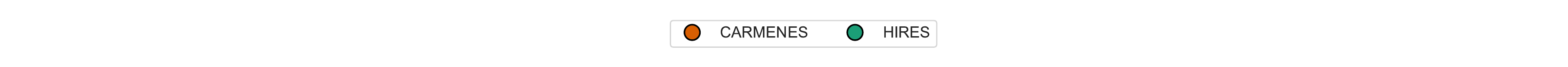}

\includegraphics[scale=0.24]{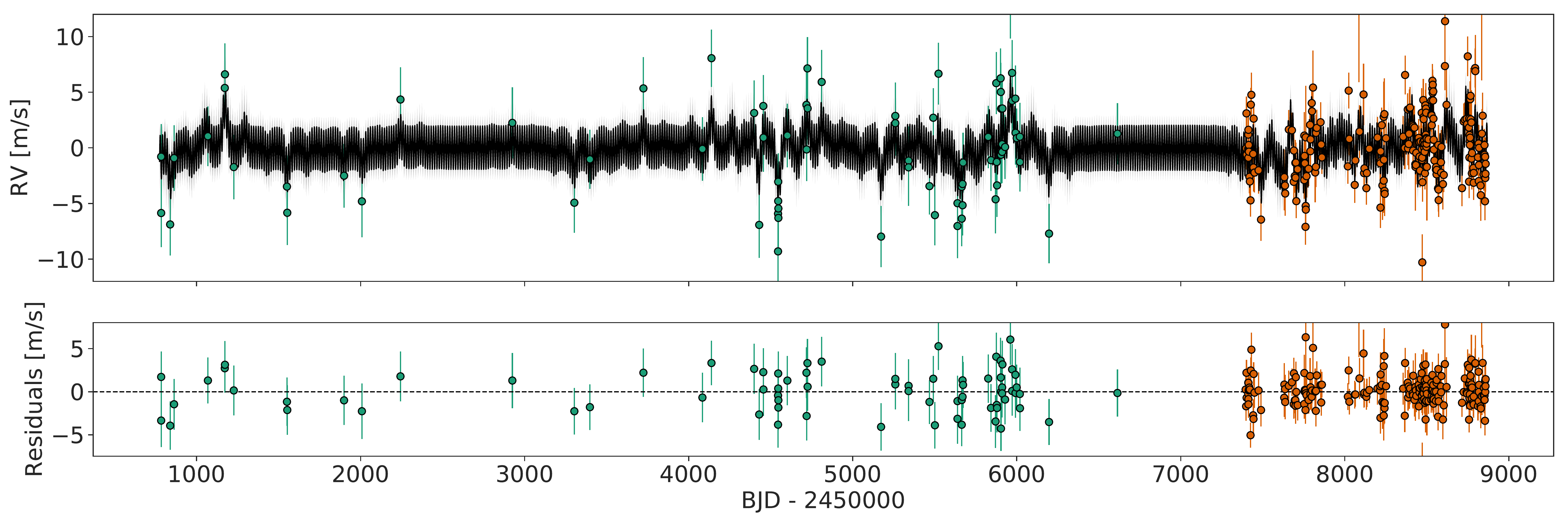}
\includegraphics[scale=0.255]{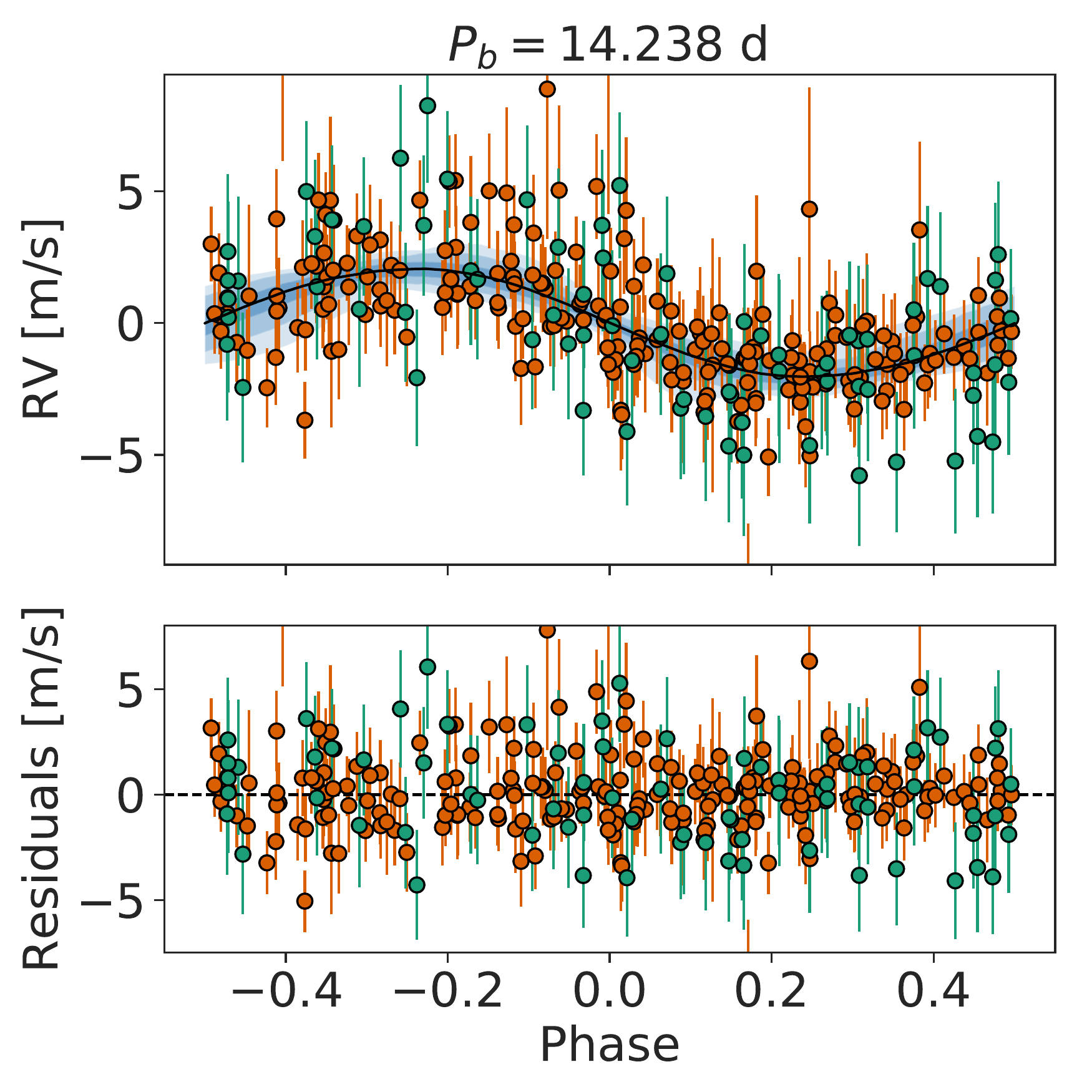}
\caption{\textit{Left:} Radial velocity data with a combined model of one-planet and stellar activity using a Keplerian model and a GP. \textit{Right:} Plot phased to the orbital period of GJ~251~b \Steph{without a GP component. \Stephb{The bottom plots show the residuals after the fit.}}}
\label{Fig: GJ251b_phased}
\end{figure*}

We found no evidence in the CARMENES and HIRES data for the planetary candidate \Stephb{claimed} by \cite{Butler2017} at 1.74\,d. Neither did we observe the 604 d signal in CARMENES data, which was highly significant in the HIRES data. We used the s-BGLS to verify whether these signals were more significant in the past but might have decayed over the time of observations. \Stephb{While we did not find any indication of the 1.74\,d signal during the period of our observations, the 604\,d signal showed variability in its signal probability (see Fig.~\ref{fig: GJ251_BGLS}).} The 604\,d signal already slightly lost coherence during the last HIRES observations. However, especially the CARMENES data show noncoherence of the signal. \Stephb{The fluctuation} is a strong indication for a nonplanetary origin \citep{Mortier2017}.

\Stephb{We performed a statistical test using model comparison in the framework of Bayesian evidence with \texttt{juliet}.} We compared one-planet ($P$ = 14.24\,d) to two-planet models. Our results showed that the two-planet model with periods of 14.24\,d and 1.74\,d is not supported by the \Steph{individual} data sets or by their combination because its log-evidence is weaker than that of the simpler one-planet model. 
Fitting the 600\,d signal as a second planet resulted in a significant model improvement compared to the one-planet model \Steph{alone} for the HIRES data. 
The same two-planet model (14.24\,d, and 604\,d) fit to the CARMENES data brought no significant improvement either compared to the one-planet model. 
Fitting the two-planet model to the combined CARMENES and HIRES data \Stephc{resulted in almost the same} log-evidence \Stephc{as the one-planet model}. \Stephc{Additionally,} the derived planetary period at $629.2^{+20.2}_{-8.4}$\,d deviates significantly from the $601.9^{+6.5}_{-5.2}$\,d obtained from the fit on the HIRES data, even though we used a prior with an informative \Steph{Gaussian distribution}, \Steph{hereafter referred to as normal prior}, with mean of 604\,d and $\sigma=30$\,d. These values were informed by the GLS periodogram peak in the HIRES data and its $3\sigma$ uncertainty. 

\Stephc{We searched for any additional planetary signal hidden behind the stellar activity by once sampling a second Keplerian with a log-uniform prior between 15\,d to 8000\,d and then sampling with a log-uniform prior between 0.5\,d to 14\,d, while simultaneously modeling the stellar activity with a constrained GP model (see Sect.~\ref{Sect.: GJ251_GP_model}). We divided the two-planet model search into two runs for technical reasons: \texttt{juliet} needs a chronological order of the planetary periods. The two two-planet models combined with the GP showed no significant improvement compared to the one-planet and GP combined model, that is, the data suggest that all periodic variations except for the 14.2\,d period are better or equally well described by a GP. }

\Stephc{In the 600\,d signal,} we identified a periodicity with an $\mathrm{FAP}<10^{-2}$ in the H$\alpha$ indicator at a period of $660\pm21$\,d. The Na~{\sc i} doublet lines and the CRX showed a significant peak with an $\mathrm{FAP}<10^{-3}$ at 300\,d. The s-BGLS of the dLW shows that a signal close to 600\,d was more significant in past observations around observation 150, which corresponds to April 2018 (see Fig.~\ref{Fig: GJ251_GLS_act} again). \Stephb{Comparing the activity s-BGLS of the dLW} to the s-BGLS of the RV data showed that this is about the same time at which the 600\,d signal was most significant in the RV data. 
These results, along with our photometric results, suggest that the signals at 73\,d, 119\,d, and 600\,d are not caused by Keplerian motion.

\subsection{Simultaneous Keplerian and GP modeling}
\label{Sect.: GJ251_GP_model}

% FINAL RESULTS TABLE
\begin{table*}[!ht]
    \centering
    \caption{Posterior parameters of the final fits obtained for GJ~251\,b, HD~238090\,b, and Lalande~21185\,b using \texttt{juliet}. }
    \label{Tab: Posteriors}
    \begin{tabular}{l c c c} 
        \hline
        \hline
        \noalign{\smallskip}
        Parameter$^{a}$ &GJ~251\,b& HD~238090\,b & Lalande~21185\,b \\
        \noalign{\smallskip}
        \hline
        \noalign{\smallskip}
                        \multicolumn{4}{c}{orbital parameters} \\
                                \noalign{\smallskip}
        $P$ (d)                         &   $14.238^{+0.002}_{-0.002}$    & $13.671^{+0.011}_{-0.010}$ & $12.946^{+0.005}_{-0.005}$ \\[0.1 cm]
        $t_0 - 2450000$ (BJD)      & $8626.69^{+0.34}_{-0.35}$           & $8630.09^{+0.52}_{-0.55}$& $8622.23^{+0.48}_{-0.45}$\\[0.1 cm]
        $K$ ($\mathrm{m\,s^{-1}}$)   &  $2.11^{+0.21}_{-0.20}$      & $2.85^{+0.38}_{-0.39}$ & $1.39^{+0.14}_{-0.14}$ \\[0.1 cm]
        $\mathcal{S}_{1,b} = \sqrt{e_{b}}\sin \omega_{b}$ & $0.20^{+0.16}_{-0.22}$ & $0.44^{+0.16}_{-0.25}$ & $0.07^{+0.19}_{-0.20}$ \\[0.1 cm]
        $\mathcal{S}_{2,b} = \sqrt{e_{b}}\cos \omega_{b}$ & $0.05^{+0.21}_{-0.22}$& $-0.25^{+0.23}_{-0.18}$ & $-0.27^{+0.25}_{-0.19}$ \\[0.1 cm]
        $e$                                 & $0.10^{+0.09}_{-0.07}$& $0.30^{+0.16}_{-0.17}$ & $0.12^{+0.12}_{-0.09}$ \\[0.1cm]
        $\omega$ (deg)                        &    $78.8^{+47.6}_{-44.7}$    & $119.3^{+22.8}_{-24.8}$ & $140.7^{+27.3}_{-53.0}$\\[0.1cm]
        \noalign{\smallskip}
        \multicolumn{4}{c}{RV parameters} \\[0.1cm]
        \noalign{\smallskip}
                $\gamma_{\textnormal{CARMENES}}$ ($\mathrm{m\,s^{-1}}$) &  $-0.06^{+0.56}_{-0.56}$      &$-0.03^{+26}_{-0.27}$ & $-0.19^{+0.45}_{-0.45}$\\[0.1 cm]
        $\sigma_{\textnormal{CARMENES}}$ ($\mathrm{m\,s^{-1}}$) & $1.05^{+0.17}_{0.16}$  & $1.57^{+0.28}_{-0.25}$  &$1.10^{+0.15}_{-0.14}$\\[0.1cm]
        $\gamma_{\textnormal{HIRES}}$ ($\mathrm{m\,s^{-1}}$)           &$0.18^{0.58}_{-0.60}$ &\ldots & \ldots \\[0.1 cm]
        $\sigma_{\textnormal{HIRES}}$ ($\mathrm{m\,s^{-1}}$)        &$1.85^{0.71}_{-0.77}$&\ldots & \ldots\\[0.1cm]
        $\gamma_{\textnormal{SOPHIE}}$ ($\mathrm{m\,s^{-1}}$)            &\ldots&\ldots &$0.45^{+0.46}_{-0.45}$\\[0.1 cm]
        $\sigma_{\textnormal{SOPHIE}}$ ($\mathrm{m\,s^{-1}}$)       &\ldots&\ldots &$1.26^{+0.20}_{-0.19}$\\[0.1 cm]
        \noalign{\smallskip}
        \multicolumn{4}{c}{GP (constrained) hyperparameters} \\
        \noalign{\smallskip}
        $\sigma_\mathrm{GP,RV}$ ($\mathrm{m\,s^{-1}}$)             & $2.27^{+0.40}_{-0.34}$& $1.92^{+1.42}_{-0.82}$ &  $1.62^{+0.31}_{-0.25}$\\[0.1 cm]
        $\alpha_\mathrm{GP,RV}$ ($10^{-5}\,\mathrm{d^{-2}}$)   & $11.4^{+7.4}_{-4.5}$          & $10^{-20}$ (fixed) & $5.9^{+3.7}_{-1.9}$  \\[0.1 cm]
        $\Gamma_\mathrm{GP,RV}$    &  $4.6^{+2.7}_{-2.0}$        & $1.1^{+2.5}_{-0.7}$ & $1.3^{+1.1}_{-0.6}$ \\[0.1 cm]
        $P_\mathrm{rot;GP,RV}$ (d)    &$124.2^{+4.8}_{-5.1}$ & $105.9^{+1.07}_{-0.93}$ & $56.2^{+0.7}_{-0.7}$ \\[0.1 cm]
        \noalign{\smallskip}
                \multicolumn{4}{c}{derived planetary parameters} \\
        \noalign{\smallskip}
        $M_{\rm p} \sin{i}$ ($M_\oplus$)    &  $4.00^{+0.40}_{-0.40}$  & $6.89^{+0.92}_{-0.95}$  &  $2.69^{+0.25}_{-0.25}$  \\[0.1 cm]
        $a_{\rm p}$ ($10^{-2}$\,au)                    & $8.18^{+0.11}_{-0.12}$ & $9.32^{+0.11}_{-0.11}$   & $7.890^{+0.068}_{-0.077}$  \\[0.1 cm]
        $T_\textnormal{eq}$ (K)\tablefootmark{b}          &  $351.0^{+1.4}_{-1.3}$   & $469.6^{+ 2.3}_{- 2.6}$ & $370.1^{+ 5.8}_{- 6.8}$ \\[0.1 cm]
        $S$ ($S_\oplus$)            &  $2.53^{+0.04}_{-0.04}$ &$8.10^{+ 0.16}_{- 0.18} $& $3.13^{+ 0.20}_{- 0.22} $  \\[0.1 cm]
        \noalign{\smallskip}
        \hline
    \end{tabular}
    \tablefoot{
        \tablefoottext{a}{Error bars denote the $68\%$ posterior credibility intervals. }     
        \tablefoottext{b}{Equilibrium temperatures estimated assuming zero Bond albedo.}\\
        Priors and descriptions for each parameter can be found in Table~\ref{Tab: Priors_planets_instruments} and Table~\ref{Tab: Priors_GP_RV}. Results of the derived parameters also take the stellar parameter uncertainties (e.g., Gaussian uncertainty) into account.
    }
\end{table*}

We performed a simultaneous fit of a one-planet Keplerian model together with the QP GP \Stephb{(Equation~\ref{Eq: kernel})} \Stephb{to account for activity-induced RV variations. For the first GP model we used wide uninformative priors, which are shown in Table \ref{Tab: Priors_GP_RV}, while we kept the same  planetary and instrumental priors as for the one-planet fit (Table \ref{Tab: Priors_planets_instruments})}. The posterior samples of this \Stephc{unconstrained} GP can provide indications for the stellar rotational period given only the RV data \citep[see also][]{Angus2018,Stock2020b}. 

Including the GP as a model for activity significantly improved the log-evidence ($\Delta\log Z=41$ on the combined CARMENES+HIRES data) compared to the one-planet fit \Stephc{alone}. We show the $\alpha_{\rm GP}$ versus period diagram of the unconstrained GP posterior samples in the top plot of Fig.~\ref{fig: GJ251_GP_RV}. Around 125\,d we identified a region of \Steph{higher posterior density}, higher likelihood, and lower values of $\alpha$ compared to the rest of the posterior solutions. This indicates a more strongly correlated periodic signal. The \Stephc{derived} median GP rotational period based on the CARMENES and HIRES RV data is 
$P_{\text{rot,RV}}=125^{+44}_{-59}$\,d,
%$P_{\text{rot,RV}}=124.6^{+43.8}_{-59.1}$\,d,
which is consistent with the results from photometric data. 
For the final one-planet and GP simultaneous fit to the combined data, we applied several additional constrains on the GP \Stephc{to reduce the posterior volume of the} model, which could lead to \Stephc{fitting incorrect residual signals} \citep{Angus2018}. We applied a normal prior to the GP rotational parameter based on the \Stephc{stellar rotation period derived from the photometry} and its 3$\sigma$ uncertainty.

For $\Gamma$, which can be interpreted as the overall number of inflection points per function period, we applied a log-uniform prior between $10^{-1}$ and $10^1$. This prior is consistent with about one to three local maxima per rotation period. \Stephc{\cite{Jeffers2009} showed that this assumption is to first approximation valid for any stellar surface, independent of the number of starspots and their distribution}. Similar, but even more informative priors \Stephc{on $\Gamma$}, have been applied in several studies that used the QP GP kernel \citep[see][and references therein]{Nava2020}.   
The timescale parameter of the QP GP kernel, $\alpha_{\tx{GP}}$, is crucial for modeling a meaningful rotational signal. For instance, if it is large, then the squared exponential term of the kernel dominates, which allows for good fits to the data without requiring any periodic covariance structure, even when the data show clear periodicities \citep[see also][ ]{Angus2018}. This effect is visible in the top plot of Fig.~\ref{fig: GJ251_GP_RV}, where a plateau of posterior samples at high $\alpha_{\tx{GP}}$ values populates the entire prior volume of the GP \Stephc{rotation parameter}. 
\Stephc{As discussed in Sect.~\ref{SubSect: GJ251_photometry}, it should not be strictly assumed that photometric and RV GP timescales are similar. A prior on $\alpha_{\tx{GP}}$ based on photometry, as used for example for the stellar rotation, therefore needs future verification.}
For the moment, the upper boundary of the $\alpha_{\tx{GP}}$ prior needs to be assessed for each target individually. \cite{Angus2018} proposed \Stephc{that this hyperparameter should be} larger than the observed stellar rotation period. In the case of GJ~251, \Stephc{with a derived photometric rotation period of about 120\,d}, the suggested rule by \cite{Angus2018} would translate into $\alpha_{\tx{GP}}<3.5~10^{-5}$. 

However, it is not clear whether this general rule can be applied to slowly rotating stars like GJ~251 because \cite{Angus2018} did not discuss such stars. \Stephc{For example, because of active longitudes \citep{Jeffers2009},  a meaningful QP signal might still be detected that would be caused by starspots that decay over approximately half of the stellar rotation every time the active region points toward the observer}. \Stephc{More importantly, from the unconstrained GP posterior samples, we find that a constraint on $\alpha_{\tx{GP}}$ based on the rule by \cite{Angus2018} would mean that the overdensity of high-likelihood posterior samples detected at 120\,d, given the data, would not be included in the final activity model. As expected, a GP model using the upper boundary of \cite{Angus2018} led to a log-evidence of $-710.2$, which is \Stephf{about} four lower than the unconstrained GP. Finding the right mixture between physical priors and data-driven behavior is critical for modeling stellar activity with GPs. }  

\Stephc{We constrained the upper boundary to $\alpha_{\tx{GP}}<3~10^{-4}\,\tx{d}^{-2}$. This constraint removed the plateau of posterior samples that fit noise on short timescales, which cannot be attributed directly to the stellar rotation and is captured by the instrument jitter parameter in our case. However, the observed high-likelihood posterior sample overdensity at the derived photometric rotation period is included in the GP model. We have applied similar constrains of $\alpha_{\tx{GP}}$ with success in \cite{Stock2020b}. The priors of our constrained GP are summarized in Table~\ref{Tab: Priors_GP_RV}.}
\Stephc{The distribution of the posterior parameters of the constrained GP in the $\alpha_{\textnormal{GP}}$ versus\ $P_{\textnormal{GP}}$ diagram is shown in the bottom plot of Fig.~\ref{fig: GJ251_GP_RV}.}
\Stephc{Corner plots of of all the fit parameters are provided in Figs.~\ref{Fig: GJ251_corner_GP} and~\ref{Fig: GJ251_corner_planet}. The derived timescale parameter in our final GP model is $\alpha_{\tx{GP}}=11.4^{+7.4}_{-4.5}\cdot 10
^{-5}$\,d$^{-2}$, which translates into a decay time of $P_{\tx{dec}}=66^{+19}_{-15}$\,d
% $\tau=66.2^{+19.0}_{-14.7}$\,d ***
and is close to half the stellar rotation.} \Stephd{The GP semiamplitude is consistent within $1\sigma$ with the expected RV semiamplitude due to stellar rotation estimated based on the relations of \cite{Suarez2018}.  }

In Fig.~\ref{Fig: GJ251_GPmodel} we show our final median GP model. We calculated the GLS periodogram of the GP model to assess its temporal behavior. To calculate the GLS periodogram we chose to sample the GP model in two different ways: first, sampling identical to that of the original data, and second, sampled daily over the entire observation time. The GP model sampled as the real observations includes the true window function of the data. \Stephb{A visual inspection of} this GLS periodogram shows that the highest peak is at 73\,d, followed by another peak at 68\,d. These were the most significant signals in the residuals of the simple one-planet fit. A peak at 28\,d, about twice the planetary period and close to the lunar cycle, is also visible. The GLS periodogram of the GP model, sampled once a day over the entire observation time, shows that the GP does not model the 28\,d period. The peak can be explained by the convolution of the GP model with the window function of the observations. 

\Stephc{From the daily sampled periodogram decomposition of the GP models, the tuned GP does not model any periods close to the planet (or twice the planetary period). It models primarily the activity related signals at 63\,d, 122\,d, and 600\,d.} \Stephc{The unconstrained GP modeled the 73\,d signal more prominently than the signals at 120\,d and 63\,d, while the constrained GP modeled 120\,d and 63\,d more strongly. Nevertheless, the tuned GP was capable of producing the same strong peak at 73\,d, given the data.} This result shows in practice the conclusions and caveats given by \cite{Nava2020} that QP models can contain signals ``unrelated to their true period''. The 73\,d signal can be explained by an alias based on a sampling frequency of $\sim 365^{-1}$\,d$^{-1}$ of the first harmonic at 63\,d of the 120\,d rotation period.

Finally, we show the combined Keplerian and tuned GP fit to the RV data, and a plot phased to the orbital period of GJ~251~b in Fig.~\ref{Fig: GJ251b_phased}. \Steph{We display the final posterior solution of the planetary and GP parameters in Table~\ref{Tab: Posteriors}. We derived a semiamplitude of $K=2.11^{+0.21}_{-0.20}$\,m\,s$^{-1}$, a period of $P=14.238\pm0.002$\,d, and an eccentricity of $e=0.10^{+0.09}_{-0.07}$. The eccentricity of the system is consistent with zero because fits without this parameter provided similar log-evidence with fewer parameters. \Stephc{Based on our posterior samples and the stellar parameters (see Table~\ref{Tab: stellar_parameters}), we derived further planetary parameters, which we also show in Table~\ref{Tab: Posteriors}. According to this analysis, GJ~251~b has a minimum mass of $4.00\pm0.40\,M_\oplus$ and a semimajor axis of $0.0818^{+0.0011}_{-0.0012}$\,au.}}

\subsection{Transit search \Stephb{and analysis} with \emph{TESS}}

\begin{figure}
\centering
\includegraphics[width=9cm]{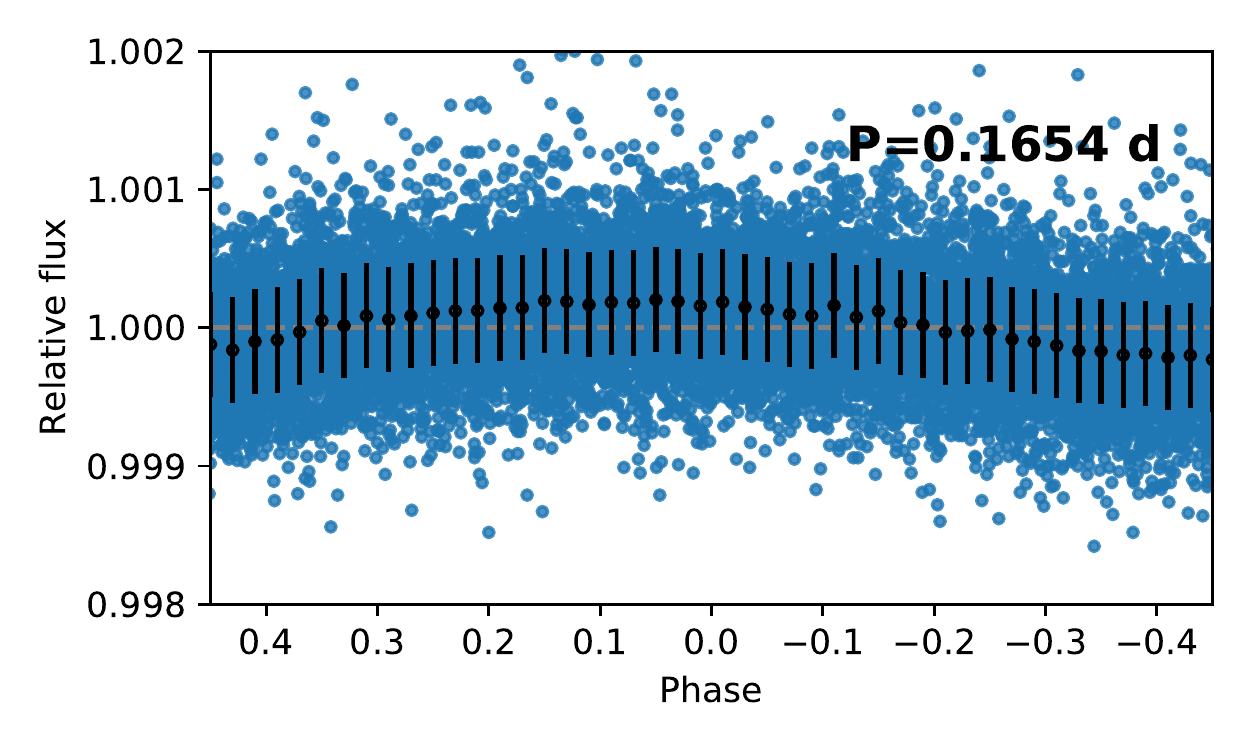}
\caption{\Stephb{\emph{TESS} PDCSAP light curve of GJ~251 folded to the 0.165\,d signal. The black dots and uncertainty bars represent binned TESS data. }}
\label{Fig: GJ251_TESS}
\end{figure}

GJ~251 was observed with the \emph{TESS} satellite \citep{Ricker2015} in sector 20, in the period from 24 December 2019 to 21 January 2020, with a total of 16556 data points, but was not marked as a \emph{TESS} object of interest (TOI). We independently searched for a transit signal using the transit least-squares (TLS; \citealt{Hippke&Heller2019}) algorithm on PDCSAP light curve. We did not identify any TLS signal that could be attributed to any possible transit for
GJ~251. \Steph{However, we identified a significant sinusoidal-like signal with a frequency $f\approx6$\,d$^{-1}$ (period 0.165\,d) and $(1.96\pm0.4)~10^{-4}$ relative flux amplitude variation, equivalent to about 0.2\,mmag), as well as its harmonics $f/2$ and $2f$ with lower amplitudes. We show a phase plot of this signal in Fig.~\ref{Fig: GJ251_TESS}. This signal is not observed within the RV data of this target.}

This 4\,h signal was already present in the simple aperture photometry (SAP) light curve. We checked that it was not instrumental in origin by extracting and analyzing the light curves of the 968 objects present in the same \emph{TESS} S20 sector, Camera 1, and CCD 3 as GJ~251. \Stephc{No other star showed the same periodicity.}

We considered the possibility that the signal originated from thermodynamical excitations of p- and g-modes, as theoretically predicted by \cite{Rodriguez2014}. \Stephc{However,} the pulsation hypothesis is unable to explain the presence of subharmonics of the main frequency in the periodogram. Moreover, solar-like pulsations or granulation, which have not yet been detected in M-dwarf stars, can also be discarded \Stephc{because they} are predicted to be on the order of minutes.

\begin{figure}
\centering
\includegraphics[width=9cm]{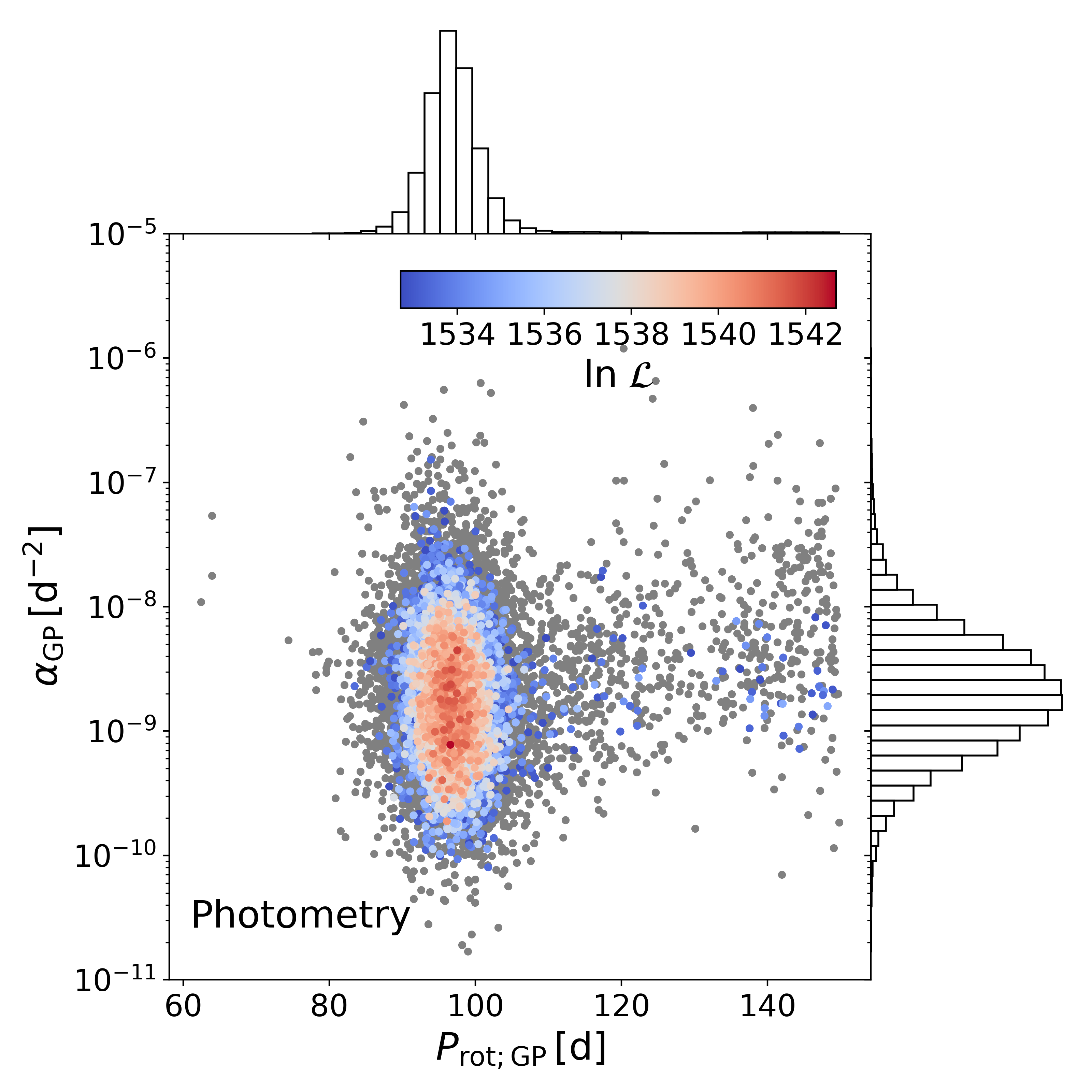}
\caption{Posterior distribution of the GP fit to the photometric data in the $\alpha_{\mathrm{GP}}$ vs.\ $P_{\mathrm{rot}}$ plane for HD~238090. The color-coding shows the log-likelihood normalized to the highest value within the posterior sample. Gray samples indicate solutions with a $\Delta \ln{L}>10$ compared to the best solution.}
\label{Fig: GP_HD238090}
\end{figure}

\Stephd{Lucky-imaging observations with FastCam \citep{Cortes2017} and Robo AO images \citep{Lamman2020} have not detected any resolved visual companion.}
The \emph{TESS} aperture includes several objects. In particular, the two brightest stars in the aperture mask are only 4.87\,mag and 5.95\,mag fainter in the $G$ band, \Stephc{corresponding to a flux contribution of 1.1\,\% and 0.4\,\%, respectively.} A 18\,mmag sinusoidal amplitude variation in the former or a 50\,mmag amplitude variation in the latter could account for the detected 4\,h signal. \Stephc{We chose different subapertures} to extract the light curve from different regions of the \emph{TESS} full-frame images. \Stephc{It did not affect the amplitude of the short-period signal, making it unlikely that the periodicity originated in background contamination. }
\Stephc{With our analysis, we} cannot draw any final conclusion on the origin of the 4\,h signal for GJ~251.

\begin{figure}
\centering
\includegraphics[width=9cm]{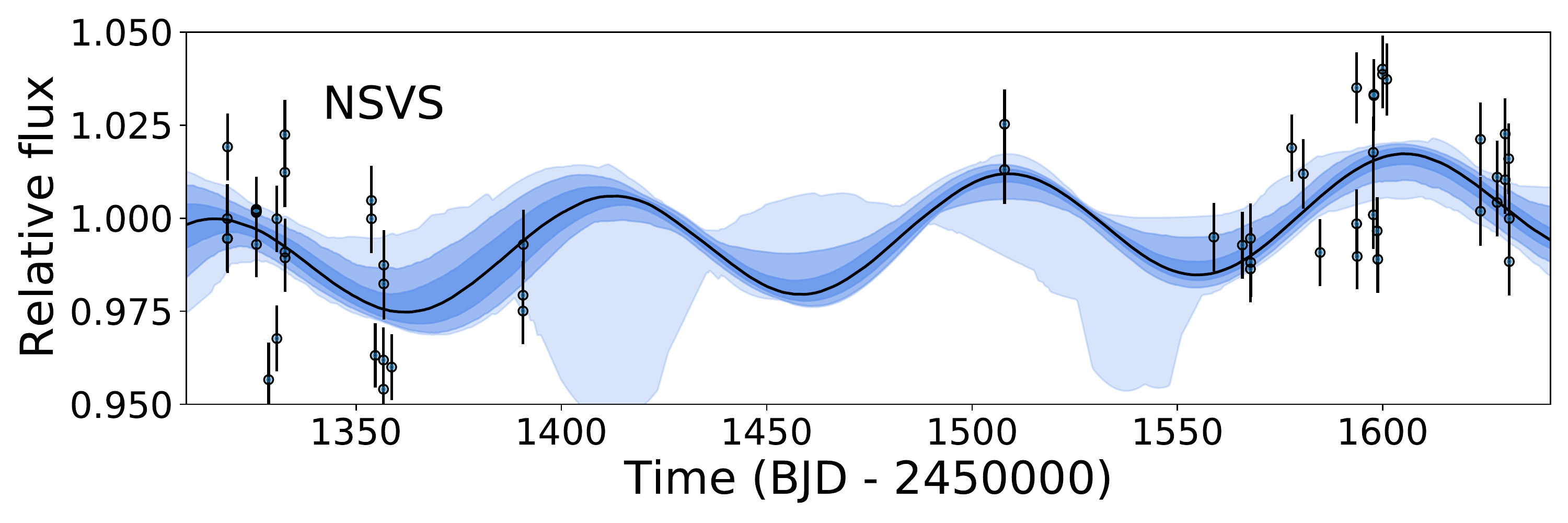}

\includegraphics[width=9cm]{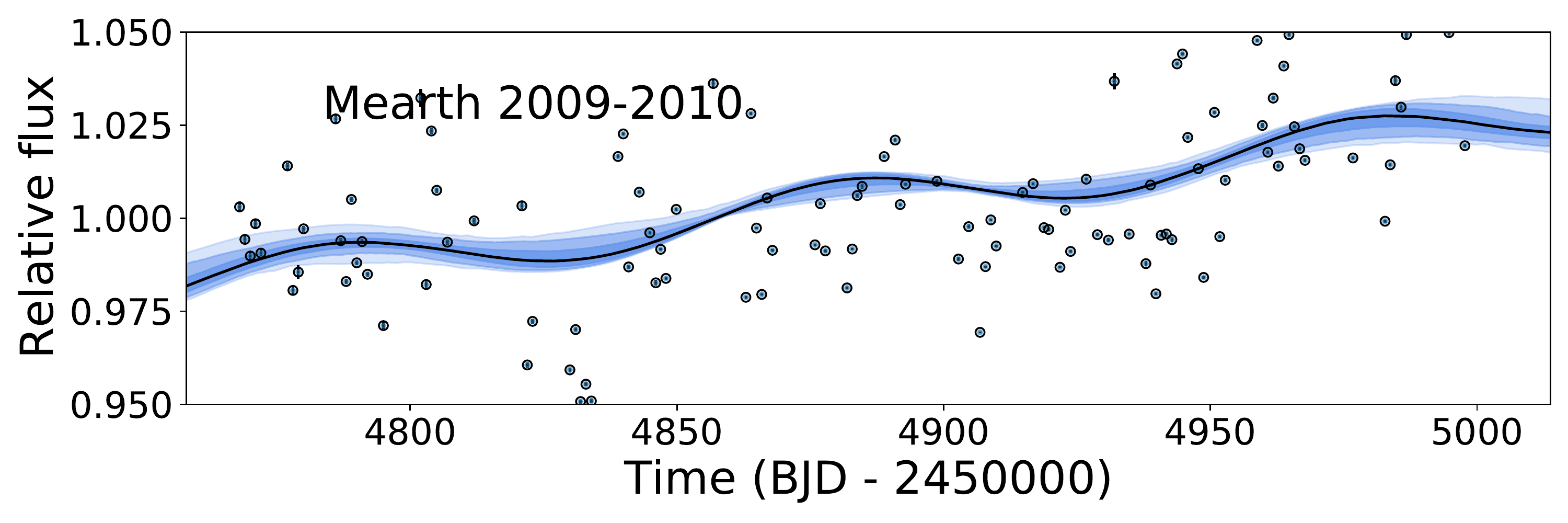}

\includegraphics[width=9cm]{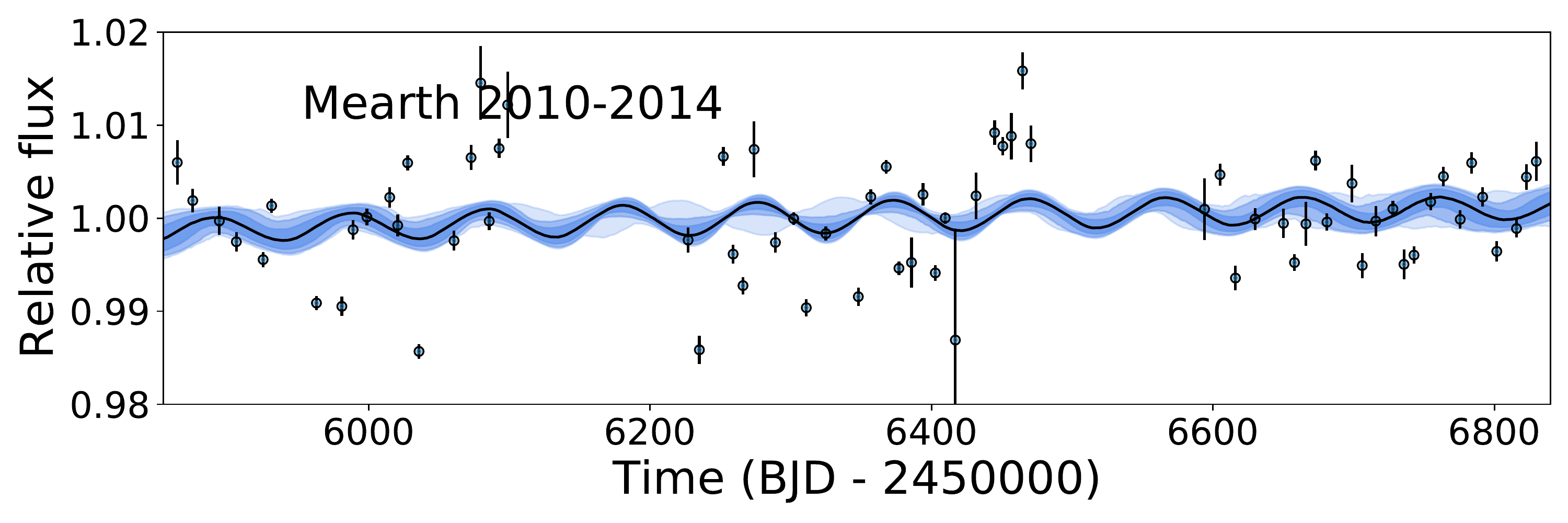}

\includegraphics[width=9cm]{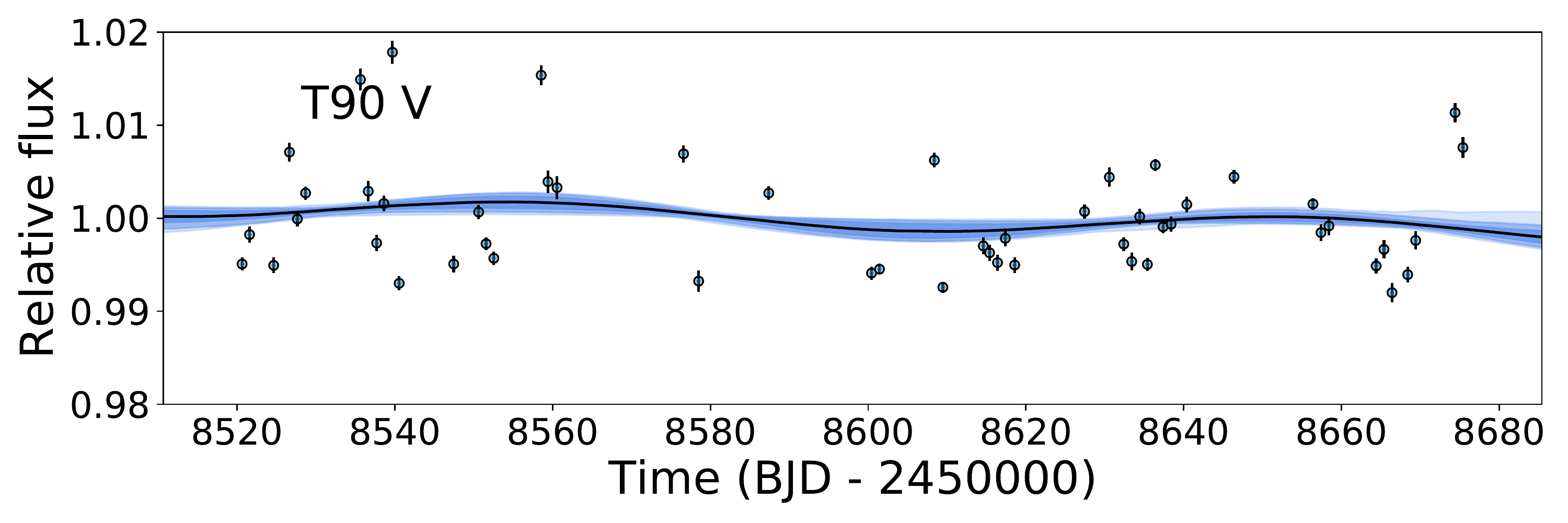}

\includegraphics[width=9cm]{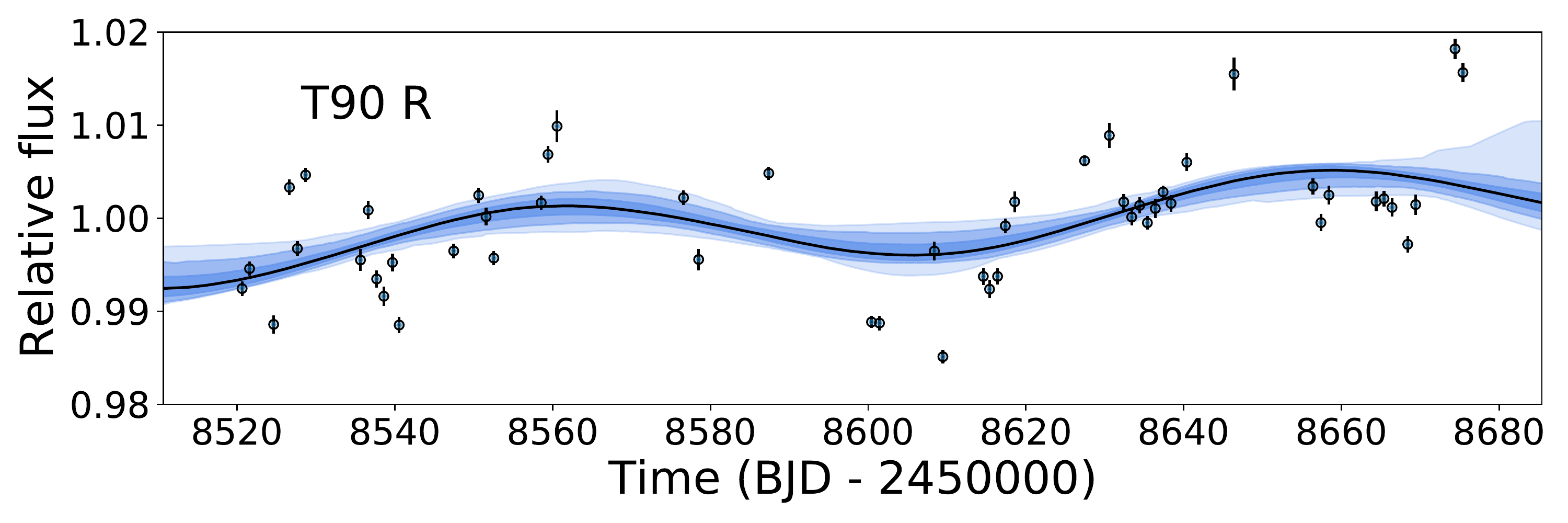}

\includegraphics[width=9cm]{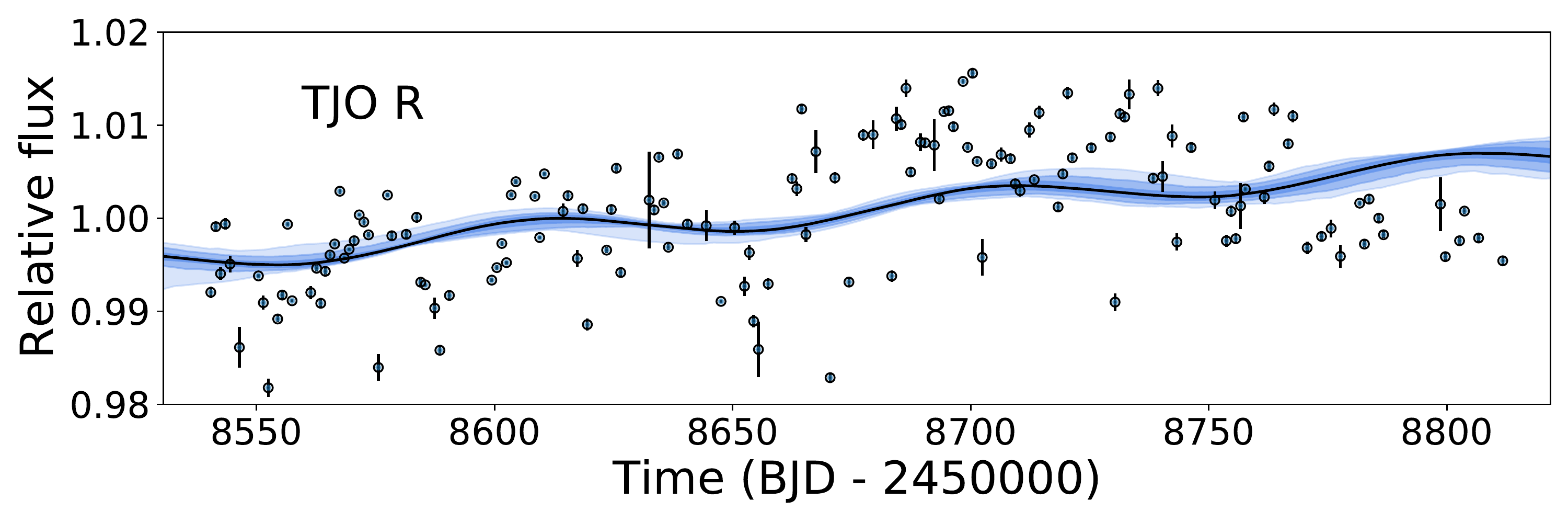}
\caption{Gaussian-process model for each nightly binned photometric data set of HD~238090. From top to bottom: NSVS, MEarth 2009 to 2010, MEarth 2010 to 2014, SNO V, SNO R, and TJO. For each instrument, we fit individual GP hyperparameters for the amplitudes $\sigma_{GP_i}$ and $\Gamma_i$, but we used global GP hyperparameters for the timescale of the amplitude modulation and the rotation period.}
\label{Fig: HD238090_photometric_GP}
\end{figure}

\Steph{We estimated the radius of GJ~251~b with the mass-radius relation of \cite{Zeng2016} \Stephb{and assumed an Earth-like core-mass fraction of 0.26} to be approximately 1.48\,$R_\oplus$, which translates into a transit depth of roughly 1.4\,ppt for GJ~251~b.} 
%given the derived minimum mass of the planet from the RV analysis.} 
Such a signal should be detectable by \emph{TESS} in case of a full transit.
\Stephd{However, we were unable to detect any transit in the light curve, in particular given the estimated $t_0$ and the orbital period $P$ of GJ~251~b by the RV fit and their uncertainties. } \Stephd{In particular, we ruled out a transit event within 1$\sigma$ of $t_0$, but not within 3$\sigma$, because of an observational gap in the \emph{TESS} light curve.} \Steph{Unfortunately, GJ~251 will not be observed again by \emph{TESS}.}

\section{HD~238090}
\label{Sect: HD238090}
\subsection{Photometric monitoring}

\begin{figure*}
\centering
\includegraphics[width=18cm]{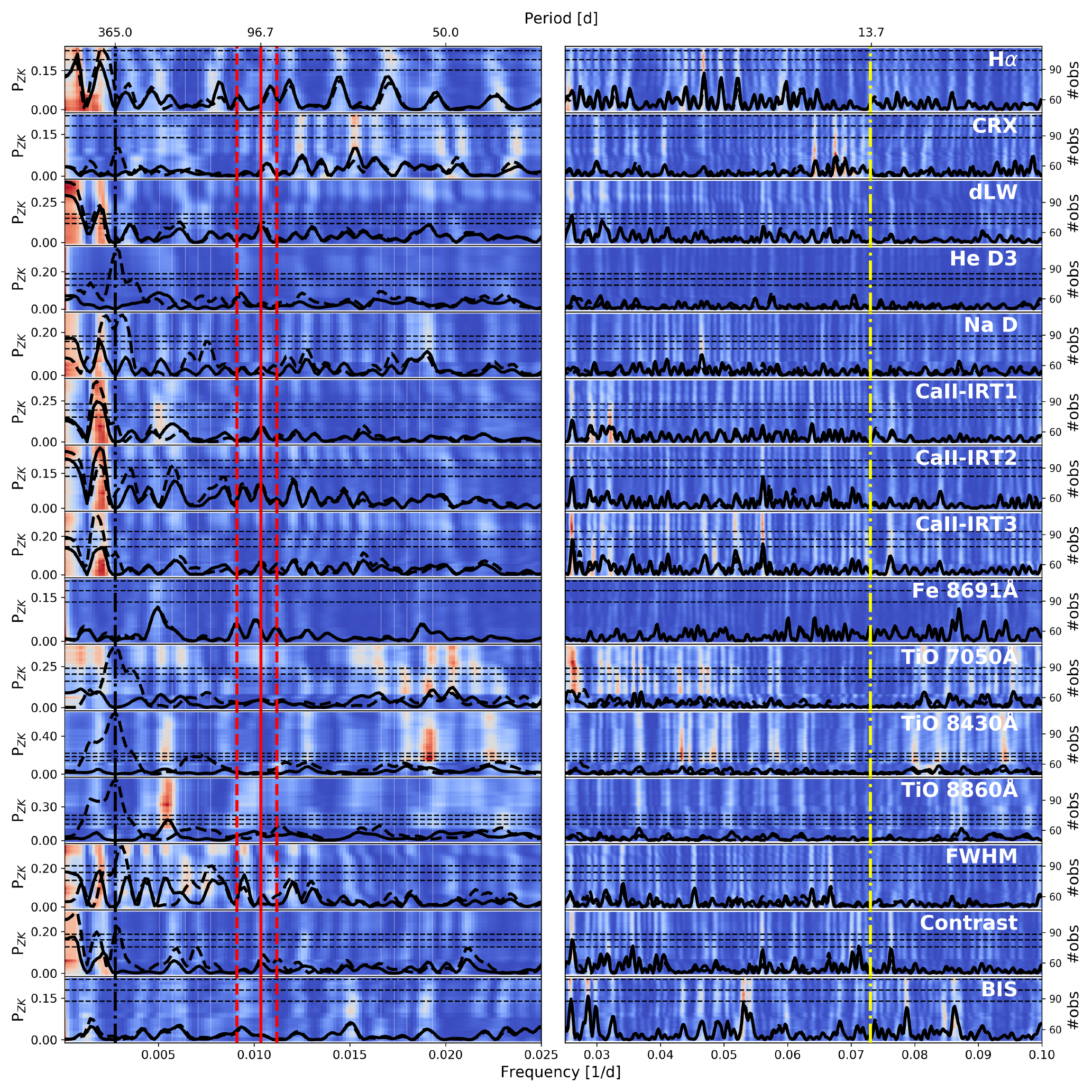}
\caption{Generalized Lomb-Scargle periodograms of several activity indicators based on spectroscopic data obtained by CARMENES for HD~238090. The dashed black periodograms represent the GLS on the activity indicators, and the solid GLS periodogram represents the residuals from which a 365\,d sinusoidal signal was subtracted. For the residuals from which the 365\,d signal was subtracted, we also overplot the s-BGLS periodogram, \Stephc{where the probability increases from blue to
white to red}. The \Steph{solid red} lines mark the rotation period estimated by photometric data, and the \Steph{dashed red} lines show the 3$\sigma$ uncertainties. The dashed black line marks a period of 365\,d, and the dashed yellow line marks the period of the planetary signal published in this work.}
\label{Fig: HD238090_act}
\end{figure*}

\begin{figure}
    \centering
    \includegraphics[width=9cm]{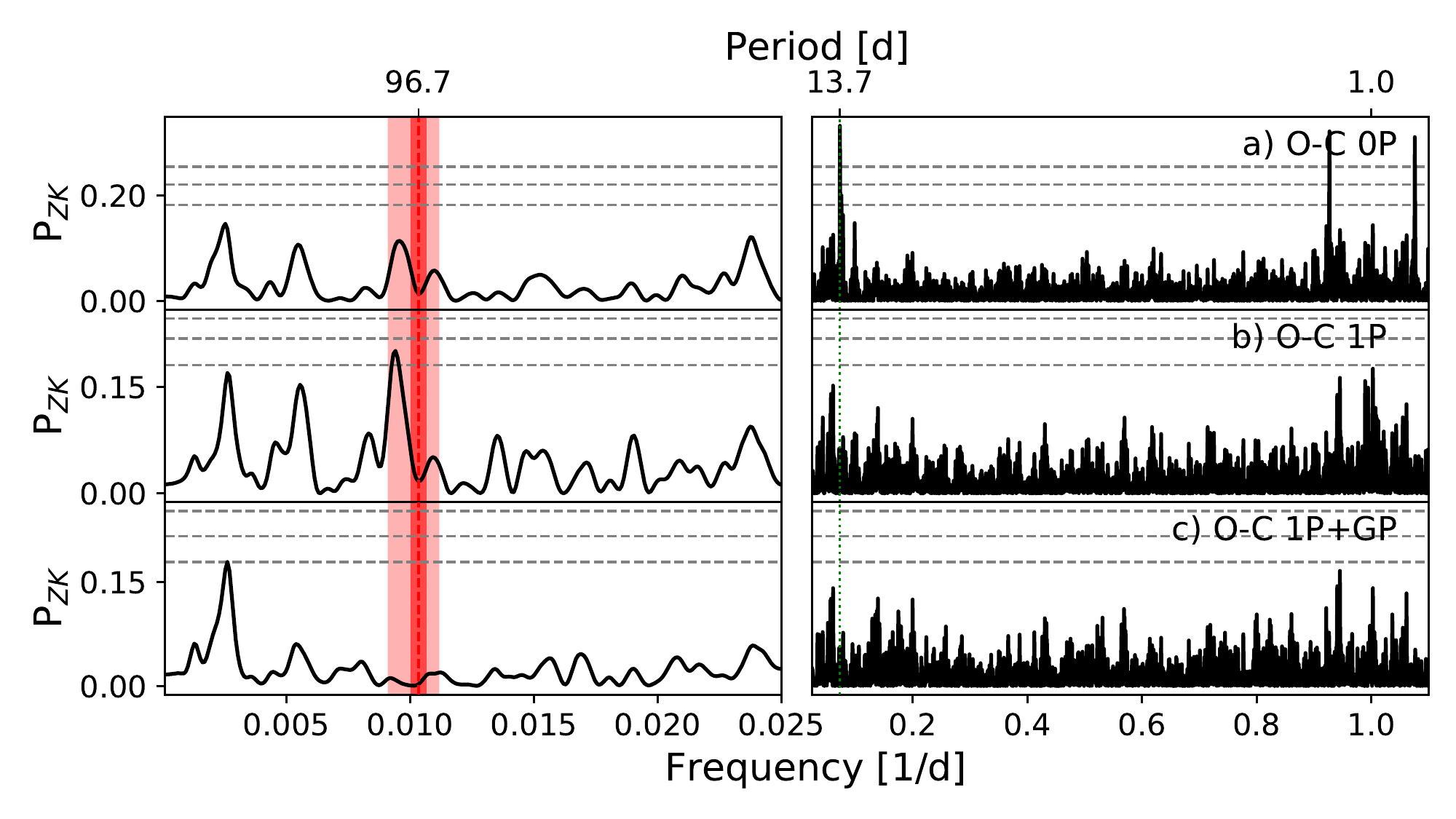}
    \caption{Generalized Lomb-Scargle periodogram of \Steph{RV data of} HD~238090 of the zero-planet fit, one-planet fit, and one-planet GP simultaneous fit. The stellar rotational period derived by photometry is plotted as the dashed red line, and $1\sigma$ and $3\sigma$ uncertainties are highlighted in red. The green line indicates the planet period at about 13.7\,d.}
    \label{fig: Model_comp_HD238090}
\end{figure}

We took ground-based photometry for HD~238090.
We combined our T90 and TJO data with public data from MEarth taken in 2009, 2010, and 2014, as well as data from NSVS taken between May 1999 and March 2000. A periodogram analysis of the combined data sets indicated periodicities around 100\,d. 
Using \texttt{juliet}, we fit for an offset and jitter terms for each instrument and filter. We used the same GP kernel and priors as for the photometric analysis of GJ~251 and separated the amplitude hyperparameters for each instrument while keeping global hyperparameters for the timescale and rotation. The priors are given in Table~\ref{Tab: Priors_photometry}. From our GP analysis, we derived that HD~238090 has a rotation period of $96.7^{+3.7}_{-3.2}$\,d. Figure~\ref{Fig: GP_HD238090} shows the distribution of the posterior samples in the $\alpha_{\tx{GP}}$-$P_{\tx{rot.}}$ space, and Fig.~\ref{Fig: HD238090_photometric_GP} shows the median GP model of each photometric data set together with the data and uncertainties. \Stephd{Following the same approach as for GJ~251 and applying the relations by \cite{Suarez2018}, we derived the median $\tx{log}(R'_{HK})$ and expected semiamplitude to $\tx{log}(R'_{HK})=-5.65^{+0.52}_{-0.58}$ (mean at $-5.69$) and $K_{\tx{exp.}}=0.83^{+4.40}_{-0.70}\,\mathrm{m\,s^{-1}}$ (mean at 4.25\,$\mathrm{m\,s^{-1}}$).}

\subsection{Spectroscopic activity indicators}

The periodogram analysis of our spectroscopic activity indicators is displayed in Fig.~\ref{Fig: HD238090_act}. \Stephc{We identified signals with an $\mathrm{FAP}<10^{-3}$ in some indices close to 1\,d and 365\,d (see Sect.~\ref{SubSect: GJ251_activity} for a discussion of these signals).}
After subtracting the yearly signal, the GLS periodograms of the residuals of many indicators show a signal around 480\,d with $\mathrm{FAP}<10^{-3}$ and a long-term trend. \Stephc{We also identified a signal in the FWHM CCF with an $\mathrm{FAP}<10^{-2}$ at 106.1\,d, which is close to the derived stellar rotation period.} Within the s-BGLS of the residual activity indicators of TiO, which we show in Fig.~\ref{Fig: HD238090_act}, we found various signals around 50\,d, which is roughly half the photometrically derived stellar rotation period. \Stephb{These signals were more significant in previous observations between July 2018 and February 2019 (CARMENES observations 60 to 90).} \Steph{Overall, the activity indicators show that the star exhibits no significant level of activity at \Stephb{periods shorter than 20\,d} over the time of RV observations.}

\subsection{Periodogram analysis}

The GLS periodogram of the RV data for HD~238090 is shown in Fig.~\ref{fig: Model_comp_HD238090}. A significant peak with an $\mathrm{FAP}<10^{-5}$ is visible at 13.68\,d. \Stephc{Two additional signals with almost the same GLS power accompany this signal at periods close to 0.93\,d and 1.08\,d.} No additional signals were significant in the data. 
\Stephc{The investigation with \texttt{AliasFinder} led to the conclusion that the 13.68\,d period is the \Steph{most probable} true period of the sampled signal because simulated periodograms based on this period fit the observed periodogram better than the periods close to one day. We show the corresponding plot obtained by \texttt{AliasFinder} in Fig. \ref{Fig: HD238090_alias1}.  }

\begin{figure}
\centering
\includegraphics[width=4.4cm]{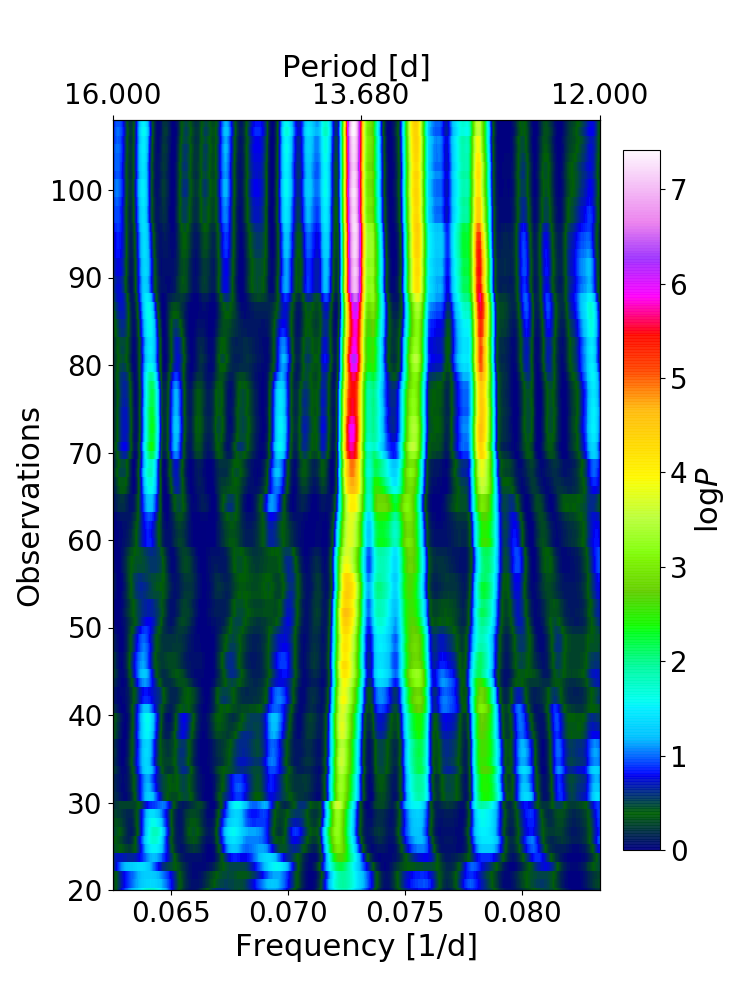}
\includegraphics[width=4.4cm]{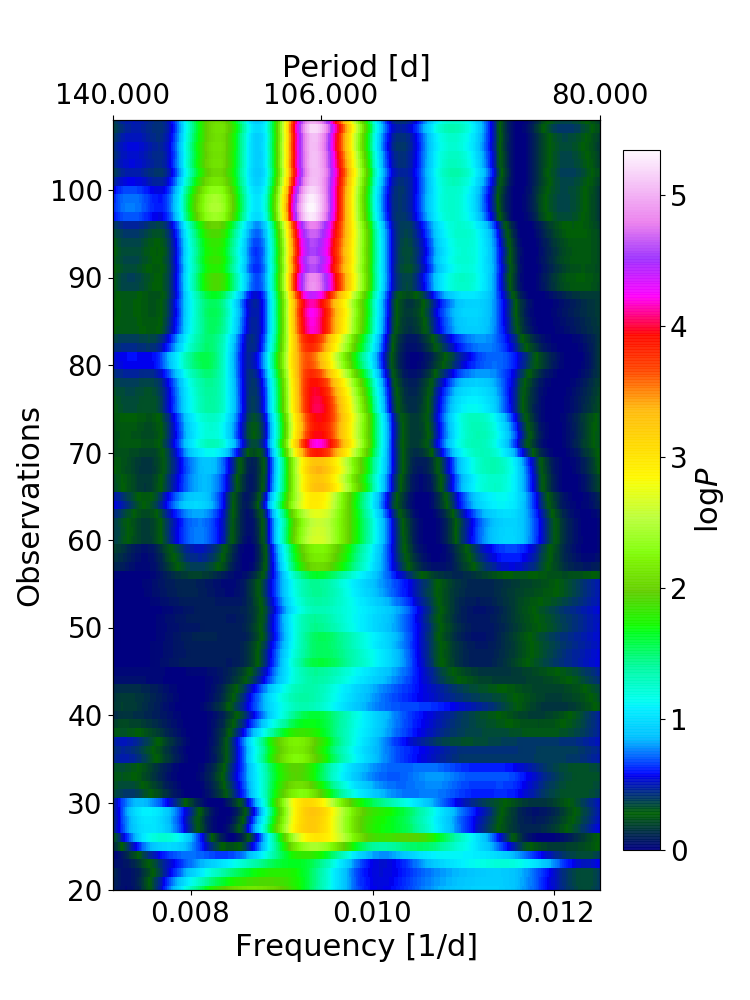}
\caption{\Stephf{Stacked-Bayesian} GLS of the planetary signal (left) and a signal close to the estimated stellar rotation (right).}
\label{Fig: HD238090_sbgls}
\end{figure}

An investigation of the 13.68\,d signal with the s-BGLS showed that the signal was coherent and increased in probability over the observation time. We display the s-BGLS of the signal in Fig.~\ref{Fig: HD238090_sbgls}.

\begin{figure}
\centering
\includegraphics[width=9cm]{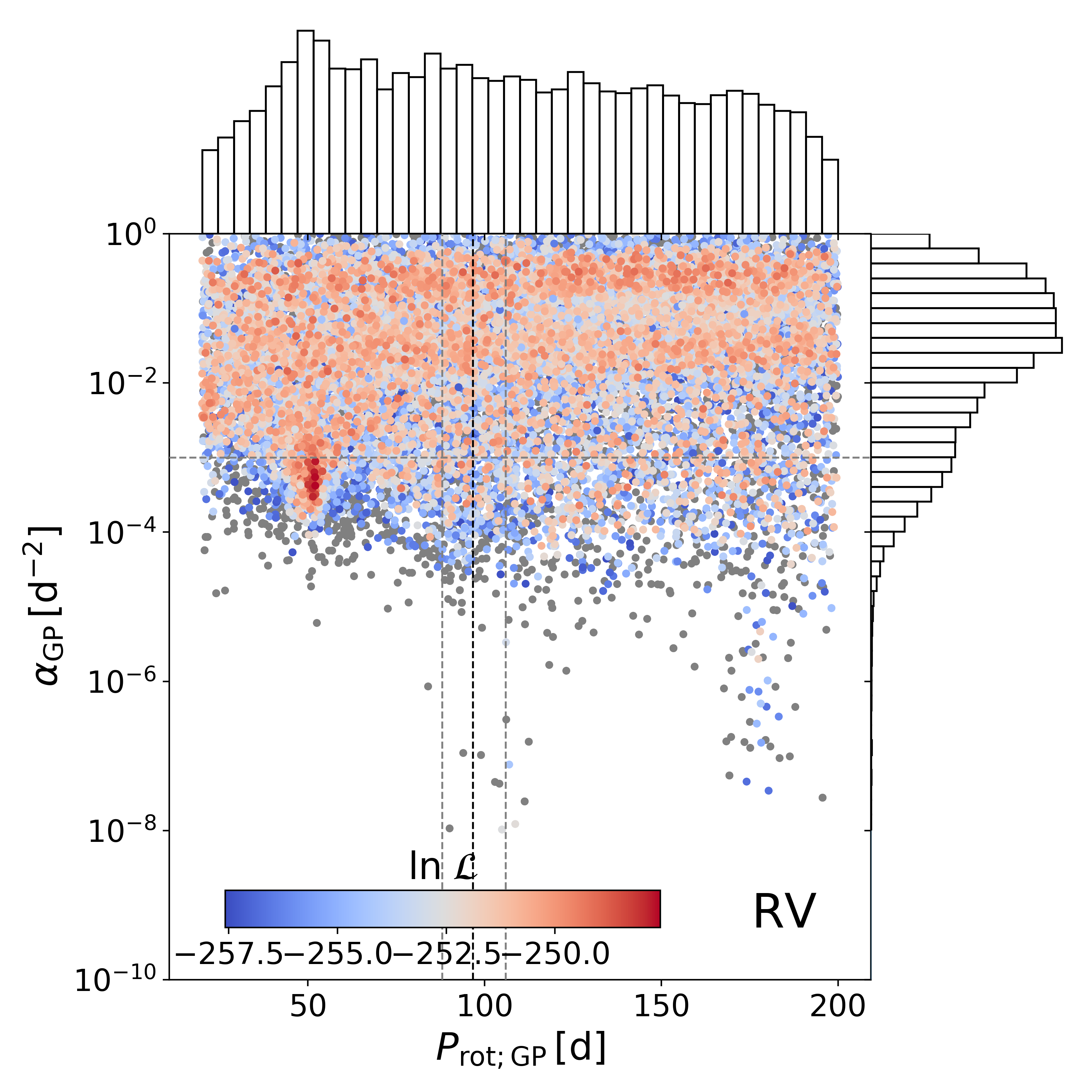}//
\includegraphics[width=9cm]{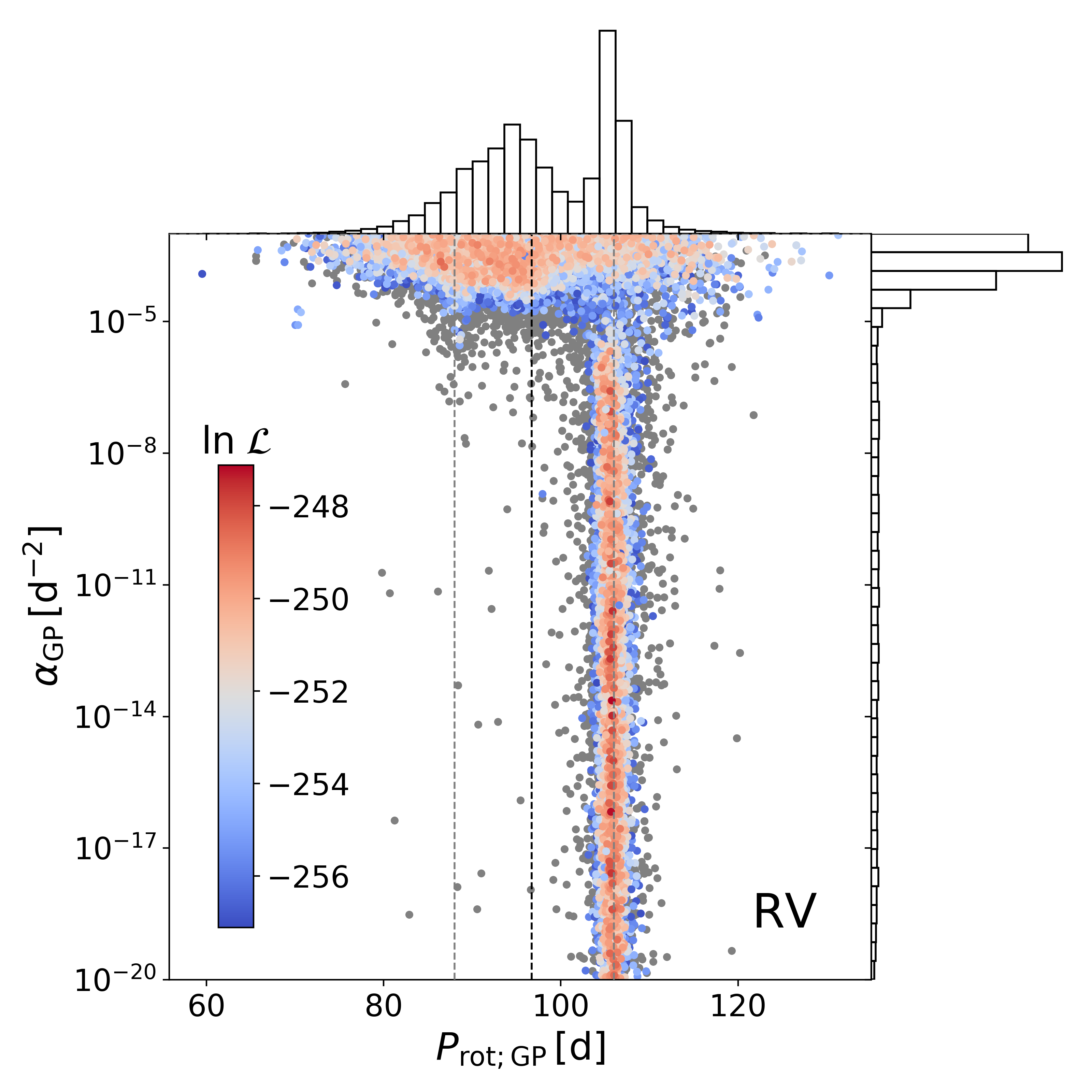}
\caption{Posterior distribution of the GP fit to the RV data in the $\alpha_{\mathrm{GP}}$ vs.\ $P_{\mathrm{rot}}$ plane for HD~238090. The color-coding shows the log-likelihood normalized to the highest value within the posterior sample. Gray samples indicate solutions with a $\Delta \ln{L}>10$. \textit{Top:} GP fit to the RV data with a wide uniform prior to the rotational period. \textit{Bottom:} GP fit to the RV data with an informative normal prior based on the photometric GP results and an upper $\alpha_{\tx{GP}}$ constraint.}
\label{Fig: GP_HD238090_RV}
\end{figure}

\begin{figure}
\centering
\includegraphics[width=9cm]{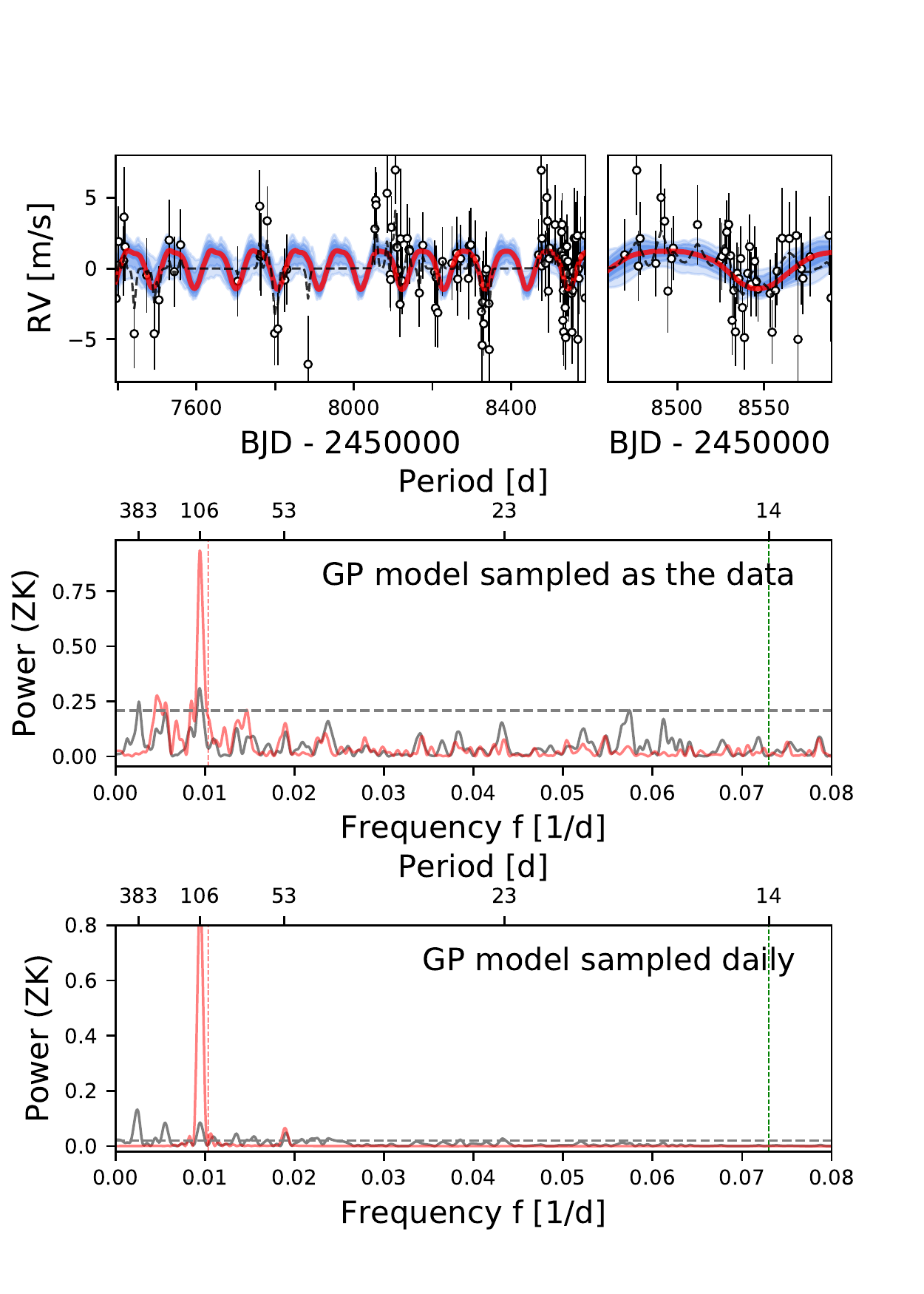}
\caption{Gaussian-process model for the RV data of HD~238090. The planetary signal is not included in the model and subtracted from the RV data. The constrained GP model is shown in red. The blue regions shows the 1$\sigma$, 2$\sigma$, and 3$\sigma$ uncertainties. We show a zoom into some CARMENES observations (top right). The GLS is evaluated on the GP model at each observed data point (top GLS) and daily (bottom GLS). The dashed line in the GLS periodograms indicates an FAP of 0.001. We also show the unconstrained GP model as the dashed black line in the upper plots and the gray periodograms in the lower plots. }
\label{Fig: HD238090_Model_components}
\end{figure}

\begin{figure*}
\centering
\includegraphics[scale=0.5]{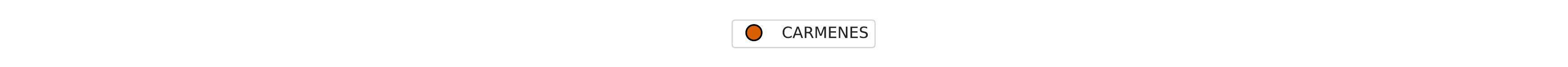}

\includegraphics[scale=0.24]{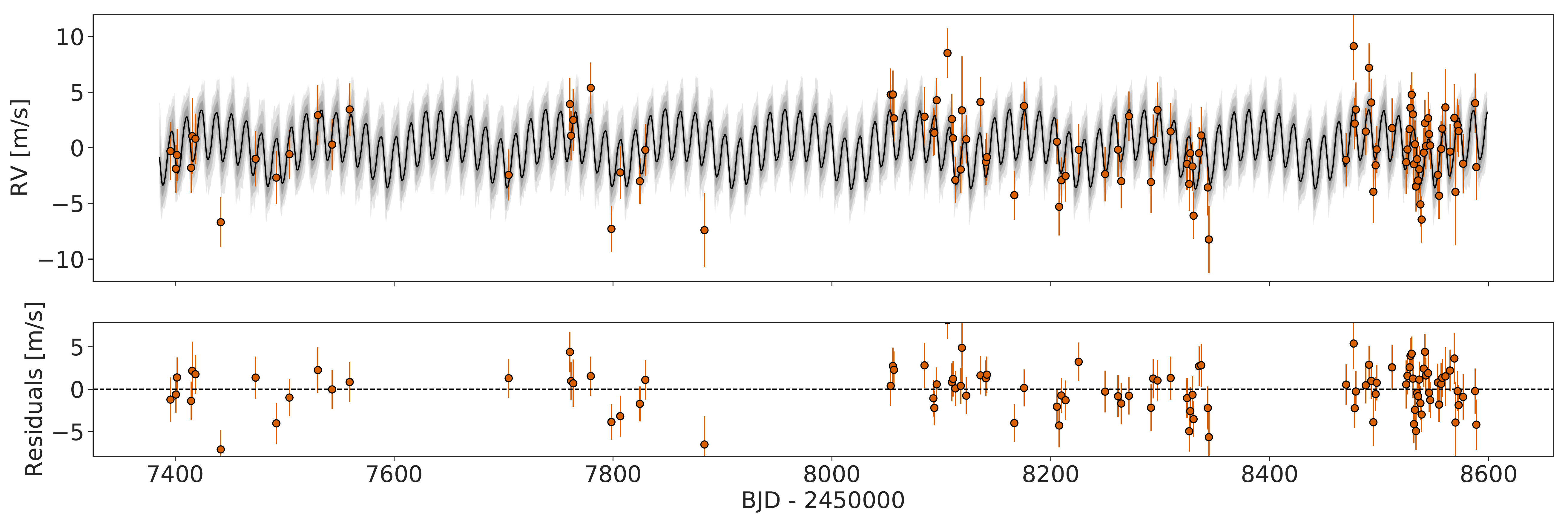}
\includegraphics[scale=0.255]{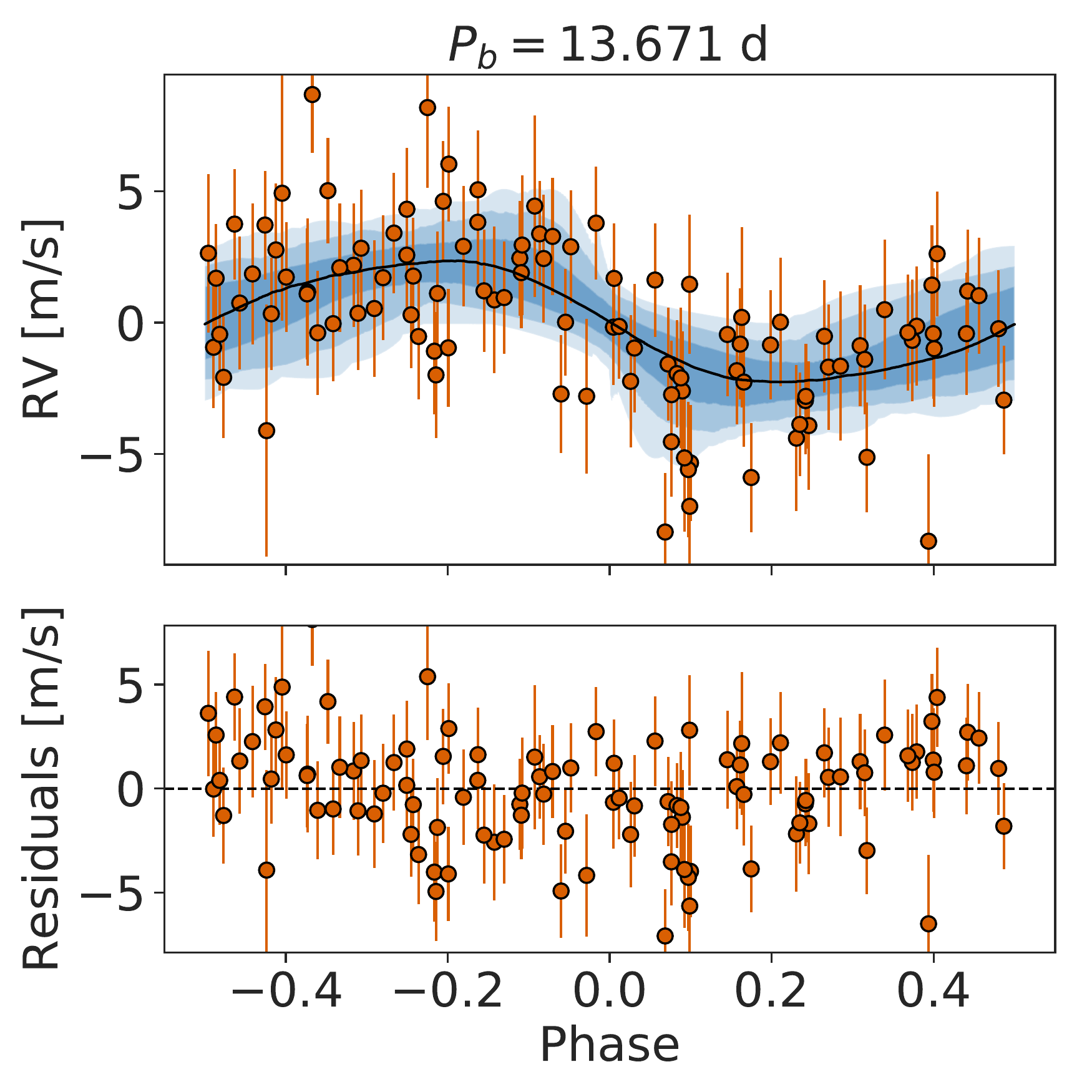}
\caption{\textit{Left:} Radial velocity data with a combined model for one planet and activity using a Keplerian model and a GP. \textit{Right:} Plot phased to the orbital period of HD~238090~b \Steph{without a GP component.}}
\label{Fig: HD238090_model}
\end{figure*}

\subsection{RV modeling}

\begin{table}
\caption{Bayesian log-evidence for HD~238090 and a number for different models based on CARMENES data.}
\label{Tab: HD238090_fit_evidence}
\centering
\begin{tabular}{l c c r}
\hline\hline
Model$^{a}$ & Periods\,[d] &$\ln{\mathcal{Z}}$ & $\Delta\ln{\mathcal{Z}}$ \\
\hline
0p & \ldots & $-287.3\pm0.1$ & 0\\
\Stephc{GP} & \Stephc{\ldots} & \Stephc{$-284.9\pm0.1$} & \Stephc{2.4}\\
1p &  13.7 & $-273.0\pm0.2$ & 14.3\\
1cp &  13.7 & $-273.3\pm0.2$  & 14.3\\
1p+1cp &  13.7,106.6 & $-267.8\pm0.2$  & 19.5\\
\Stephc{2p+GP} & \Stephc{7.0, 13.7} & \Stephc{$-267.7\pm0.2$} & \Stephc{19.4}\\
1p+uGP &  13.7  & $-267.0\pm0.2$ & 20.3\\
1p+GP &   13.7 & $-267.1\pm0.2$ & 20.2\\
1cp+GP &   13.7 & $-267.0\pm0.2$ & 20.3\\
\Stephc{2p+GP} & \Stephc{13.7, 389.8} & \Stephc{$-265.1\pm0.2$} & \Stephc{22.2}\\
\hline
\end{tabular}

 \tablefoot{
 \tablefoottext{a}{Planetary models based on CARMENES RV data.
        0p: 0 planets, 1p: one planet, 1cp: one planet in circular orbit ($e = 0$).
        GP and uGP: additional constrained and unconstrained GPs, respectively. Sin: additional sinusoidal model.
        Orbital periods rounded to one decimal.}
      
    }
\end{table}

\Steph{Because the signal at 13.68\,d showed long-term coherence and had no counterparts in the activity indicators, we fit a Keplerian model to the 13.68\,d signal, hereafter referred to as HD~238090~b.}
 Table~\ref{Tab: HD238090_fit_evidence} shows that a one-planet fit is significantly favored by the data compared to a flat model.

The residual periodogram of a one-planet fit to the 13.68\,d signal showed peaks with an FAP of almost $10^{-2}$ for 106.4\,d.  The signal at 106.4\,d is below our optimal detection criterion by the GLS analysis, which is $\mathrm{FAP}<10^{-3}$. \Stephc{Additionally, it resides within the 3$\sigma$ uncertainty of the derived rotation period of the star by photometry and has a counterpart in the FWHM CCF, as discussed before}. An s-BGLS analysis of the 106\,d signal indicated that the probability of the signal decreased by almost two magnitudes after 30 observations before reappearing in later CARMENES epochs. We show the s-BGLS in Fig.~\ref{Fig: HD238090_sbgls}.   

\Stephc{These results indicate that the signal is probably of non-Keplerian origin. Nevertheless, we performed an additional statistical model comparison where we compared a second circular Keplerian to a GP for this signal. 
The two-planet fit resulted in a log-evidence improvement compared to a one-planet model of $\Delta\ln Z = 5.2$. A wide unconstrained GP to account for the 106\,d signal performed equally well. Statistically, there is no clear tendency for a GP or Keplerian model. Future observations of HD~238090 might be warranted to completely exclude the possibility of a Keplerian signal with a period of 106\,d, especially because such a planet would reside inside the optimal habitable zone described by \cite{Kopparapu2013}. Based on the current data and our photometry, s-BGLS, and activity indicator analysis, we regard the 106\,d signal as caused by stellar activity.}

\Stephc{As for GJ~251, we improved the GP modeling by constraining the prior volume (and therefore the posterior). The GP alpha-period diagram ($\alpha_{\textnormal{GP}}$ versus\ $P_{\textnormal{GP}}$) of the unconstrained GP is shown in the top plot of Fig.~\ref{Fig: GP_HD238090} and the priors are given in Table~\ref{Tab: Priors_GP_RV}. 
We found a posterior overdensity with a marginally higher likelihood around 50\,d, which is close to half the derived photometric rotation period for the unconstrained GP.
However, the distribution of posterior samples showed no peculiar structure overall and suggested that the RV data of HD~238090 are not significantly affected by a strong correlated quasi-periodic signal with decay-timescales of more than several days.}

\Stephc{In the lower plot of Fig.~\ref{Fig: GP_HD238090} we show a GP for which we constrained the rotation parameter $P_{\textnormal{GP}}$ to be Gaussian distributed around the derived stellar rotation period and the timescale parameter $\alpha_{\textnormal{GP}}$ to be between $10^{-3}$ and $10^{-20}$. With the new priors, the dynamic nested sampling algorithm found posterior solutions around 106\,d that reached a similar maximum likelihood as the unconstrained GP posterior samples. The log-evidence of this model was equal to the unconstrained GP. That the distribution of $\alpha_{\textnormal{GP}}$ reached the prior boundary at $10^{-20}$ implies that this parameter converged to zero because this decay timescale is orders of magnitudes longer than the observation time. }

\Stephc{These GP results motivated us to fix the $\alpha$ value to $10^{-20}$, which is consistent with zero. This resulted in only three free hyperparameters to be fit by the GP. \Stephc{This simpler rotational GP performed similarly in terms of log-evidence as the two previously described QP-GPs}. We applied this GP as the final activity model of the system, hereafter called constrained GP, and show the model in Fig.~\ref{Fig: HD238090_Model_components}. For comparison, we plot the unconstrained quasi-periodic GP model, which included high $\alpha_{\textnormal{GP}}$ values within its posterior, in the same figure. %One can see that ***
The constrained GP model represents a more realistic fit of the data. \Stephd{The GP semiamplitude is consistent within $1\sigma$ with the expected semiamplitude that is due to the stellar rotation, estimated based on the relations of \cite{Suarez2018}, and the GP rotation period is within the $3\sigma$ uncertainty of the photometric estimate of the stellar rotation.} The GLS periodogram \Stephc{decomposition} of the constrained GP model sampled daily indicates that the GP only models the 106\,d period and its first harmonic.} 

\Stephc{We performed a search for any additional planetary signal hidden behind the stellar activity with the tuned activity model. For this, we included a second Keplerian signal using a log-uniform prior between 0.5\,d, and 13\,d and then 14\,d to 1000\,d for its period, while simultaneously fitting for GJ~251~b and the stellar activity with the constrained GP.  We found that the two two-planet models together with the GP did not perform significantly better than the one-planet plus GP model. We therefore preferred the one-planet model for its simplicity.}

We show the final one-planet and activity model to the RV data in Fig. \ref{Fig: HD238090_model} and the posterior parameters in Table~\ref{Tab: Posteriors}. We derived a nonsignificant eccentricity of $0.30^{+0.16}_{-0.17}$, as a circular model resulted in similar log-evidence. \Stephc{We derive the minimum mass of HD~238090~b to be $6.89^{+0.92}_{-0.95}$\,$M_\oplus$.} We provide corner plots of all posterior samples in the appendix in Fig.~\ref{Fig: HD238090_corner_GP} and Fig.~\ref{Fig: HD238090_corner_planet}, respectively.

\subsection{Transit search with TESS}

\Stephc{\emph{TESS} observed HD~238090 (TIC 224289449) in sectors 15, 21, and 22.} Based on the mass-radius relation by \cite{Zeng2016} and applying a core-mass fraction of 0.26, which corresponds to Earth-like composition, we estimated a planetary radius of 1.69\,R$_{\oplus}$ for HD~238090~b. Given the stellar parameters of HD~238090, the transit depth \Stephc{was approximated to} 0.72\,ppt, which should be detectable in the \emph{TESS} light curve. We \Stephd{investigated} the \emph{TESS} light curve around the estimated $t_0$ from the RV fit. 
\Stephd{We did not identify any transit event.}

\Stephb{We calculated the TLS periodogram of the light curve and found three signals with a signal detection efficiency (SDE)\,$>$\,7, which we regard as significant \citep{Hippke&Heller2019}, at periods of 30.00\,d, 27.38\,d, and 13.67\,d.} The first periods are close to the \emph{TESS} sector length, \Stephc{while the 13.67\,d period is about half the TESS sector length, but represents exactly the period of HD~238090~b.} \Stephc{However, the time of transit center derived for the 13.67\,d signal detected in the TLS is incompatible with the value obtained from the RV analysis ($t_{0,\tx{transit}}=2458638.30\pm0.05$\,BJD versus $t_{0,\tx{RV}}=2458630.35\pm0.07$\,BJD). Inspection of the \emph{TESS} light curve showed that the presumed transits were fit at the edges or inside observational gaps of the light curves of the \emph{TESS} sectors.} We concluded that these TLS signals, although significant and close to the planetary period, represent false positives caused by the observational sampling \Stephc{and that no transits of HD~238090~b are detected.}

\section{Lalande~21185}
\label{Sect: Lalande}

\begin{figure*}
\centering
\includegraphics[width=18cm]{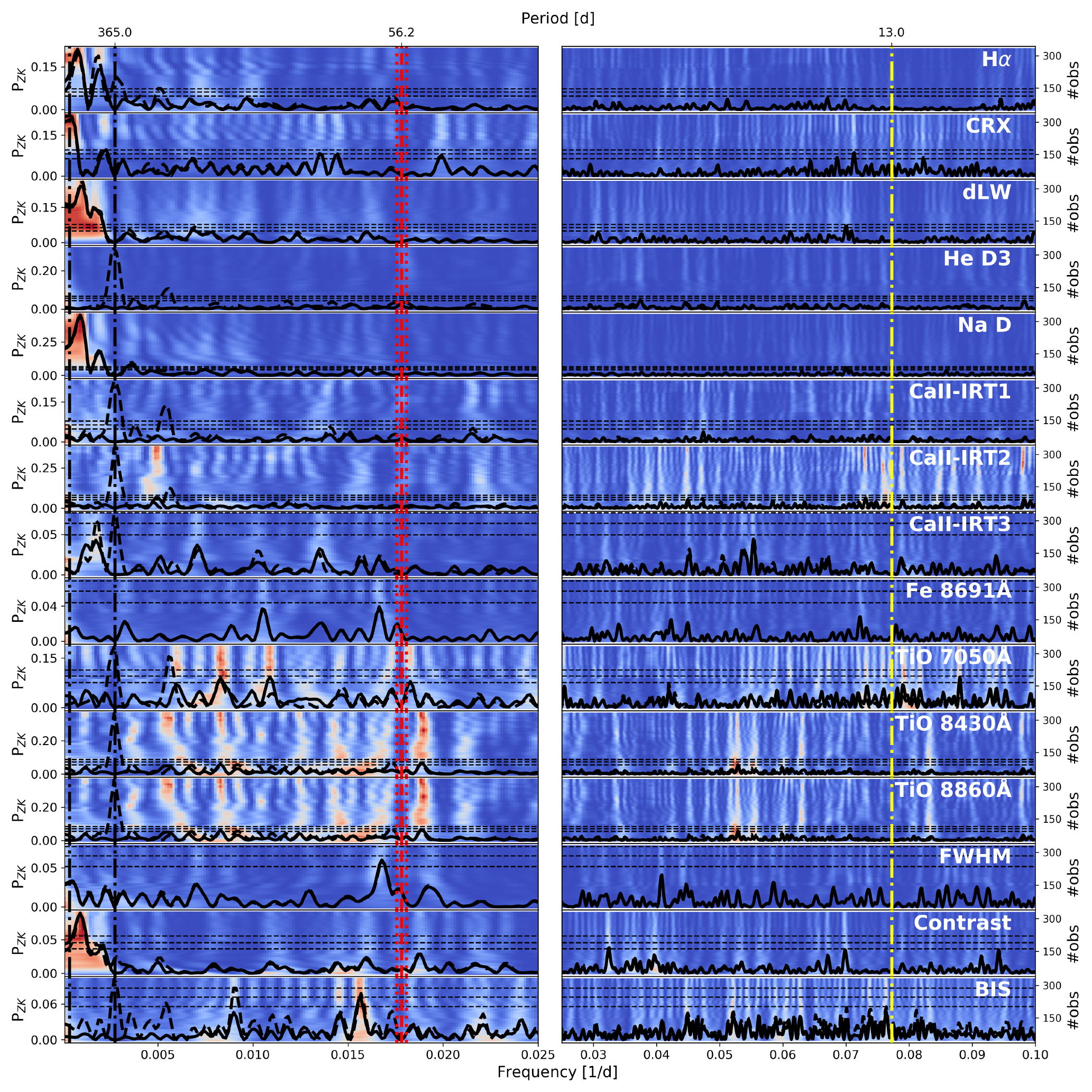}
\caption{Generalized Lomb-Scargle periodograms of several activity indicators based on spectroscopic data obtained by CARMENES for Lalande~21185. The thinner dashed black periodograms represent the GLS on the activity indicators, and the solid GLS periodogram represents the residuals from which a 365\,d sinusoidal signal was subtracted. For the residuals from which the 365\,d signal was subtracted, we also overplot the s-BGLS periodogram, \Stephc{where the probability increases from blue to white to red}. The dashed red lines mark the rotation period and the first harmonic estimated by photometric data, and the dotted red lines show the 3$\sigma$ uncertainties. The dashed black lines mark a period of 365\,d, 1400\,d, and 2800\,d, and the dashed yellow line marks the period of the planetary signal published in this work.}
\label{Fig: Lalande_act}
\end{figure*}

\begin{figure}
\centering
\includegraphics[width=9cm]{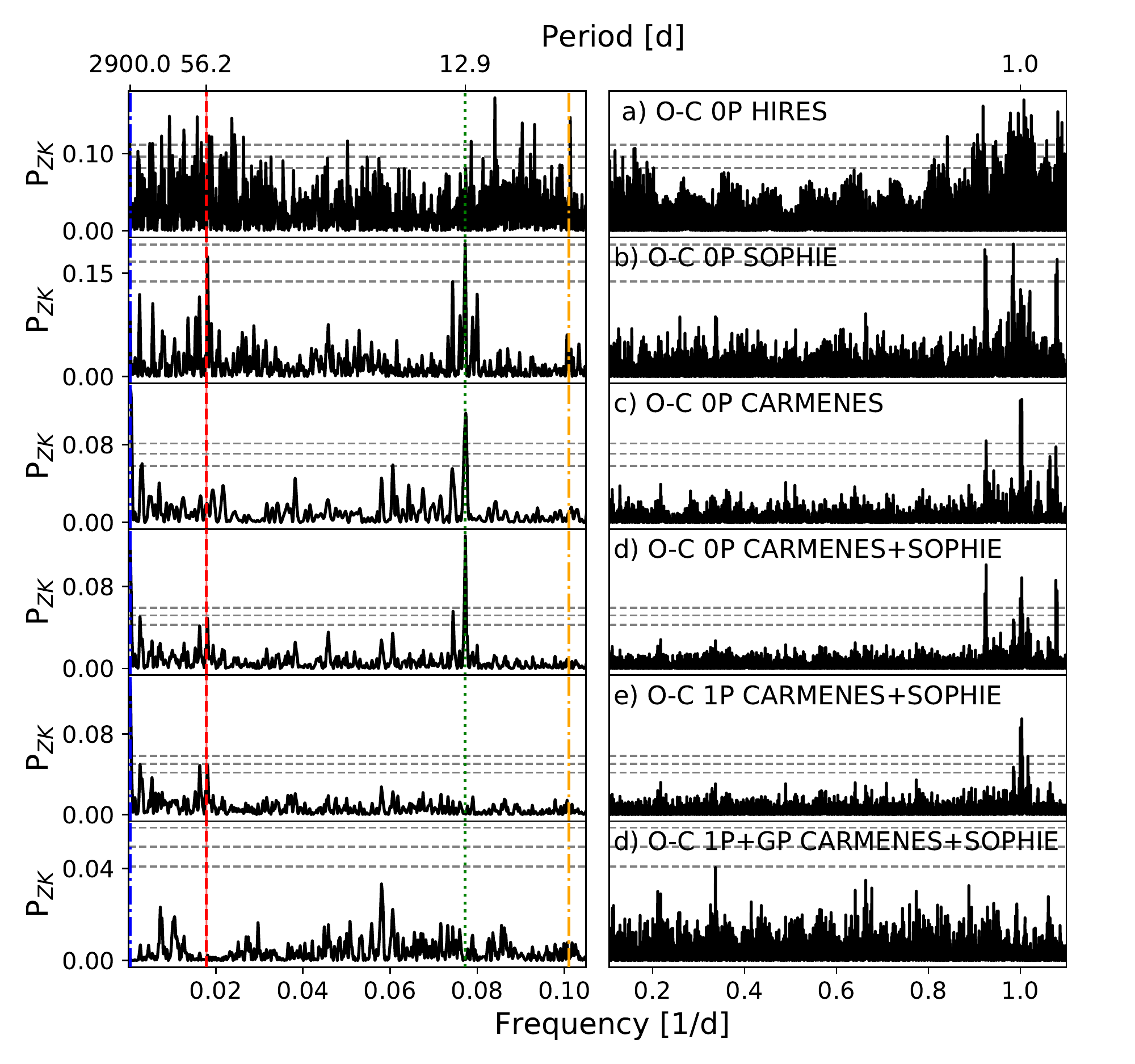}
\caption{Generalized Lomb-Scargle periodograms of Lalande~21185 for the different data sets and the combined CARMENES and SOPHIE data. Residual periodogram of the one-planet fit and the one-planet GP simultaneous fit for the combined CARMENES and SOPHIE data are also shown. Marked frequencies represent the claimed planet at 12.95\,d by \cite{Lalande2019} (green), photometric stellar rotational period (red), the claimed planet candidate by \cite{Butler2017} (orange), and the long-term period (blue).}
\label{Fig: Lalande_GLS}
\end{figure}

\cite{Lalande2019} reported the discovery of a temperate super-Earth orbiting Lalande~21185 with a period of 12.95\,d or, less likely, 1.08\,d, but the period could not be determined unambiguously because of aliasing. \cite{Lalande2019} did not find evidence for a planetary candidate reported previously by \cite{Butler2017}, which was supposed to orbit the star at 9.9\,d. We reanalyzed the HIRES and SOPHIE data together with our CARMENES observations.

\subsection{Photometry and spectroscopic activity indicators.}
\cite{Lalande2019} took extensive photometric observations between 2011 and 2018 with the Tennessee State University T3 0.40\,m automatic photoelectric telescope at Fairborn Observatory in southern Arizona. Based on their analysis, they reported a stellar rotation period of $56.15\pm0.27$\,d for Lalande~21185. This rotation period is consistent with values previously obtained by \citet[][48\,d]{Noyes1984} and by \citet[][54\,d]{Olah2016}, although these studies did not provide uncertainties. 

\Stephd{Based on the derived stellar rotation period by \cite{Lalande2019}, we made use of the relations by \cite{Suarez2018} and derived for the median $\tx{log}(R'_{HK})$ and median expected RV semiamplitude $\tx{log}(R'_{HK})=-5.34^{+0.53}_{-0.58}$ (mean at $-5.37$) and $K_{\tx{exp.}}=1.50^{+7.50}_{-0.1.27}\,\mathrm{m\,s^{-1}}$ (mean at 7.27\,$\mathrm{m\,s^{-1}}$), respectively.}

\Stephc{The periodograms of the activity indicators are provided in Fig.~\ref{Fig: Lalande_act}.}
After subtracting the yearly signal (see Sect. \ref{SubSect: GJ251_activity} for discussion of this signal), we found several signals in the CARMENES activity indicators \Stephc{close to the reported} stellar rotational period of Lalande~22185, e.g., TiO at 7050\,\AA\, with an $\mathrm{FAP}< 10^{-1}$, TiO at 8430\,\AA\, and at 8860\,\AA\, with an $\mathrm{FAP}<10^{-2}$, in the FWHM with an $\mathrm{FAP}<10^{-1}$ and in BIS with an $\mathrm{FAP}<10^{-2}$. The s-BGLS of these signals show the instability of these signals over the observation time.  For the activity analysis of the SOPHIE data, we refer to \cite{Lalande2019}. 

In addition to these \Stephc{stellar rotation related signals, we observed two significant signals ($\mathrm{FAP}<10^{-3}$) in H$\alpha$ at periods of 1313\,d and 539\,d, and around 1400\,d in the dLW, the Na lines, and the contrast CCF. Additionally, we observed a linear trend in the CRX. }
In the CRX, dLW, and the Na lines, we identified signals at roughly 14\,d (with FAPs of $<0.1$, $<0.01$, and $\sim0.1$, respectively), which is about $1/4$ of the stellar rotational period, but close to the claimed planetary signal by \cite{Lalande2019}. Owing to the long CARMENES time baseline, the GLS periodogram resolution allows us to separate these peaks from the planetary signal at 12.95\,d.

\begin{figure*}[ht!]
\centering
\subfloat[]{\includegraphics[width=16cm]{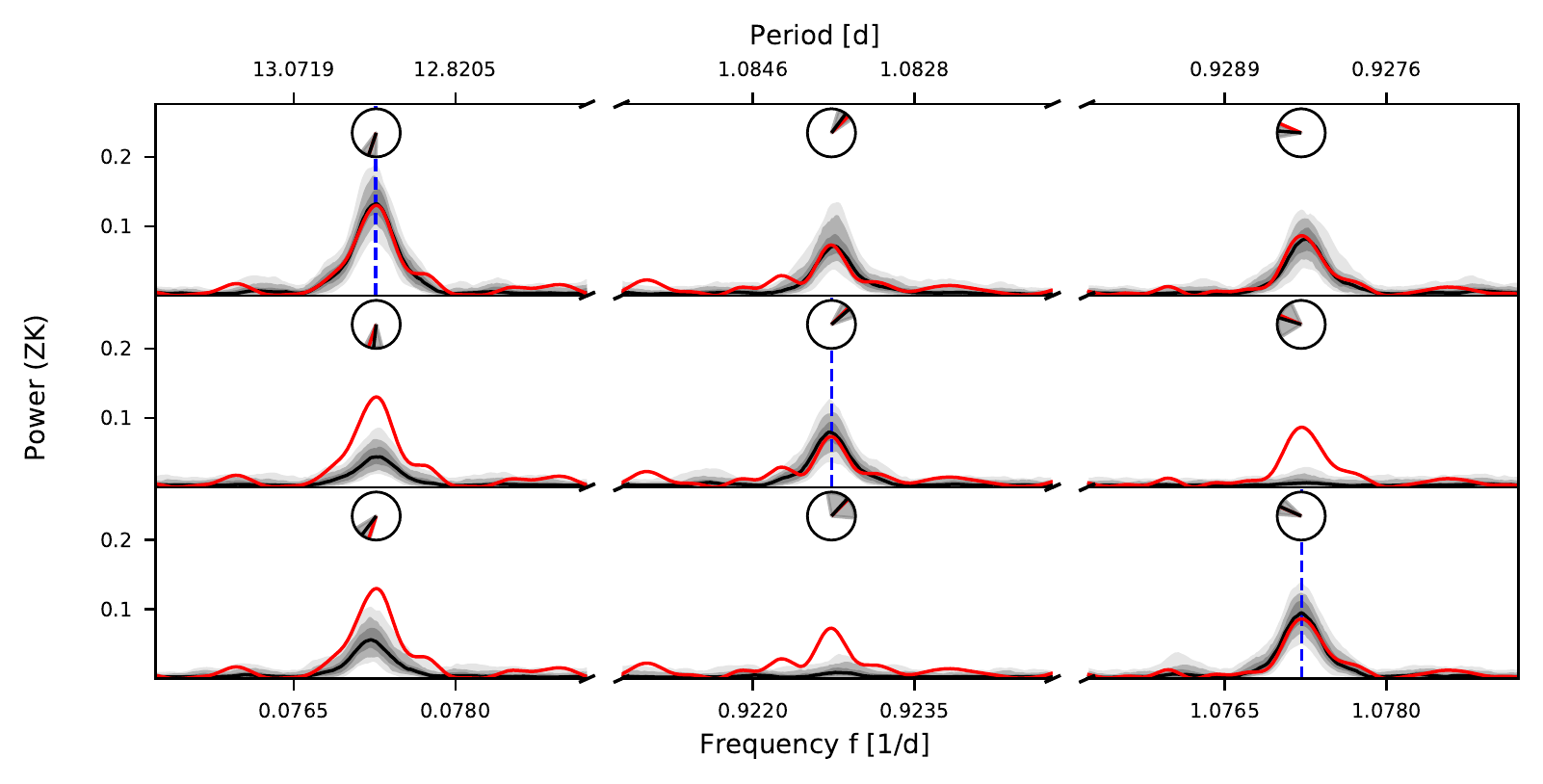}}\\
\subfloat[]{\includegraphics[width=16cm]{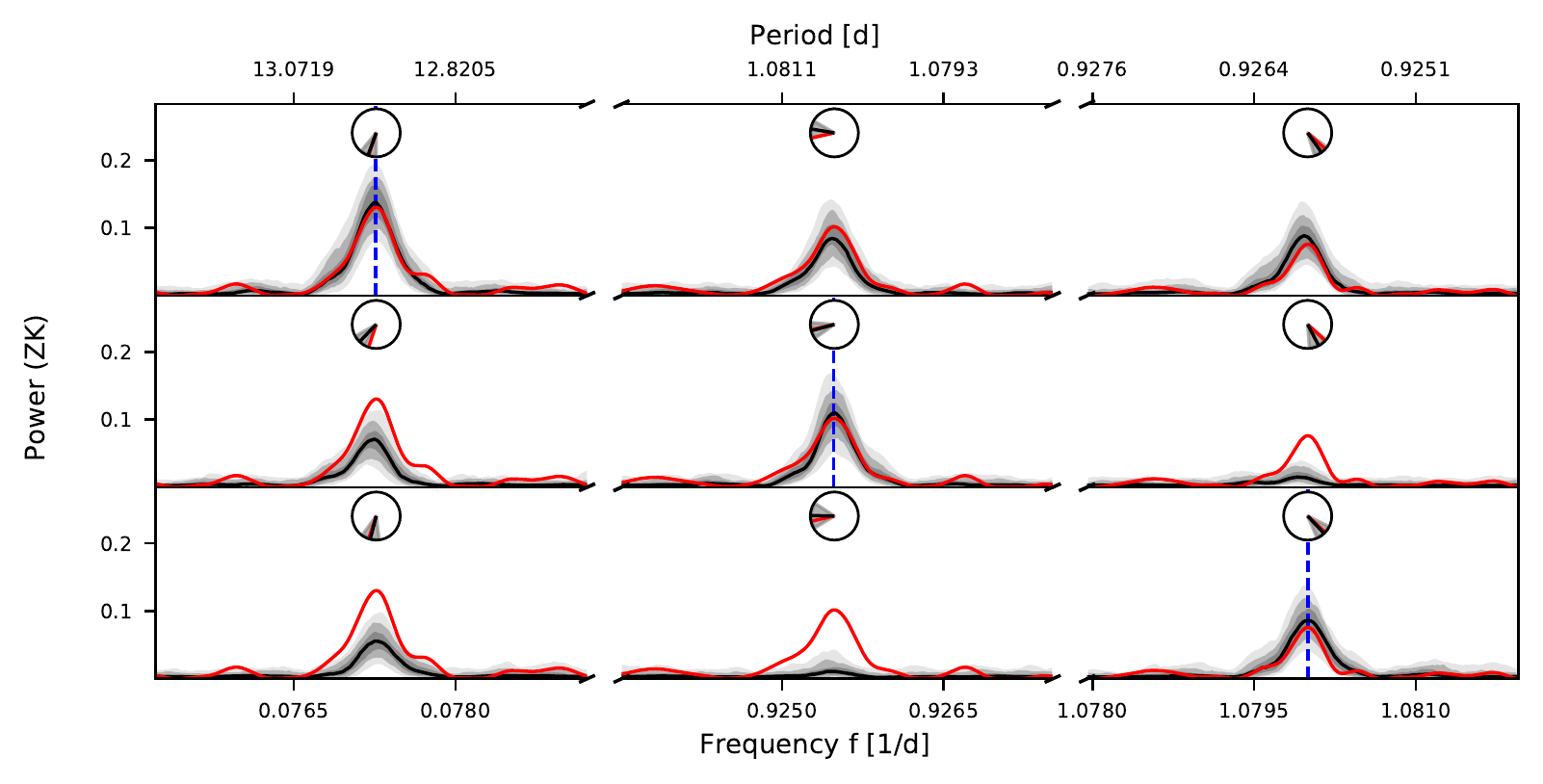}}
\caption{Alias tests for Lalande~21185. The top plot (a) shows simulations motivated by a sampling frequency of $f_{s_1}=1.0000\,{d^{-1}}$. The bottom plot (b) shows simulations motivated by a sampling frequency of $f_{s_2}=1.0027\,{d^{-1}}$. Each row in these plots corresponds to one set of simulations for which the frequency of the injected signal is indicated by a vertical dashed \Stephf{blue line}. The first row shows simulations with a period of $12.95$\,d, and the second and third row show the simulations in which the first-order aliases of $12.95$\,d, regarding the investigated sampling frequency, were injected. Each column shows informative ranges of the periodograms based on the assumed sampling frequency and can be used to compare data and simulations. From 1000 simulated data sets, we show the median of the obtained periodograms (solid black line), the interquartile range, and the ranges of 90\% and 99\% (gray shades). The periodogram of the observed data is plotted with a solid red line. The angular mean of the phase and the standard deviation is shown in the clock diagrams (black line and gray shades) and can be compared to the phase of the signals in the observed periodogram (red line).}
\label{Fig: Lalande21185_alias}
\end{figure*}

\begin{figure*}
\centering
\includegraphics[width=5.6cm]{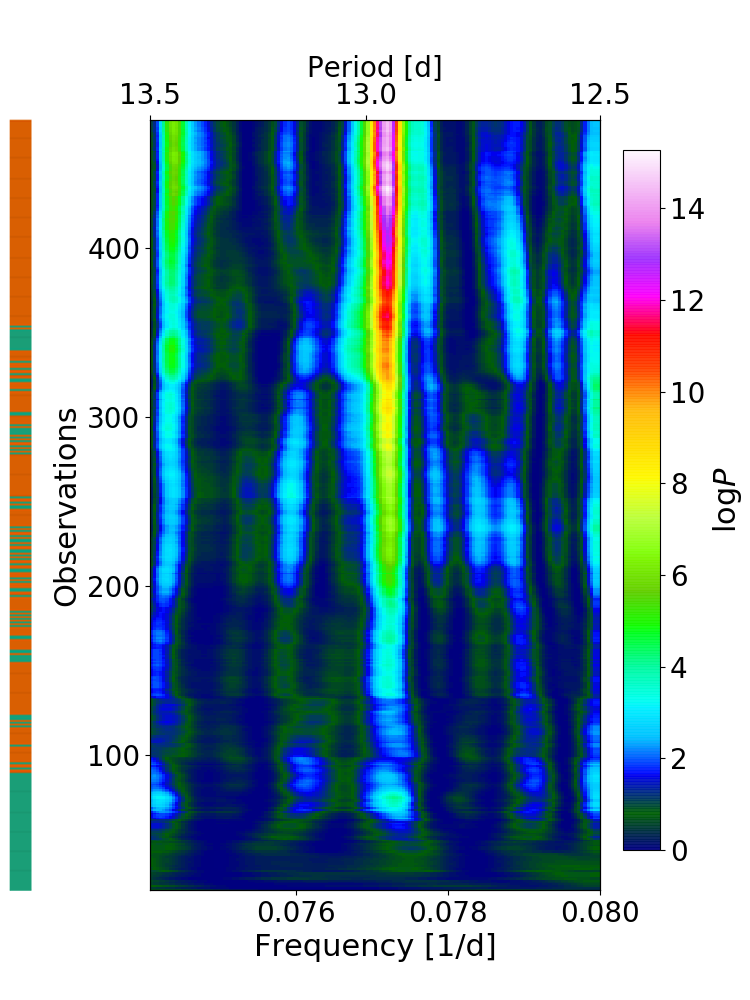}
\includegraphics[width=5.6cm]{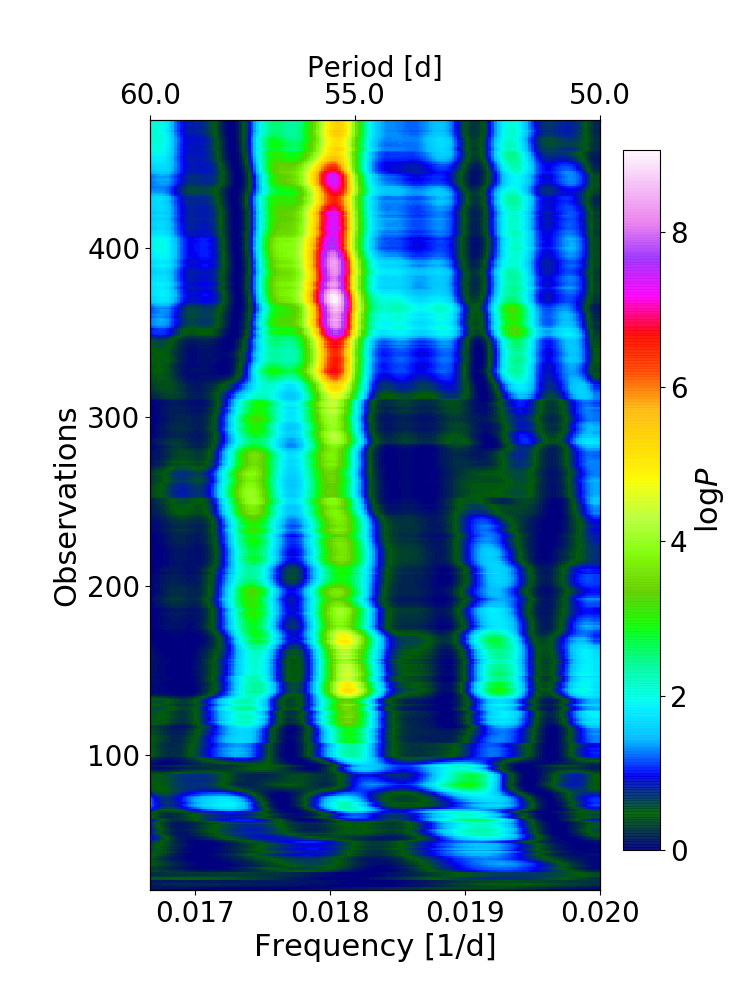}
\includegraphics[width=5.6cm]{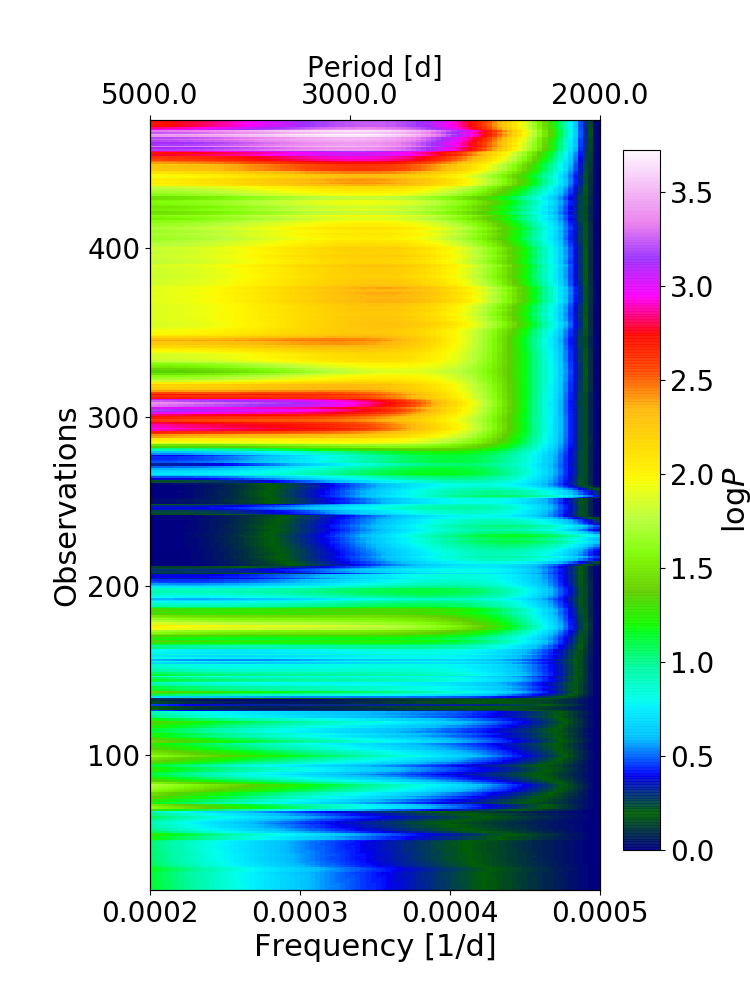}
\caption{Stacked-Bayesian GLS periodogram of Lalande~21185. The color bar to the left color-codes each observation with the associated spectrograph (\textit{orange}: CARMENES, and \textit{teal:  SOPHIE}). \textit{Left:} s-BGLS of the zero-planet model at the period of the planetary signal. \textit{Middle}: s-BGLS on the residuals of the one-planet fit with a period of 12.95\,d shown around a period of 55\,d, which is the rotational period determined by \cite{Lalande2019}. \textit{Right:}  s-BGLS on the residuals of the one-planet fit around the observed long-period signal. } 
\label{Fig: BGLS_Lalande}
\end{figure*}

\begin{figure}
\centering
\includegraphics[width=9cm]{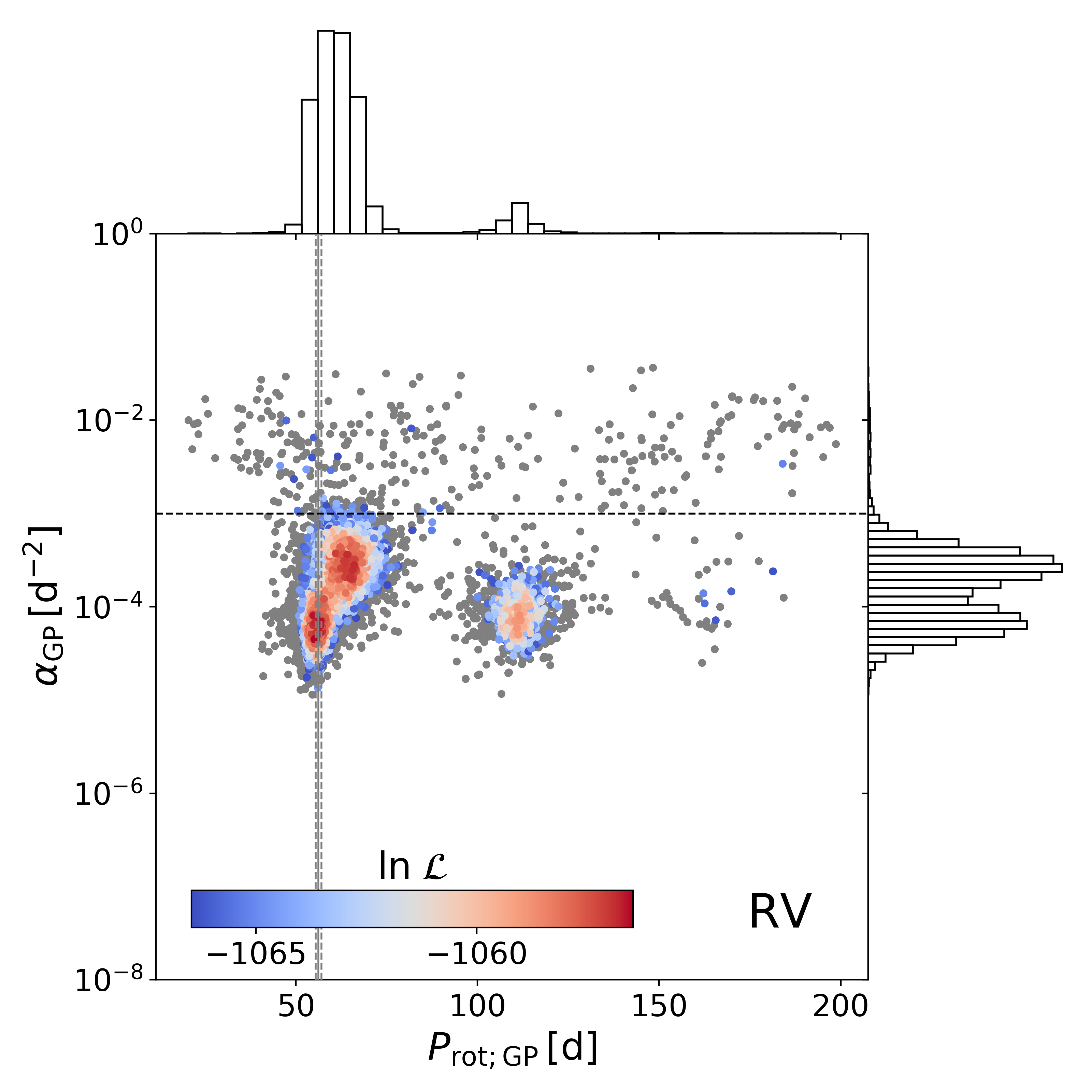}

\includegraphics[width=9cm]{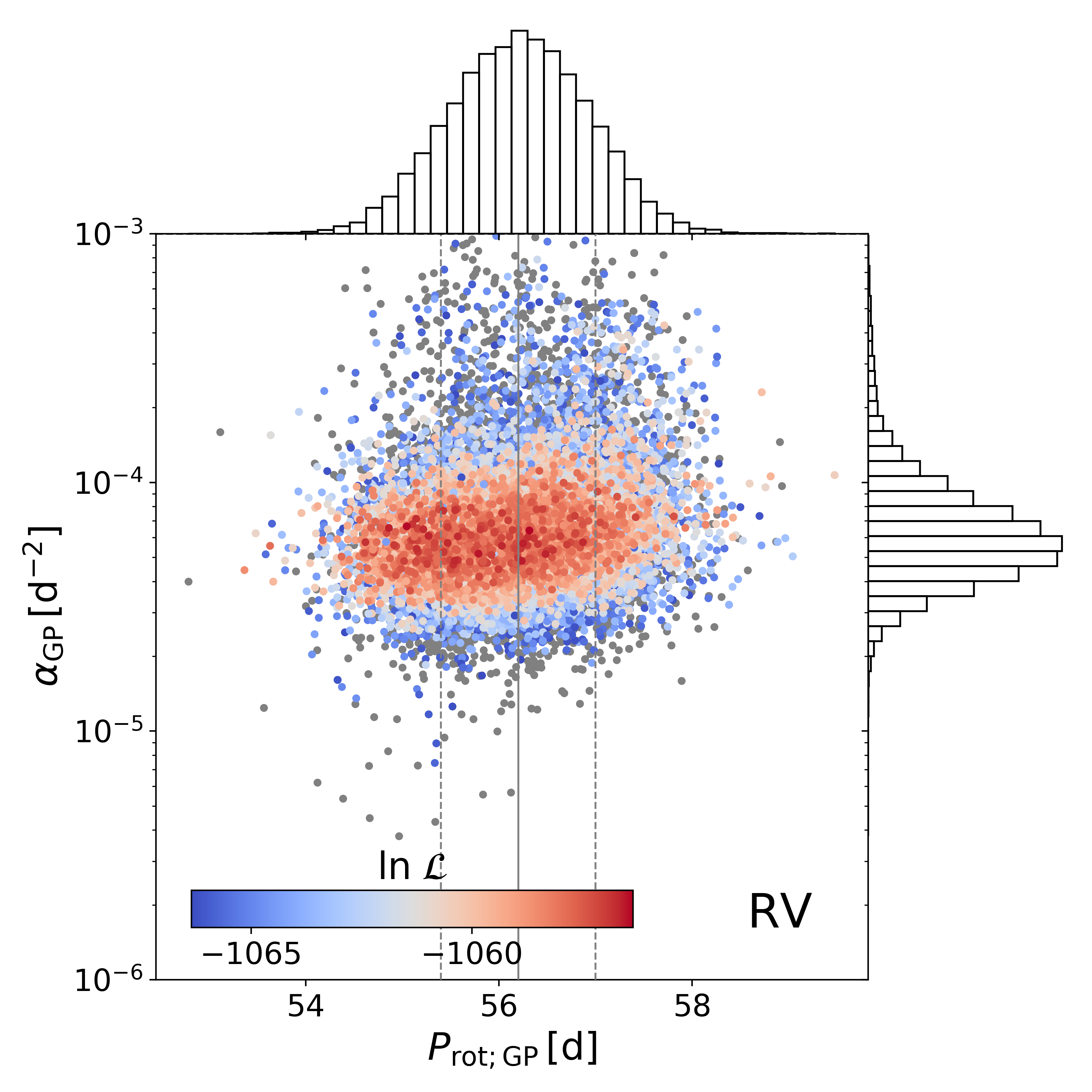}
\caption{Posterior distribution in the $GP_\alpha$ vs. $P_{\mathrm{rot}}$ plane for Lalande~21185. The color-coding shows the log-likelihood normalized to the highest value within the posterior sample. Gray samples indicate solutions with a $\Delta \ln{L}>10$. \textit{Top:} GP fit to the RV data with a wide uniform prior to the rotational period. \textit{Bottom:} GP fit to the RV data with an informative normal prior based on the photometric rotational period proposed by \cite{Lalande2019}.}
\label{Fig: GP_Lalande}
\end{figure}

\begin{figure}
\centering
\includegraphics[width=9cm]{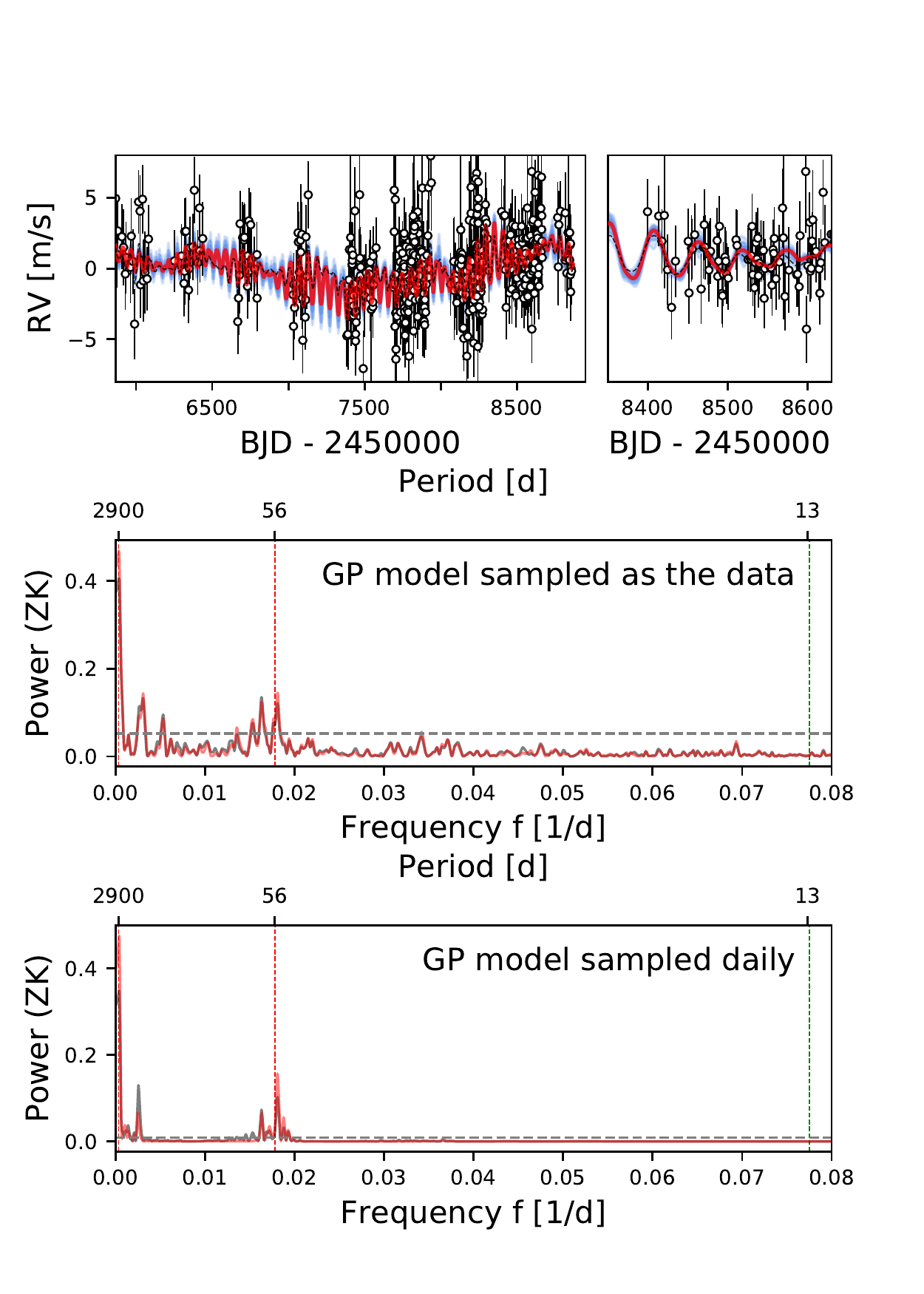}
\caption{Gaussian process model for the RV data of Lalande~21185. The planetary signal is not included in the model and subtracted from the RV data. The constrained GP model is shown in red. The blue regions shows the 1$\sigma$, 2$\sigma$, and 3$\sigma$ uncertainties. We show a zoom into some CARMENES observations (top right). The GLS is evaluated on the GP model at each observed data point (top GLS) and daily (bottom GLS). The dashed line in the GLS periodograms indicates an FAP of $10^{-3}$. We also show the unconstrained GP model as the dashed black line in the upper plots and as the gray periodograms in the lower plots. }
\label{Fig: Lalande_model_comp}
\end{figure}

\begin{figure*}
\centering
\includegraphics[scale=0.5]{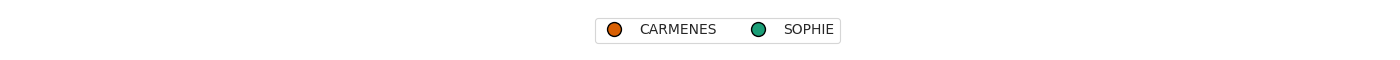}

\includegraphics[scale=0.24]{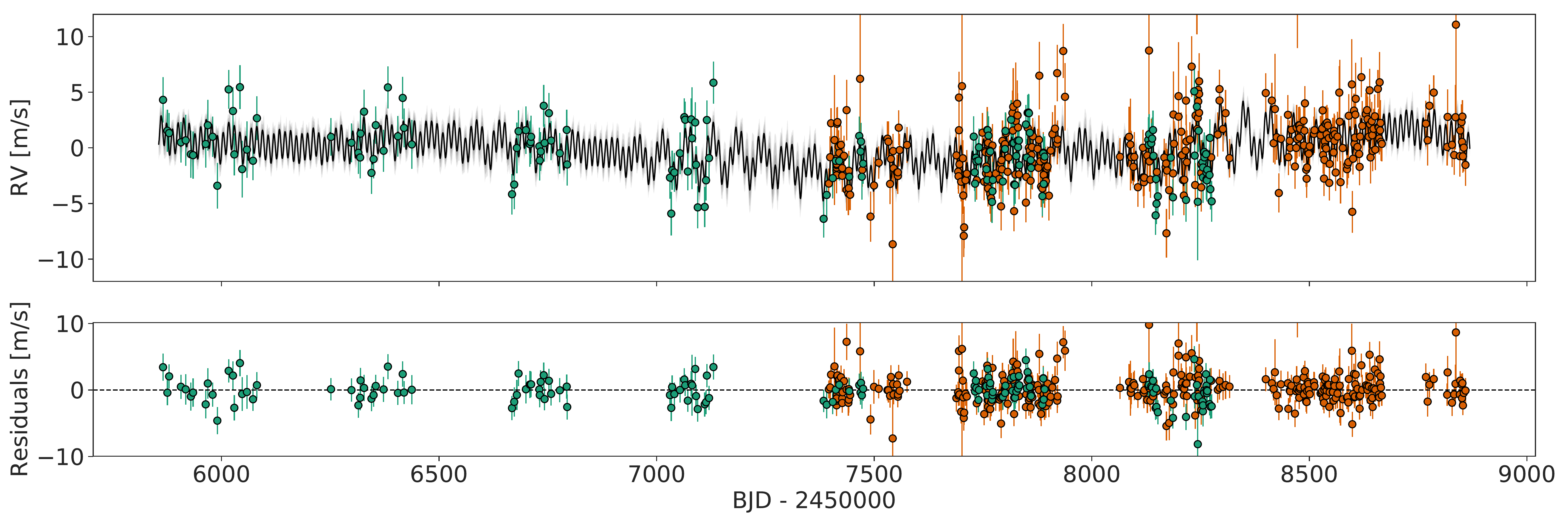}
\includegraphics[scale=0.255]{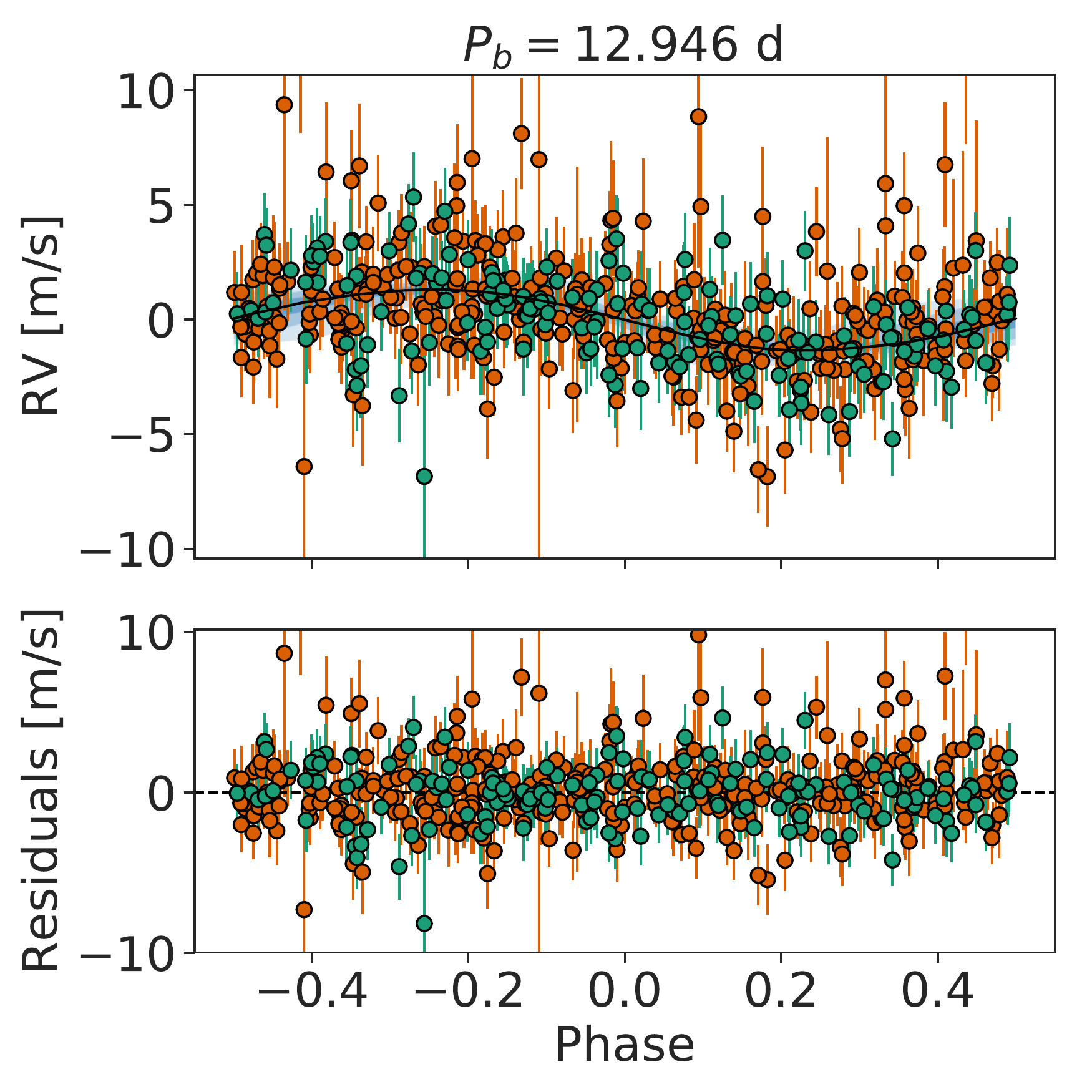}
\caption{\textit{Left:} Radial velocity data with a combined model for one planet and activity using a Keplerian model and a GP. \textit{Right:} Plot phased to the orbital period of Lalande~21185~b \Steph{without a GP component.}}
\label{Fig: Lalande_model}
\end{figure*}

\subsection{Periodogram analysis}
Fig.~\ref{Fig: Lalande_GLS} displays our GLS periodograms for the CARMENES, SOPHIE, and HIRES data, as well as residual periodograms for different models fit to the combination of CARMENES and SOPHIE data.
The CARMENES and the SOPHIE GLS periodograms show a significant peak at a period of 12.95\,d, which is the period of the planetary signal published by \cite{Lalande2019}. With our CARMENES data alone, we can confirm this signal. In addition to the 12.95\,d signal, the SOPHIE data show a signal at 55\,d, with an $\mathrm{FAP}<10^{-2}$, which is consistent with the stellar rotation period of 56.15$\pm$0.27\,d. The CARMENES data also show a significant long-period signal at 2677\,d. \Stephc{A linear trend has been reported by \cite{Lalande2019} from the SOPHIE data.  With the addition of the CARMENES data, we now observed a long period.} 

Neither the CARMENES data nor SOPHIE data or their combination shows a significant signal at \Stephc{9.9\,d, where a signal was claimed by \cite{Butler2017} using HIRES data.} This signal is visible with an $\mathrm{FAP}<10^{-3}$ in the periodogram of the HIRES data, but we find many more signals with similar or higher significance in the periodogram of the HIRES data, with amplitudes of a few $\mathrm{m\,s^{-1}}$. All these signals are absent in the SOPHIE and CARMENES data, however. The time baseline of the combined CARMENES and SOPHIE observations is about 8.2\,yr, and the precision of both instruments should be appropriate to identify these signals if they were still present in the RVs of Lalande~21185. The highly significant presumably planetary signal at 12.95\,d, with an amplitude of roughly 1.4\,$\mathrm{m\,s^{-1}}$ in the CARMENES and SOPHIE data, cannot be identified in the periodogram of the HIRES data. The amplitude of the signal might not be large enough for HIRES because it is at the limit of the long-term precision, which has been about 1--2\,$\mathrm{m\,s^{-1}}$ since 2004 \citep{Butler2017}.
\Stephc{The sampling of the HIRES data shows many nights with multiple observations. After applying a nightly binning scheme on the HIRES data, the GLS periodogram showed no peak with an $\mathrm{FAP}<10^{-2}$.}
Because the 12.95\,d signal is absent in the HIRES data and the forest of significant but spurious signals at various frequencies, we restrict the RV analysis to the combined SOPHIE and CARMENES data sets and treat the HIRES data individually.

\cite{Lalande2019} reported that the SOPHIE data did not allow determining the orbital period of the planetary signal unambiguously because of aliasing. By adding our CARMENES data and using the \texttt{AliasFinder}, we can confirm that the sampled signal has a period of 12.95\,d because the \texttt{AliasFinder} simulations were able to reproduce the properties of the observed periodogram only when this period was assumed to be the correct one. We show the relevant plots in Fig.~\ref{Fig: Lalande21185_alias}. 

\subsection{RV modeling with CARMENES and SOPHIE}

\begin{table}
\caption{Bayesian log-evidence for Lalande~21185 and a number of different models based on CARMENES, SOPHIE, and combined data.}
\label{Tab: Lalande_fit_evidence}
\centering
\begin{tabular}{l c c r}
\hline\hline
Model$^{a}$ & Periods &$\ln{\mathcal{Z}}$ & $\Delta\ln{\mathcal{Z}}$ \\
\hline
\noalign{\smallskip}
\multicolumn{4}{c}{CARMENES}\\
0p & \ldots & $-797.5\pm0.1$ & 0\\
1p &  12.9 & $-783.7\pm0.2$ &13.8\\
1p+uGP &  12.9  & $-748.1\pm0.2$ & 49.4\\
\noalign{\smallskip}
\noalign{\smallskip}
\multicolumn{4}{c}{SOPHIE}\\
0p & \ldots & $-371.6\pm0.1$ & 0\\
1p &  12.9 & $-362.4\pm0.2$ &9.2\\
1p+uGP &  12.9  & $-349.0\pm0.2$ & 22.6\\
\noalign{\smallskip}
\noalign{\smallskip}
\multicolumn{4}{c}{CARMENES + SOPHIE}\\
0p & \ldots & $-1169.5\pm0.2$ & 0\\
\Stephc{2p+GP} & \Stephc{1.5,12.9} & \Stephc{$-1161.8\pm0.3$} & \Stephc{7.7}\\
1p  &  12.9 & $-1143.0\pm0.2$ & 26.5\\
1cp  &  12.9 & $-1142.2\pm0.3$  & 27.3\\
%2 Planets & 12.9, 2600 &$-1110.0\pm0.3$ & 21.3\\
\Stephc{GP} & \Stephc{\ldots} & \Stephc{$-1125.4\pm0.2$} & \Stephc{}\\
1p+uGP &  12.9  & $-1092.3\pm0.3$ & 77.2\\
\Stephc{2p+GP} & \Stephc{12.9, 364.5} & \Stephc{$-1086.2\pm0.3$} & \Stephc{83.3}\\
1p+GP  &   12.9 & $-1085.7\pm0.3$ & 83.8\\
1cp+GP &   12.9 & $-1085.3\pm0.3$ & 84.2\\
\noalign{\smallskip}
\noalign{\smallskip}
\multicolumn{4}{c}{HIRES}\\
0p & \ldots & $-777.7\pm0.2$ & 0\\
1p$^{b}$ &  12.5 & $-761.5$ &16.2\\
1p$^{c}$ &  12.9 & $-771.8$ &5.9\\
\noalign{\smallskip}
\hline
\end{tabular}
 \tablefoot{
 \tablefoottext{a}{Planetary models based on CARMENES, SOPHIE, HIRES, and combined CARMENES and SOPHIE RV data.
        0p: 0 planets, 1p: one planet, 1cp: one planet on a circular orbit ($e = 0$).
        GP and uGP: additional constrained and unconstrained GPs, respectively. 
        Orbital periods rounded to one decimal.}
        %\tablefoottext{a}{Rounded to first digit.}
        %\tablefoottext{b}{Model with circular orbits.}
        %\tablefoottext{c}{Unconstrained GP.}
        %\tablefoottext{d}{Constrained GP.}
        \tablefoottext{b}{Priors as in Table~\ref{Tab: Priors_planets_instruments}.}
        \tablefoottext{c}{Gaussian priors for planetary parameters based on posterior solution from Table~\ref{Tab: Posteriors}. }
    }
\end{table}

We fit a Keplerian model to the 12.95\,d signal (see Table~\ref{Tab: Lalande_fit_evidence} for the log-evidence).
The GLS periodogram of the residuals of the combined data set reveals additional significant signals at periods of $2852\pm568$\,d with an $\mathrm{FAP}<10^{-3}$ and at $55.3\pm0.2$\,d, $61.3\pm0.3$\,d and $383.8\pm10.6$\,d with an $\mathrm{FAP}<0.01$.  \Steph{The long-period signal is highly significant in the combined data set, but we find significant variations at similar periods or half this period in some of the activity indicators, mainly in CRX and H$\alpha$.}

The s-BGLS was used to assess the coherence of all these signals. We show the s-BGLS of the combined data set around the orbital period of the planet signal at 12.95\,d and the s-BGLS of the one-planet fit residuals around the long-periodic signal and the stellar rotational period in Fig.~\ref{Fig: BGLS_Lalande}. The data are ordered chronologically. The signal probability of the suspected planetary signal at 12.95\,d increases and shows coherence over the entire observation time. The signal at 55\,d, which we related to the stellar rotational period, shows an increase in signal probability until roughly 360 observations and has decreased since then by about four orders of magnitude in probability. A similar pattern but anticorrelated to the rotational signal in terms of signal probability over time is visible for the long-period signal.  

Based on the results on the activity indicators and the s-BGLS analysis, we find that the period around 2800\,d can be best explained by a long-term activity cycle. 
We fit a simple sinusoid to this signal \Steph{in order to search for additional signals}. The residual periodogram of the one-planet + sinusoid fit has one remaining significant signal at 55.3\,d. A fit of this signal, which represents the rotational period, with a second sinusoid, resulted in a flat periodogram; no peaks in the GLS periodogram reach an $\mathrm{FAP}<0.1$. Based on CARMENES and SOPHIE RV data, the Lalande~21185 system can be explained by one Keplerian model with an orbital period of 12.95\,d, and two sinusoids that model the activity contribution. 

We fit the system using a one-planet model simultaneously with a GP model, which accounts for these activity-related signals. In a first step, we fit a rather unconstrained GP to the data simultaneously with the one-planet Keplerian model. In the top plot of Fig.~\ref{Fig: GP_Lalande} we show the posterior sample distribution in the $\alpha_{\tx{GP}}$ versus $P_{\mathrm{rot}}$ plane for this fit to Lalande~21185. Most of the posterior samples peak with a bimodal distribution at a period of 55\,d and 65\,d and at rather low $\alpha$-values representing a stable quasi-periodic signal. We also see fewer posterior samples around 100\,d but with a lower likelihood. The 65\,d signal was already visible in the residuals of a simple one-planet fit and belonged to an alias of the rotational period based on a yearly sampling frequency $f_s=1/365.25\,d^{-1}$. 
\Steph{In the next step,} we fit a more constrained GP with a normal prior on the GP rotational period based on the photometric estimates and its 3$\sigma$ uncertainty and \Steph{additional constraints in $\Gamma$ and $\alpha$ deduced from the posterior distribution of the unconstrained fit as before.} The priors of the applied GP models are provided in Table \ref{Tab: Priors_GP_RV}. This GP fit resulted in a significant improvement of the log-evidence compared to the unconstrained GP and represented the best model we derived for this system (see Table~\ref{Tab: Lalande_fit_evidence}). \Stephc{This GP fit is compatible with the estimated RV semiamplitude from the stellar rotation period given the relations by \cite{Suarez2018}.}

We show the GP model and a GLS periodogram analysis to assess its temporal behavior in Fig. \ref{Fig: Lalande_model_comp}. \Stephc{We identify that in addition to the 56\,d and long-term period, signals at about 380\,d and 60\,d are modeled.}
\Stephc{As for the other two targets, we searched for planetary signals hidden behind the stellar activity by sampling for a second Keplerian, using a log-uniform prior between 0.5\,d, and 12\,d and then  a log-uniform prior from 13\,d to 3000\,d for the second Keplerian, while simultaneously fitting the stellar activity with our final GP model. The two two-planet models combined with the GP performed worse in log-evidence (see Table~\ref{Tab: Lalande_fit_evidence}\Stephe{)} than the one-planet model combined with the GP; this means that no additional Keplerian signal is statistically supported by the data.} 
A plot of the final one-planet and GP model, the RV data, and a phase plot to the planetary period of Lalande~21185~b are provided in Fig.~\ref{Fig: Lalande_model}. 
The updated orbital parameters of Lalande~21185~b based on the posterior solutions of the fit to the combined CARMENES and SOPHIE data are given in Table~\ref{Tab: Posteriors}. \Stephc{Additionally, we list further derived planetary parameters, such as the planetary minimum mass, which is estimated to $2.69\pm0.25\,M_\oplus$.}

\subsection{HIRES RV data}

We excluded the HIRES data from our final analysis of Lalande~21185 because of spurious frequencies, noisy data and the absence of an obvious planetary signal. We fit a one-planet model with the same priors as for the SOPHIE and CARMENES data to the HIRES data. For example, for the period, we used $\mathcal{U}(12.5,13.5)$. While this improved the log-evidence significantly, it resulted in an inconsistent orbital period of $12.57^{+0.002}_{-0.001}$\,d and an extremely high eccentricity of $0.9^{+0.04}_{-0.04}$. 

We also tested more informed priors. For instance, we applied normal priors to every planetary parameter based on the solution of the fit to the CARMENES and SOPHIE data. This fit performed reasonably well because it was still significantly better than a flat model, which could indicate that the planetary signal is apparent in the HIRES data. \Stephc{However, the same model fit to the daily-binned HIRES data did not result in a log-evidence improvement compared to a zero-planet model.} The noise level of the HIRES data compared to the other data sets and the small planetary amplitude, which is at the limit of the HIRES long-term precision, justifies the exclusion of this data set. Nevertheless, we used the extended HIRES time baseline to analyze the long-period signal. With HIRES, we have a total of 737 RV observations for Lalande~21185. We find that the long-period signal is also apparent in the HIRES data. However, the HIRES data between BJD 2452200 and 2453200 indicate a possible phase shift of that signal, consistent with our analysis that this signal is caused by a long-term activity cycle. We fit our simplistic model of one planet at 12.95\,d, and two sinusoids, to the combined CARMENES, SOPHIE, and HIRES data. The residual GLS periodogram of this fit had no peaks with an $\mathrm{FAP}<10^{-3}$, indicating that the HIRES data do not indicate an additional coherent signal when combined with CARMENES and SOPHIE data.

\subsection{Transit search with TESS}

\emph{TESS} observed Lalande~21185 in sector 22, but the star was not announced as a TOI. We ran our independent signal search with the TLS on the PDCSAP light curve. We find only one peak with SDE\,$>$\,7 \citep{Hippke&Heller2019}, which is at 13.0\,d,  very close to the planetary orbital period derived from the RV fit, just as for HD~238090. However, peaks in this period range are visible in many light curves and are often caused by the observational gap of the \emph{TESS} observations. Nevertheless, as for HD~238090, we performed a transit fit to this signal using \texttt{juliet} to evaluate how such a fit performs compared to a nontransit model. The log-evidence of a transit model was not significantly different ($\Delta\log{Z}\approx0.5$), indicating that a simpler nontransit model is better than a transit model.
The derived minimum mass of the planet was used to estimate the radius with the mass-radius relation of \cite{Zeng2016}. With an approximated radius of 1.33\,$R_\oplus$, the transit depth was approximated to 0.97\,ppt for Lalande~21185~b. 
\Stephd{We were unable to identify any transit events around the RV estimated time of transit center for Lalande~21185~b.}  

\begin{figure}
\centering
\includegraphics[width=9cm]{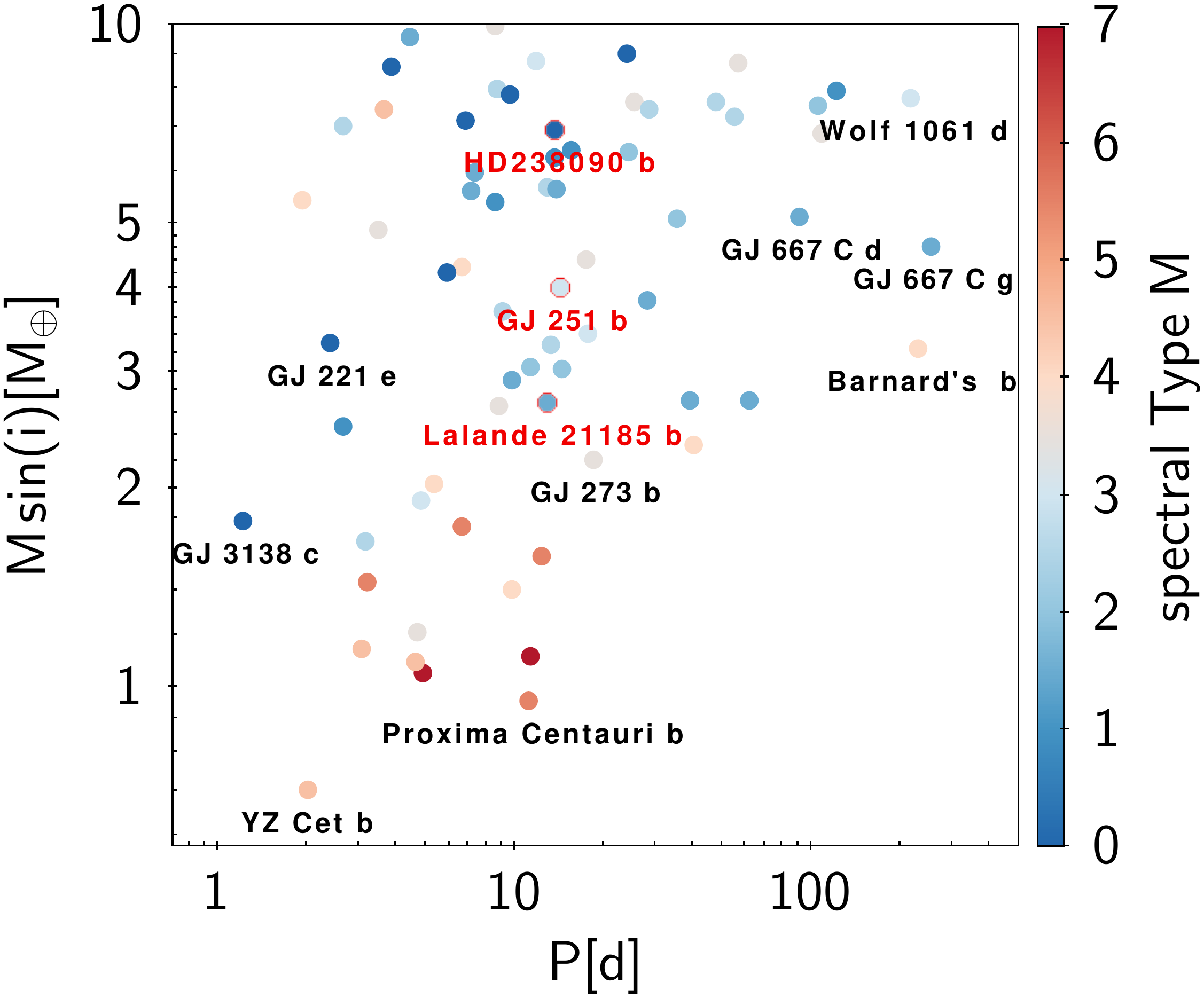}
\caption{Confirmed exoplanets around M-dwarf host stars, detected with the RV method and with minimum masses below $10\,M_\oplus$ and orbital periods between 1\,d and \Steph{400}\,d. The color-coding represents the spectral type. GJ~251~b, HD~238090~b, and Lalande~21185~b are marked in red. Data taken from \url{http://exoplanet.eu}.}
\label{Fig: overview}
\end{figure}

\section{Discussion}
\label{Sect: Discussion}

\subsection{GJ~251}

Based on our analysis of CARMENES and HIRES RV data, we report the discovery of GJ~251~b, a planet that orbits its host star with a period of 14.24\,d. \Stephc{The posterior sample median eccentricity is $0.10^{+0.09}_{-0.07}$, which is not significant and consistent with zero, as shown by a log-evidence comparison with circular Keplerian models.} In Fig.~\ref{Fig: overview} we place GJ~251 into context with other confirmed exoplanets around M dwarfs detected with the RV method and the other two planets discussed in this work. The equilibrium temperature of GJ~251~b, assuming a zero Bond albedo, is $351.0^{+1.4}_{-1.3}$\,K \Stephc{(see Table \ref{Tab: Posteriors})}. Our minimum mass and temperature estimates add GJ~251~b to the family of temperate super-Earths. The planet receives about 2.5 times the flux of Earth. Of the three planets discussed in this work, GJ~251~b \Steph{is the planet with the most moderate temperature.} However, according to \cite{Kopparapu2013}, the planet is too close to its host star to be in the habitable zone, that is, to allow liquid water on its surface.

\cite{Butler2017} previously claimed a planetary candidate for GJ~251 with an orbital period of 1.74\,d. Our analysis of the CARMENES data and the combined data from CARMENES and HIRES did not confirm this claim.
An independent analysis of the HIRES data does not identify any signal close to the 1.74\,d period as significant. We note that \citet{Butler2017} used a different approach based on autocorrelation functions and a statistical model for the RVs, including a moving average to model correlated noise and information provided by the activity index based on Ca~{\sc ii} lines. 

The HIRES RV data show a significant second RV signal around 600\,d. This signal was also visible during some epochs in the CARMENES data, but because the signal is incoherent and a significant H$\alpha$ activity signal lies at the same period, we attribute this signal to nonplanetary origin.
%, e.g., a long-term activity cycle. 
Any other additional significant signals can be best explained by the stellar rotation of GJ~251 or its harmonics. In the current RV data, we find no evidence for a second companion in the GJ~251 system. There is also no significant linear trend in the RV data over the entire time baseline of the combined HIRES and CARMENES observations, which is about $15.0$\,yr. \Steph{However,} the {\em Gaia} DR2 \citep{Gaia} catalog lists a significant astrometric excess noise for this star, which could be caused by a massive companion on a wide orbit. % Nevertheless we note that a companion with much smaller semi-amplitude hiding below the activity or an additional more massive companion on a much longer orbit might be a possibility for the GJ~251 system. 

\subsection{HD~238090}

Our analysis of CARMENES RV data shows that HD~238090 is orbited by a warm super-Earth, HD~238090~b, with an orbital period of 13.69\,d. The posterior sample median eccentricity is $0.30^{+0.16}_{-0.17}$, which, as in the case of GJ~251, is not significant. 
HD~238090~b has a minimum mass of roughly 6.8\,$M_\oplus$, and orbits its host star at a separation of approximately $0.093$\,au with an equilibrium temperature of 470\,K (see Table~\ref{Tab: Posteriors}). With total insolation about eight times that of Earth, the planet is too close to the host star to sustain liquid water on its surface. Figure~\ref{Fig: overview} shows the position of the planet in the minimum mass-period plane.
%\Stephb{The final fit, which includes a GP, represents a realistic representation of the stellar activity as the GP hyper-parameters also allow for a physical interpretation.} 

\subsection{Lalande~21185}

Our analysis of the CARMENES and combined CARMENES and SOPHIE RV data for Lalande~21185 confirms the findings by \cite{Lalande2019} regarding Lalande~21185~b. With our data, we can break the degeneracy between the daily aliases and confirm that the planet orbits the star with a period of 12.95\,d. With the additional CARMENES observations, we have reduced the uncertainties of most planetary parameters by a factor of two or more compared to the estimates by \cite{Lalande2019}.

The CARMENES data of Lalande~21185 agree well with the SOPHIE data but cast doubt on the HIRES data for this system. With 476 precise high-precision RVs for the system by CARMENES and SOPHIE, we find no evidence for a second planet in the system. In particular, the planet candidate claimed by \cite{Butler2017} at a period of 9.9\,d is absent in the CARMENES and SOPHIE RV data. In addition to the 12.95\,d signal, there are two additional significant signals in the combined CARMENES and SOPHIE data set \Steph{at 55.3\,d and 2800\,d}, which can be attributed to the stellar rotation and possible long-term activity\Steph{, respectively.}

\subsection{Formation scenario}

The planetary systems presented in this work represent systems that share similar orbital properties (period, separation, insolation, and equilibrium temperature) and \Steph{minimum masses} that place them into the group of temperate or warm super-Earths.
While the available data are insufficient to draw definite conclusions on the exact formation channel of these systems, the derived orbital and planetary parameters allow for some cautious conjectures. These planets are commonly thought to form by combined accretion of planetesimals and pebbles \citep[e.g.,][]{Ormel2010,Lambrechts2012,Bitsch2019b}.

Their final masses can be best explained if the supply of pebbles was cut off during their formation, preventing their evolution into gas giants. One way to stop the supply of solid material from the outer disk is the emergence of a massive companion that stops the pebble flux by opening a gap in the protoplanetary disk \citep{Ormel2017a}. However, we do not see evidence for additional planets in any of the three systems, even though the available long-baseline data allow for a strong sensitivity for the detection of such companions in the cases of GJ~251 and Lalande~21185. 

Another proposed mechanism to terminate pebble accretion is self-isolation from the pebble flux by modulation of the gas pressure profile by the growing planet \citep{Morbidelli2012, Lambrechts2014}. For all three planets, their combination of mass and orbital separation would be consistent with this possibility. In this case, the measured masses are simply their pebble isolation masses, regardless of other planets in the system. On the other hand, systems such as those presented here can also be explained by models that grow solid cores only by accretion of planetesimals. In this scenario, planetary core growth can reach a natural stall before entering runaway gas accretion because the accretion efficiency is lower than pebble accretion \citep{NGPPS1, NGPPS2}. In a related study focusing on planet formation around low-mass stars, planets similar to our discoveries are among the most abundant in a synthetic planet population \citep[][; Burn et al. in prep.]{NGPPS3}. While \Stephf{the model by Burn et al. (in prep.)} predicts an average multiplicity higher than one for super-Earths, the singular detection in GJ~251, HD~238090, and Lalande~21185 could be explained by planets with lower masses or wider orbits that do not reach the RV semiamplitudes necessary for robust detection, especially in the case of strong activity.

\section{Summary}
\label{Sect: Summary}

% In summary, *** Redundant
We presented the discovery of two super-Earth planets around the low-mass stars GJ~251 and HD~238090 with orbital periods of 14.24\,d, and 13.67\,d, respectively, based on CARMENES VIS RV observations. For GJ~251, we additionally used RV data obtained by HIRES in order to increase the time baseline and to search for additional signals. We also confirmed the nearby temperate super-Earth Lalande~21185~b recently discovered by \cite{Lalande2019}, and we can robustly determine its orbital period to be 12.95\,d. No transits could be detected with \emph{TESS} for any of the three systems. The RV data of GJ~251 and Lalande~21185 exhibit long-term periods, which we attribute to activity. \Steph{Furthermore, all three systems show RV signals related to the stellar rotation period in the RV residuals of the planetary fits.}

\Steph{We modeled the stellar activity using GP models based on a quasi-periodic kernel \Stephc{simultaneously with the Keplerian signals. In particular, we carefully modeled the stellar activity by applying physically motivated constraints to the GP hyperparameters to ensure that the GP did not fit any signal unrelated to stellar activity. We advocate the use of classical periodograms to decompose the modeled frequencies by GPs. Such an analysis can be used as verification of the desired GP behavior on the data set. Nevertheless, the unconstrained GP posterior distribution can provide useful information on the stellar activity. In particular, we used the the GP timescale versus GP rotation plane to infer more information about possible rotation periods and lifetimes of stellar surface features.} For the analysis of the RV data, we used various advanced tools, such as \texttt{juliet} \Stephc{together with a Bayesian approach based on the log-evidence,} \texttt{AliasFinder} \Stephc{to further distinguish samples signals and their aliases}, and the s-BGLS to further constrain the planetary or stellar origin of the signals \Stephc{in addition to a classical periodogram analysis of an extensive number of activity indicators.}
 We showed that with good statistical models, priors from auxiliary data, and elaborate simulations, planetary signals can be recovered and modeled that are on the order of the noise or even slightly weaker. \Stephf{The properties of all our detections are consistent with current formation scenarios regarding super-Earth planets.}}

\begin{acknowledgements} 
\Stephc{We thank the anonymous referee, whose comments improved this work.}
This work was supported by the DFG Research Unit FOR2544 ``Blue Planets around Red Stars'', project no.~RE 2694/4-1. 
CARMENES is an instrument for the Centro Astron\'omico Hispano-Alem\'an (CAHA) at Calar Alto (Almer\'{\i}a, Spain), \Stephc{operated jointly by the Junta de Andaluc\'ia and the Instituto de Astrof\'isica de Andaluc\'ia (CSIC).}
CARMENES was funded by the  
  Max-Planck-Gesellschaft (MPG), 
  the Consejo Superior de Investigaciones Cient\'{\i}ficas (CSIC),
  the Ministerio de Econom\'ia y Competitividad (MINECO) and the \Stephc{European Regional Development Fund (ERDF) through projects FICTS-2011-02, ICTS-2017-07-CAHA-4, and CAHA16-CE-3978}, 
  and the members of the CARMENES Consortium
(Max-Planck-Institut f\"ur Astronomie, Instituto de Astrof\'isica de Andaluc\'ia, Landessternwarte Königstuhl, Institut de Ci\`encies de l'Espai, Insitut f\"ur Astrophysik G\"ottingen, Universidad Complutense de Madrid, Th\"uringer Landessternwarte Tautenburg, Instituto de Astrof\'isica de Canarias, Hamburger Sternwarte, Centro de Astrobiolog\'ia and Centro Astron\'omico Hispano-Alem\'an), with additional contributions by the Spanish Ministry of Economy,  
the German Science Foundation through the Major Research Instrumentation Program and DFG Research Unit FOR2544 ``Blue Planets around Red Stars'', the Klaus Tschira Stiftung, the states of Baden-W\"urttemberg and Niedersachsen, and by the Junta de Andaluc\'ia. 
\Stephc{We acknowledge financial support from the Agencia Estatal de Investigaci\'on of the Ministerio de Ciencia, Innovaci\'on y Universidades and the ERDF through projects 
  PID2019-109522GB-C51/2/3/4    % CAB+IAA+IAC+UCM
  PGC2018-098153-B-C33          % ICE
  AYA2016-79425-C3-1/2/3-P,     % UCM+CAB+IAA (until January 2020)
  ESP2016-80435-C2-1-R,         % ICE (until January 2020) 
and the Centre of Excellence ``Severo Ochoa'' and ``Mar\'ia de Maeztu'' awards to the Instituto de Astrof\'isica de Canarias (SEV-2015-0548), Instituto de Astrof\'isica de Andaluc\'ia (SEV-2017-0709), and Centro de Astrobiolog\'ia (MDM-2017-0737), the Generalitat de Catalunya/CERCA programme, and the NASA Grant NNX17AG24G.} 
\Steph{LCOGT observations were partially acquired via program number TAU2019A-002 of the Wise Observatory, Tel-Aviv University, Israel.}  
This paper includes data collected by the \emph{TESS} mission, which are publicly available from the Mikulski Archive for Space Telescopes (MAST).
The analysis of this work has made use of a wide variety of public available software packages that are not referenced in the manuscript: \texttt{Exo-Striker} \citep{Trifonov.2019}, \texttt{astropy} \cite{AstropyCollaboration.2018},
\texttt{scipy} \citep{Virtanen.2020}, \texttt{numpy} \citep{Oliphant.2006}, \texttt{matplotlib} \citep{Hunter.2007}, \texttt{tqdm} \citep{daCostaLuis.2019}, \texttt{pandas} \citep{Thepandasdevelopmentteam.2020}, and \texttt{seaborn} \citep{Waskom.2020}.
\end{acknowledgements}

% for the bibliography, at the end
\bibliographystyle{aa} % style aa.bst
%\bibliography{references.bib}
\bibliography{references}

\begin{appendix}
\onecolumn

\section{Priors for \texttt{juliet}}
% PRIORS TABLE
\begin{table*}[ht!]
    \centering
    \caption{Priors used within \texttt{juliet} to model the photometric data. }
    \label{Tab: Priors_photometry}
    \begin{tabular}{lccr} 
        \hline
        \hline
        \noalign{\smallskip}
        Parameter name & Prior & Units & Description \\
        \noalign{\smallskip}
        \hline
        \noalign{\smallskip}
        \multicolumn{4}{c}{GP parameters} \\
        \noalign{\smallskip}
        ~~~$\sigma_{\textnormal{GP, instrument}}$                    & $\mathcal{J}(10{^{-8}},10^8)$    & ppm                & Amplitude of GP component of instrument \\
        ~~~$\Gamma_{\textnormal{GP, instrument}}$      & $\mathcal{J}(10^{-6},10^6)$    & \ldots                & Amplitude of GP sine-squared component of instrument \\
        ~~~$\alpha_{\textnormal{GP, global}}$                  & $\mathcal{J}(10^{-10},10^0)$          &  d$^{-2}$   & Global inverse length-scale of GP exponential component of instruments \\
 ~~~$P_{\textnormal{rot, GP, global}}$    & $\mathcal{U}(1,200)$ & d  & Global period of the GP quasi-periodic component of instruments \\
        \noalign{\smallskip}
        \multicolumn{4}{c}{instrumental parameters} \\
        \noalign{\smallskip}
        ~~~$D_{\textnormal{instrument}}$        & 1 (fixed)   & \ldots & Dilution factor of instrument \\
        ~~~$M_{\textnormal{Instrument}}$   & $\mathcal{N}(0,10^5)$              & ppm & Relative flux offset of instrument \\
        ~~~$\sigma_{w,\textnormal{instrument}}$           & $\mathcal{J}(10^{-5},10^5)$    & ppm & Extra jitter term of instrument \\
        \noalign{\smallskip}
                                
        \hline
    \end{tabular}
    \tablefoot{The prior labels $\mathcal{U}$, $\mathcal{N,}$ and $\mathcal{J}$ represent uniform, normal, and Jeffrey's distributions \citep{Jeffreys1949}.}
\end{table*}

\begin{table*}[ht!]
    \centering
    \caption{Planetary and instrumental parameter priors used within \texttt{juliet}. }
    \label{Tab: Priors_planets_instruments}
    \begin{tabular}{lccr} 
        \hline
        \hline
        \noalign{\smallskip}
        Parameter name & Prior & Units & Description \\
        \noalign{\smallskip}
        \hline
        \noalign{\smallskip}
        \multicolumn{4}{c}{GJ~251~b} \\
        \noalign{\smallskip}
        ~~~$P_b$                    & $\mathcal{U}(14,15)$    & d                 & Period \\
        ~~~$t_{0,b} - 2450000$      & $\mathcal{U}(8620,8635)$    & d                 & Time of transit center \\
        ~~~$K_{b}$                  & $\mathcal{U}(0,5)$          & $\mathrm{m\,s^{-1}}$     & RV semiamplitude \\
 ~~~$\mathcal{S}_{1,b} = \sqrt{e_{b}}\sin \omega_{b}$    & $\mathcal{U}(-1,1)$ & \dots  & Parameterization for $e$ and $\omega$. \\
        ~~~$\mathcal{S}_{2,b} = \sqrt{e_{b}}\cos \omega_{b}$    & $\mathcal{U}(-1,1)$ & \dots  & Parameterization for $e$ and $\omega$.\\
        \noalign{\smallskip}
        \multicolumn{4}{c}{HD~238090~b} \\
        \noalign{\smallskip}
        ~~~$P_b$                    & $\mathcal{U}(13,14)$    & d                 & Period \\
        ~~~$t_{0,b} - 2450000$      & $\mathcal{U}(8620,8634)$    & d                 & Time of transit center \\
        ~~~$K_{b}$                  & $\mathcal{U}(0,5)$          & $\mathrm{m\,s^{-1}}$     & RV semiamplitude \\
 ~~~$\mathcal{S}_{1,b} = \sqrt{e_{b}}\sin \omega_{b}$    & $\mathcal{U}(-1,1)$ & \dots  & Parameterization for $e$ and $\omega$. \\
        ~~~$\mathcal{S}_{2,b} = \sqrt{e_{b}}\cos \omega_{b}$    & $\mathcal{U}(-1,1)$ & \dots  & Parameterization for $e$ and $\omega$.
        \\
        \noalign{\smallskip}
        \multicolumn{4}{c}{Lalande~21185~b} \\
        \noalign{\smallskip}
        ~~~$P_b$                    & $\mathcal{U}(12.5,13.5)$    & d                 & Period \\
        ~~~$t_{0,b} - 2450000$      & $\mathcal{U}(5865,5878)$    & d                 & Time of transit center \\
        ~~~$K_{b}$                  & $\mathcal{U}(0,5)$          & $\mathrm{m\,s^{-1}}$     & RV semiamplitude \\
 ~~~$\mathcal{S}_{1,b} = \sqrt{e_{b}}\sin \omega_{b}$    & $\mathcal{U}(-1,1)$ & \dots  & Parameterization for $e$ and $\omega$. \\
        ~~~$\mathcal{S}_{2,b} = \sqrt{e_{b}}\cos \omega_{b}$    & $\mathcal{U}(-1,1)$ & \dots  & Parameterization for $e$ and $\omega$.
        \\
        \noalign{\smallskip}
        \multicolumn{4}{c}{RV parameters} \\
        \noalign{\smallskip}
        ~~~$\gamma_{\textnormal{CARMENES}}$        & $\mathcal{U}(-10,10)$    & $\mathrm{m\,s^{-1}}$ &  Velocity zero-point for CARMENES \\
        ~~~$\sigma_{\textnormal{CARMENES}}$   & $\mathcal{J}(0.01,100)$              & $\mathrm{m\,s^{-1}}$ & Extra jitter term for CARMENES \\
        ~~~$\gamma_{\textnormal{HIRES}}$           & $\mathcal{U}(-10,10)$    & $\mathrm{m\,s^{-1}}$ &  Velocity zero-point for HIRES \\
        ~~~$\sigma_{\textnormal{HIRES}}$      & $\mathcal{J}(0.01,100)$                & $\mathrm{m\,s^{-1}}$ & Extra jitter term for HIRES \\
        ~~~$\gamma_{\textnormal{SOPHIE}}$           & $\mathcal{U}(-10,10)$    & $\mathrm{m\,s^{-1}}$ & Velocity zero-point for SOPHIE \\
        ~~~$\sigma_{\textnormal{SOPHIE}}$      & $\mathcal{J}(0.01,100)$             & $\mathrm{m\,s^{-1}}$ & Extra jitter term for SOPHIE \\
        \noalign{\smallskip}
                                
        \hline
    \end{tabular}
    \tablefoot{The prior labels $\mathcal{U}$ and $\mathcal{J}$ represent uniform, and Jeffrey's distributions \citep{Jeffreys1949}.}
\end{table*}

% PRIORS TABLE
\begin{table*}[ht!]
    \centering
    \caption{Gaussian process priors used within \texttt{juliet} for the RV data of GJ~251, HD~238090 and Lalande~21185. }
    \label{Tab: Priors_GP_RV}
    \begin{tabular}{lccr} 
        \hline
        \hline
        \noalign{\smallskip}
        Parameter name & Prior & Units & Description \\
        \noalign{\smallskip}
        \hline
        \noalign{\smallskip}
        \multicolumn{4}{c}{uGP (wide priors) for GJ~251, HD~238090, Lalande~21185} \\
        \noalign{\smallskip}
        \noalign{\smallskip}
        ~~~$\sigma_{\textnormal{GP, RV}}$                    & $\mathcal{U}(0,5)$    & $\mathrm{m\,s^{-1}}$               & Amplitude of GP component for RVs \\
        ~~~$\Gamma_{\textnormal{GP, RV}}$      & $\mathcal{J}(10^{-2},10^2)$    & \ldots               & Amplitude of GP sine-squared component for RVs \\
        ~~~$\alpha_{\textnormal{GP, RV}}$                  & $\mathcal{J}(10^{-8},10^0)$          &  d$^{-2}$   & Inverse length-scale of GP exponential component for RVs \\
 ~~~$P_{\textnormal{rot, GP,RV}}$    & $\mathcal{U}(20,200)$ & d  & Period of the GP quasi-periodic component for RVs \\
         \noalign{\smallskip}
        \multicolumn{4}{c}{GP (constrained) for GJ~251} \\
        \noalign{\smallskip}
        \noalign{\smallskip}
        ~~~$\sigma_{\textnormal{GP, RV}}$                    & $\mathcal{U}(0,5)$    & $\mathrm{m\,s^{-1}}$               & Amplitude of GP component for RVs \\
        ~~~$\Gamma_{\textnormal{GP, RV}}$      & $\mathcal{J}(10^{-1},10^1)$    & \ldots               & Amplitude of GP sine-squared component for RVs \\
        ~~~$\alpha_{\textnormal{GP, RV}}$                  & $\mathcal{J}(10^{-8},3\cdot10^{-4})$          &  d$^{-2}$   & Inverse length-scale of GP exponential component for RVs \\
 ~~~$P_{\textnormal{rot, GP,RV}}$    & $\mathcal{N}(122.1,6.6)$ & d  & Period of the GP quasi-periodic component for RVs \\
        \noalign{\smallskip}
                \noalign{\smallskip}
        \multicolumn{4}{c}{GP (constrained) for HD~238090} \\
        \noalign{\smallskip}
        \noalign{\smallskip}
        ~~~$\sigma_{\textnormal{GP, RV}}$                    & $\mathcal{U}(0,5)$    & $\mathrm{m\,s^{-1}}$               & Amplitude of GP component for RVs \\
        ~~~$\Gamma_{\textnormal{GP, RV}}$      & $\mathcal{J}(10^{-1},10^1)$    & \ldots               & Amplitude of GP sine-squared component for RVs \\
        ~~~$\alpha_{\textnormal{GP, RV}}$                  & $10^{-20}$ (fixed)          &  d$^{-2}$  & Inverse length-scale of GP exponential component for RVs \\
 ~~~$P_{\textnormal{rot, GP,RV}}$    & $\mathcal{N}(96.7,9.3)$ & d  & Period of the GP quasi-periodic component for RVs \\
                 \noalign{\smallskip}
        \multicolumn{4}{c}{GP (constrained) for Lalande~21185} \\
        \noalign{\smallskip}
        \noalign{\smallskip}
        ~~~$\sigma_{\textnormal{GP, RV}}$                    & $\mathcal{U}(0,5)$    & $\mathrm{m\,s^{-1}}$               & Amplitude of GP component for RVs \\
        ~~~$\Gamma_{\textnormal{GP, RV}}$      & $\mathcal{J}(10^{-1},10^1)$    & \ldots               & Amplitude of GP sine-squared component for RVs \\
        ~~~$\alpha_{\textnormal{GP, RV}}$                  & $\mathcal{J}(10^{-8},10^{-3})$          &  d$^{-2}$   & Inverse length-scale of GP exponential component for RVs \\
 ~~~$P_{\textnormal{rot, GP,RV}}$    & $\mathcal{N}(56.2,0.81)$ & d  & Period of the GP quasi-periodic component for RVs \\

        \hline
    \end{tabular}
    \tablefoot{The prior labels $\mathcal{U}$, $\mathcal{N,}$ and $\mathcal{J}$ represent uniform, Normal and Jeffrey's distributions \citep{Jeffreys1949}.}
\end{table*}

\clearpage
\section{Corner plots}

\begin{figure*}[!ht]
\centering
\includegraphics[width=18.5cm]{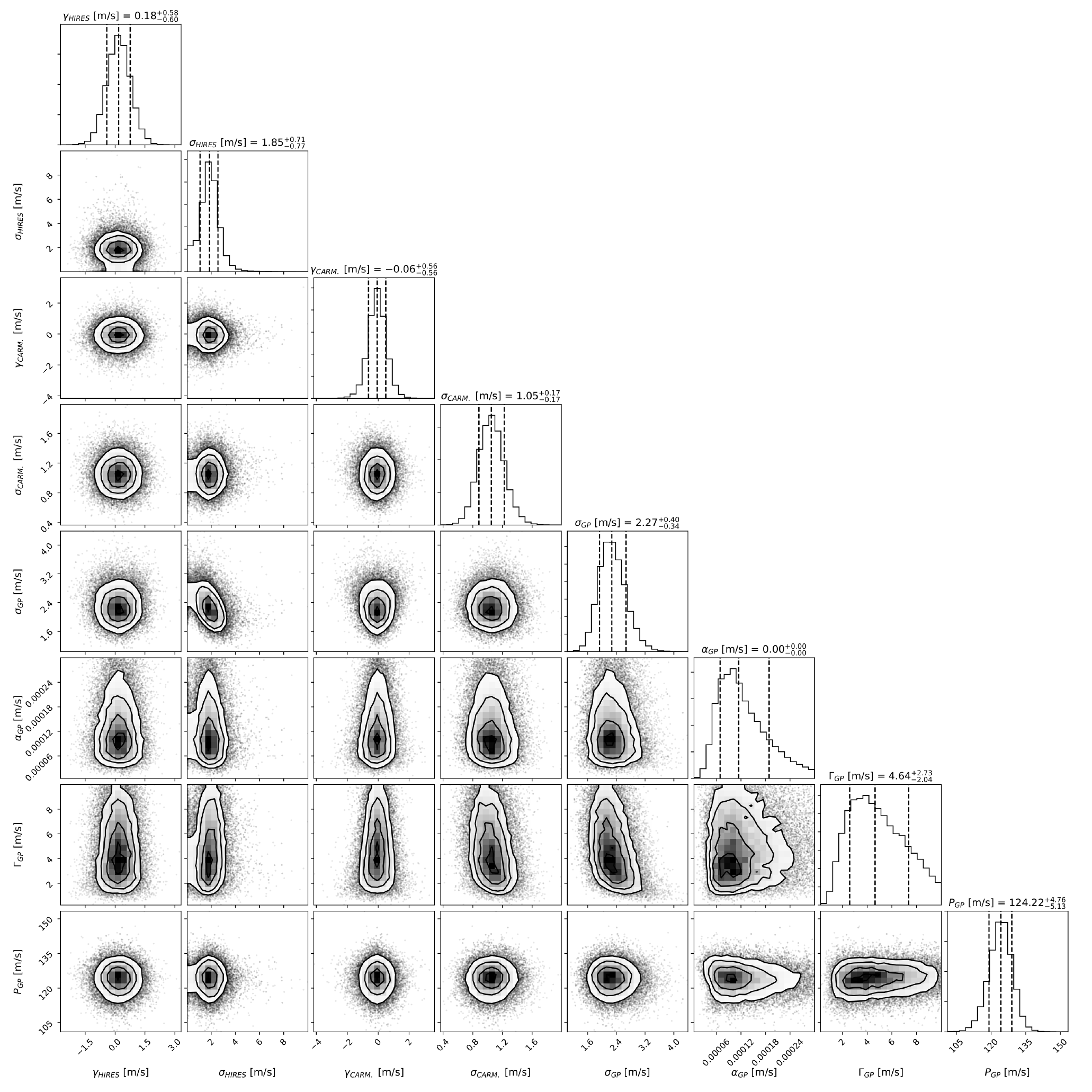}
\caption{Corner plot of the instrumental and GP parameters for GJ~251. Error bars denote the $68\%$ posterior credibility intervals. }
\label{Fig: GJ251_corner_GP}
\end{figure*}
\newpage
\begin{figure*}[!ht]
\centering
\includegraphics[width=18.5cm]{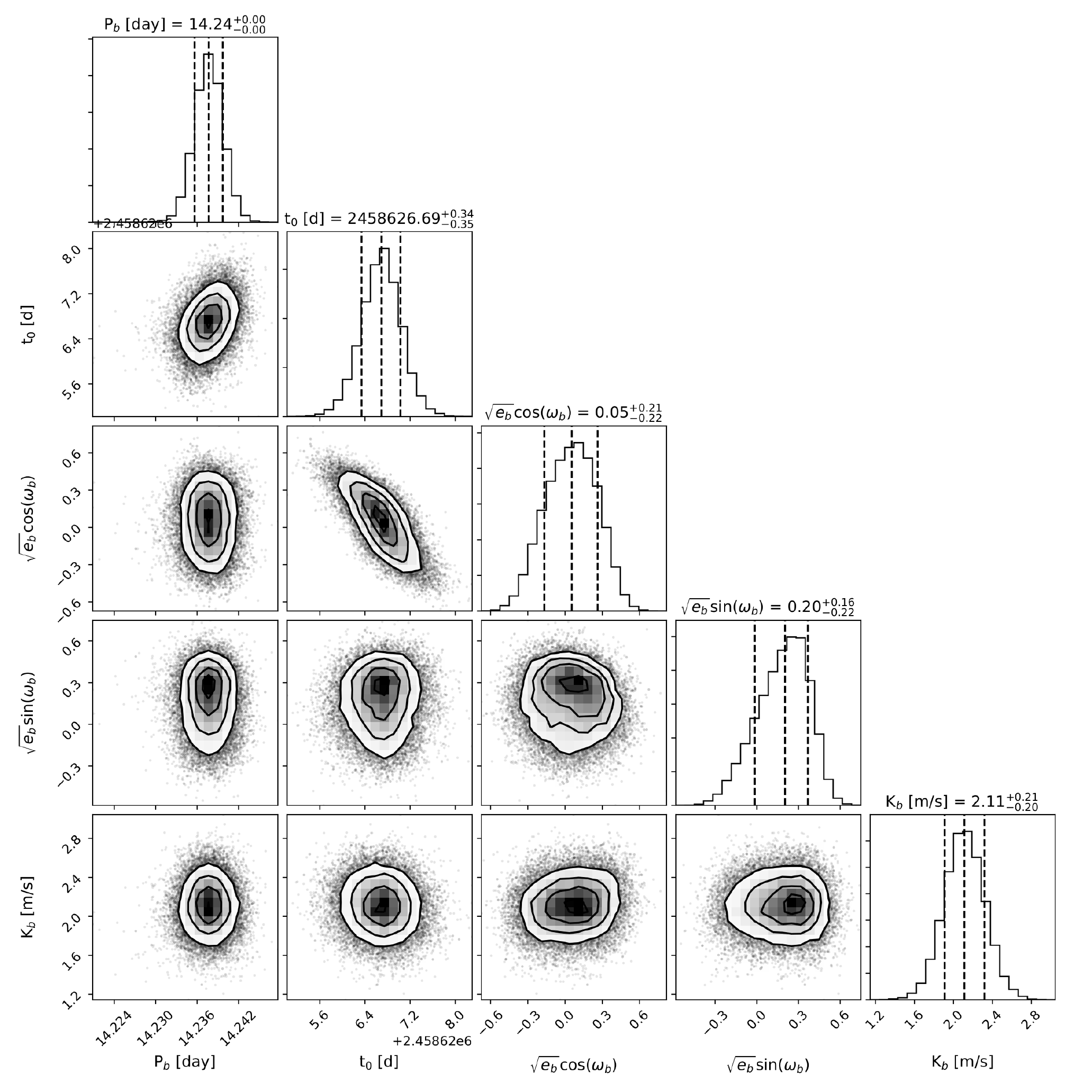}
\caption{Corner plot of the planetary parameters for GJ~251. Error bars denote the $68\%$ posterior credibility intervals. }
\label{Fig: GJ251_corner_planet}
\end{figure*}
\newpage

\begin{figure*}[!ht]
\centering
\includegraphics[width=18.5cm]{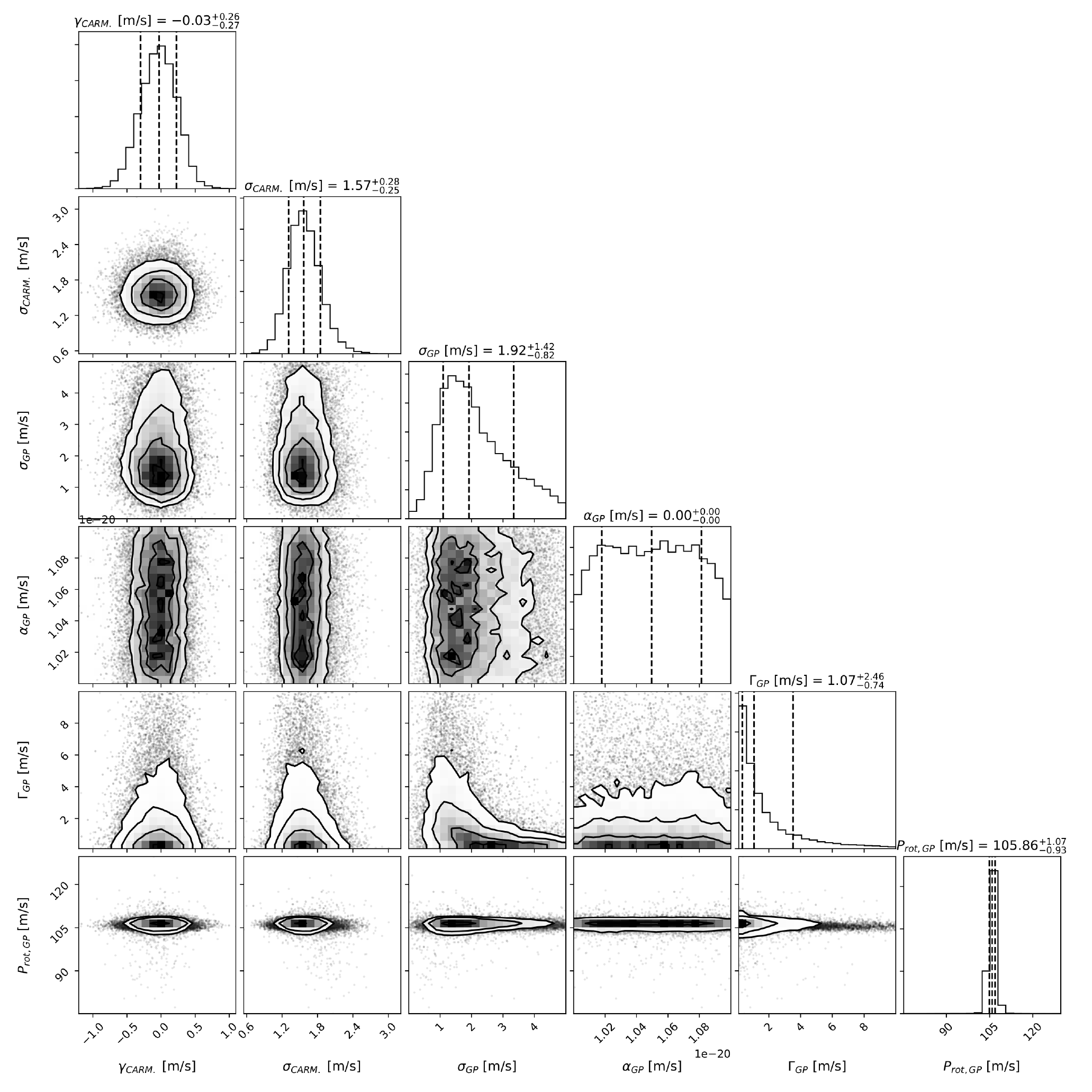}
\caption{Corner plot of the instrumental and GP parameters for HD~238090. Error bars denote the $68\%$ posterior credibility intervals. }
\label{Fig: HD238090_corner_GP}
\end{figure*}
\newpage
\begin{figure*}[!ht]
\centering
\includegraphics[width=18.5cm]{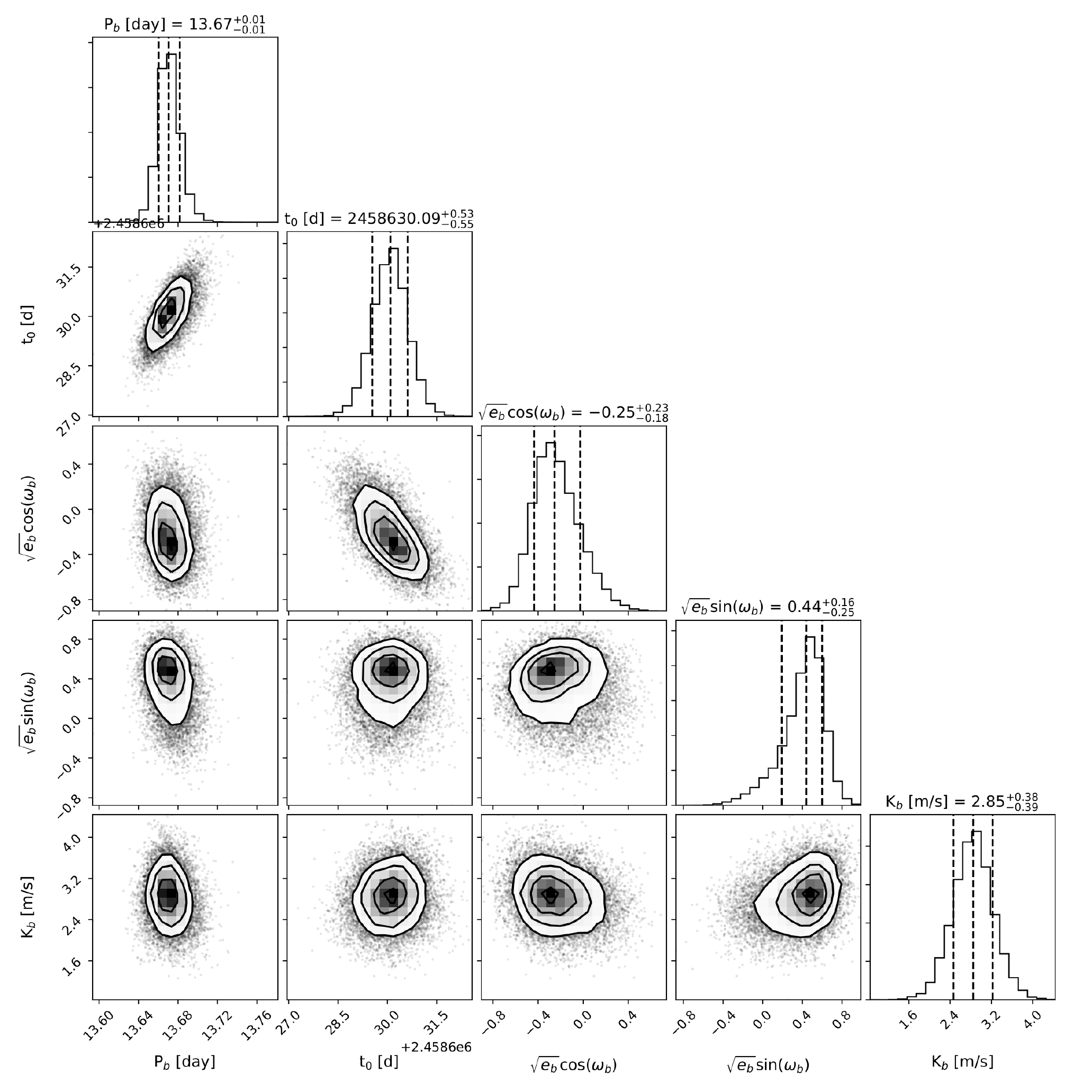}
\caption{Corner plot of the planetary parameters for HD~238090. Error bars denote the $68\%$ posterior credibility intervals. }
\label{Fig: HD238090_corner_planet}
\end{figure*}
\newpage

\begin{figure*}[!ht]
\centering
\includegraphics[width=18.5cm]{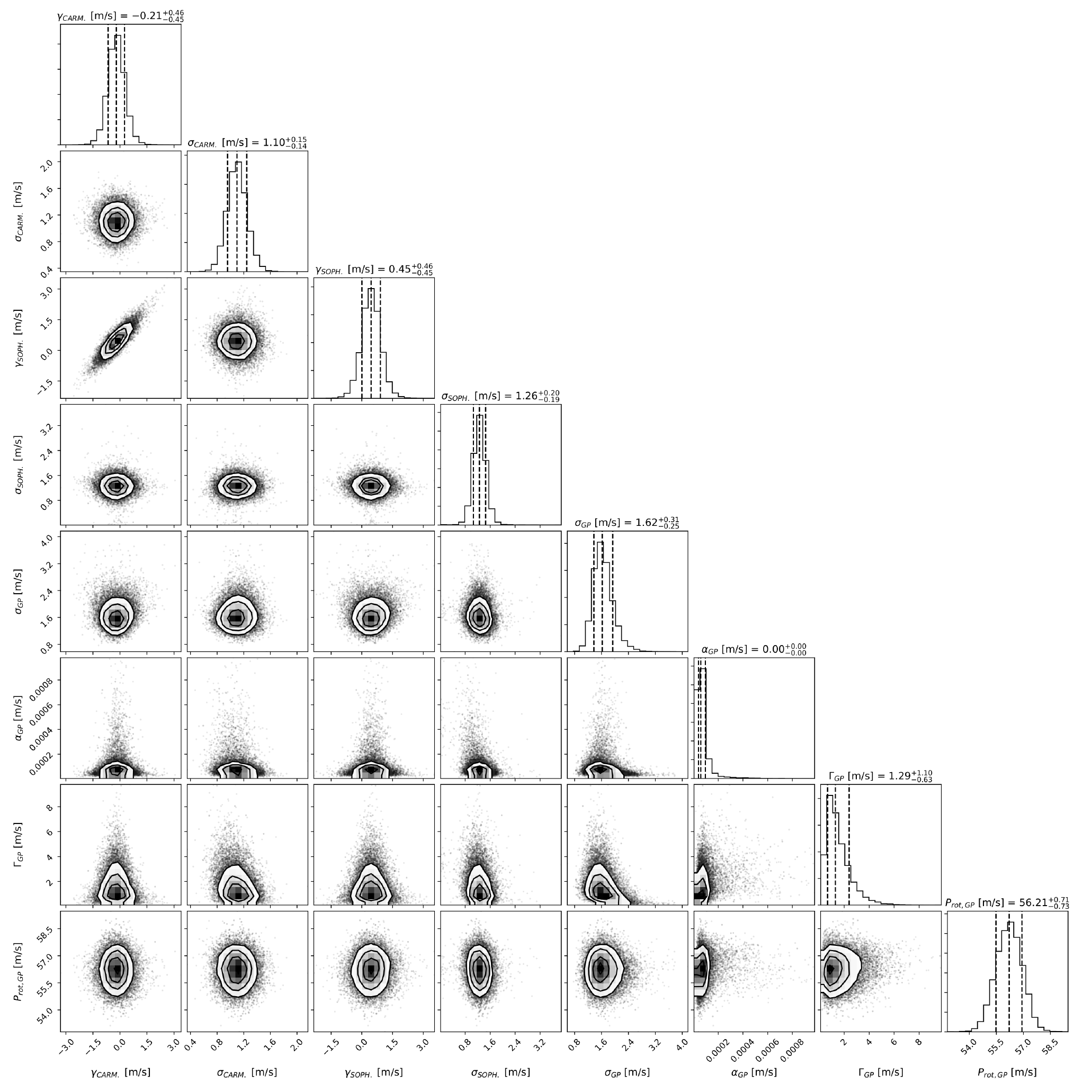}
\caption{Corner plot of the instrumental and GP parameters for Lalande~21185. Error bars denote the $68\%$ posterior credibility intervals. }
\label{Fig: Lalande21185_corner_GP}
\end{figure*}
\newpage
\begin{figure*}[!ht]
\centering
\includegraphics[width=18.5cm]{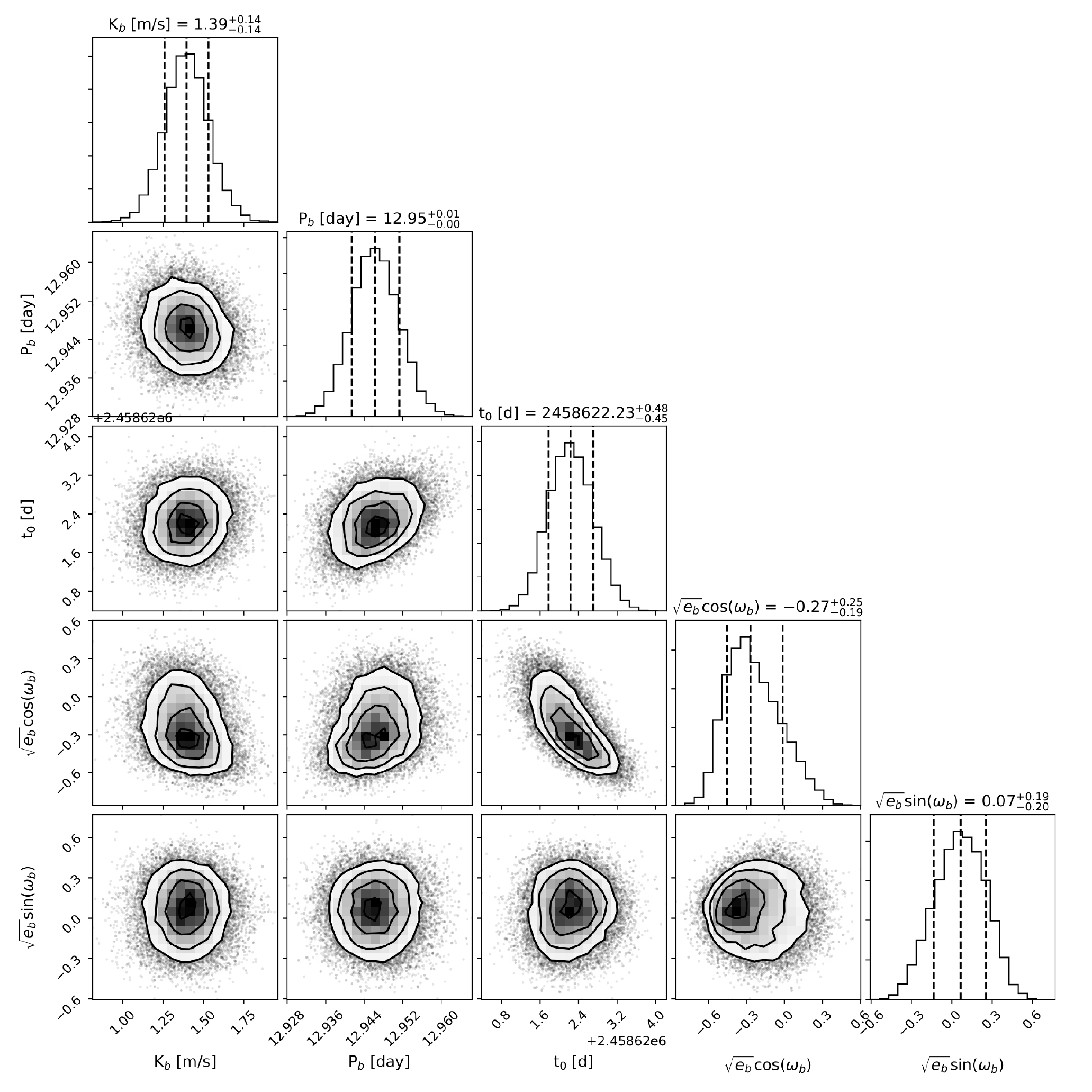}
\caption{Corner plot of the planetary parameters for Lalande~21185. Error bars denote the $68\%$ posterior credibility intervals. }
\label{Fig: Lalande21185_corner_planet}
\end{figure*}

\clearpage

\section{\texttt{AliasFinder} plot of HD~238090}
\begin{figure*}[ht!]
\centering
\subfloat[]{\includegraphics[width=15cm]{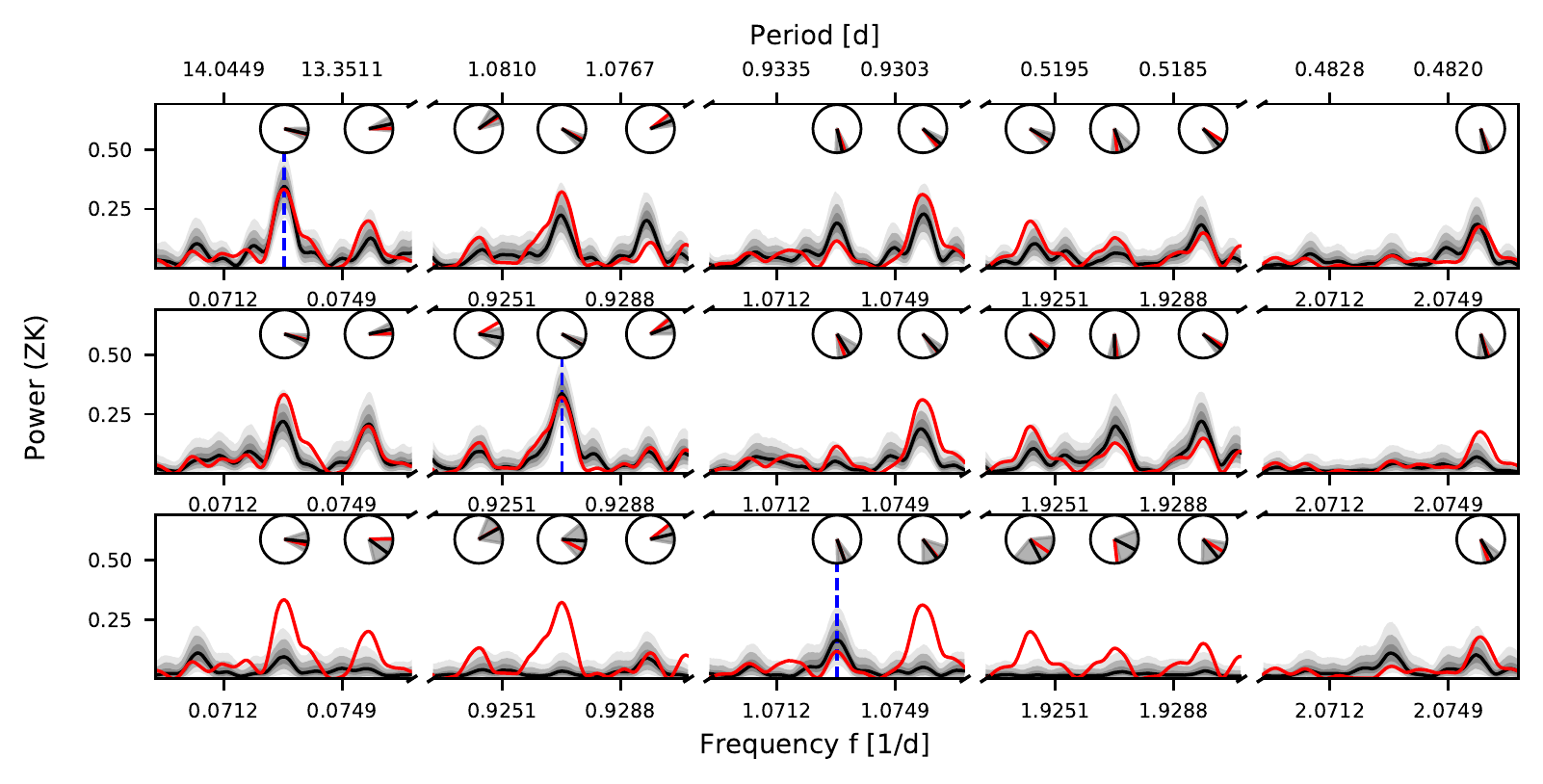}}\\
\subfloat[]{\includegraphics[width=15cm]{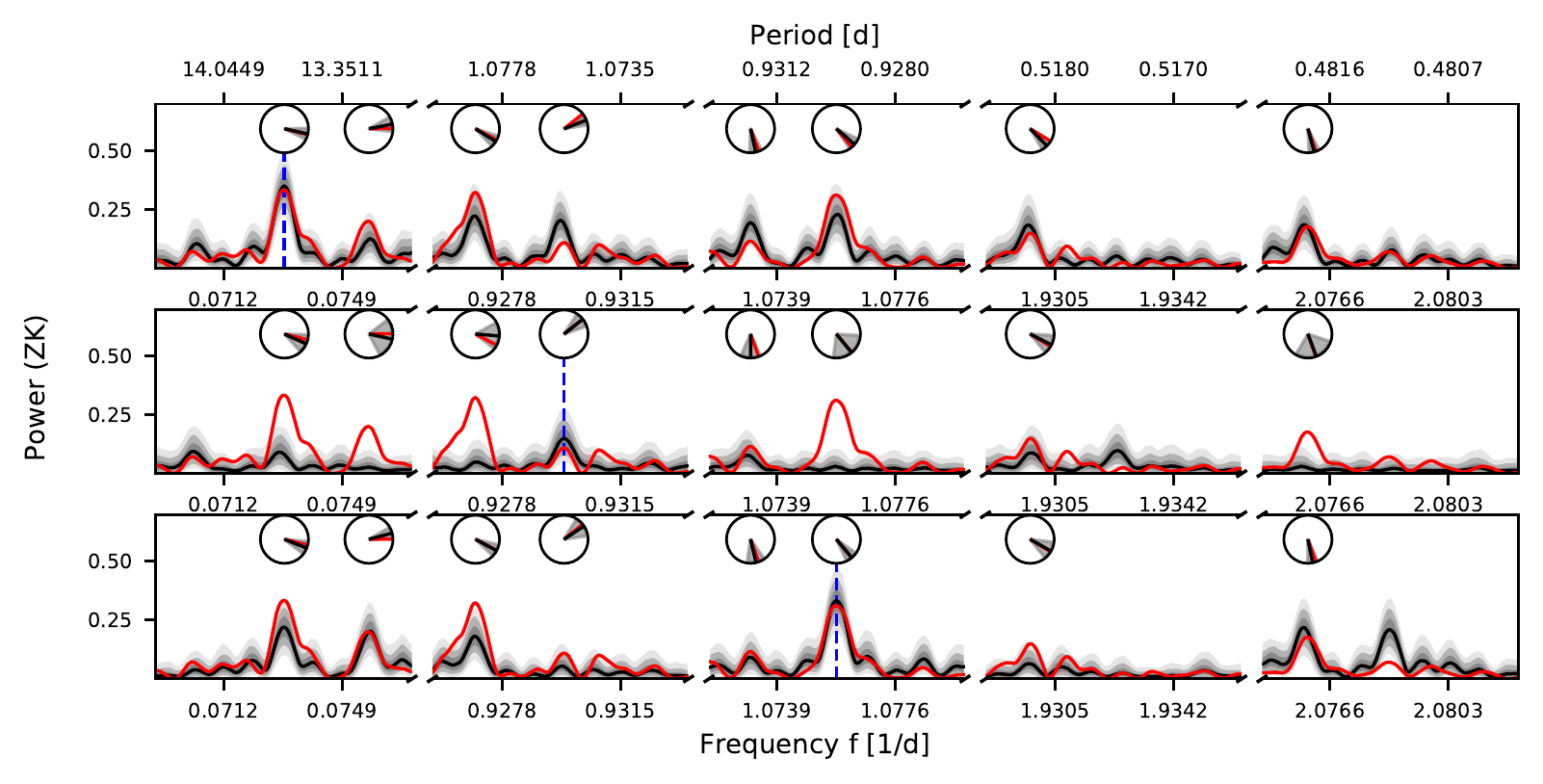}}
\caption{Alias tests for HD~238090. The top plot (a) shows simulations motivated by a sampling frequency of $f_{s_1}=1.0000\,{d^{-1}}$. The bottom plot (b) shows simulations motivated by a sampling frequency of $f_{s_2}=1.0027\,{d^{-1}}$. Each row in these plots corresponds to one set of simulations for which the frequency of the injected signal is indicated by a vertical dashed blue line. The first row shows simulations with a period of $13.6838$\,d, and the second and third row show the simulations where the first-order aliases of $13.68$\,d, regarding the investigated sampling frequency were injected. Each column shows informative ranges of the periodograms, which are based on the assumed sampling frequency, and can be used for the comparison of data and simulations. From 1000 simulated data sets each, the median of the obtained periodograms (solid black line), the inter-quartile range and the ranges of 90\% and 99\% (gray shades) are shown. For comparison, the periodogram of the observed true data is plotted with a solid red line. The angular mean of the phase of some peaks and their standard deviation are shown in the clock diagrams (black line and gray shades) and can be compared to the phase of these peaks in the observed periodogram (red line).}
\label{Fig: HD238090_alias1}
\end{figure*}

\clearpage

\section{RV data}

\begin{longtable}{l c c c c c c c c}
\caption{CARMENES VIS RVs \Stephc{and some activity indicators} of G~J251.}\\
%\label{Tab: RV_Data1_GJ251}\\
\hline
\hline
  \multicolumn{1}{l}{BJD } &
  \multicolumn{1}{c}{RV\,[m\,s$^{-1}$]} &
  \multicolumn{1}{c}{$\sigma_{\tx{RV}}$\,[m\,s$^{-1}$]} &
    \multicolumn{1}{c}{CRX\,[m\,s$^{-1}$]} &
  \multicolumn{1}{c}{$\sigma_{\tx{CRX}}$\,[m\,s$^{-1}$]} &
    \multicolumn{1}{c}{dLW\,[]} &
  \multicolumn{1}{c}{$\sigma_{\tx{dlW}}$\,[]} &
    \multicolumn{1}{c}{H$\alpha$\,[\r{A}]} &
  \multicolumn{1}{c}{$\sigma_{\tx{H\alpha}}$\,[\r{A}]} \\
  \hline
  \endhead
   \hline
  \endfoot
\hline
  2457395.52638 & -0.15 & 1.08 & -10.54 & 9.57 & -8.86 & 1.59 & -0.013 & 0.015\\
  2457398.46916 & -0.11 & 1.31 & -12.35 & 9.87 & -8.69 & 1.81 & -0.006 & 0.016\\
  2457400.56705 & 3.04 & 1.08 & 20.62 & 8.76 & -11.06 & 1.14 & -0.0 & 0.016\\
  2457401.49163 & -0.68 & 0.83 & 1.06 & 5.93 & 5.16 & 0.78 & -0.034 & 0.01\\
  2457411.45624 & -0.16 & 1.37 & 17.1 & 7.45 & -0.95 & 1.36 & 0.0 & 0.017\\
  2457411.46376 & 1.13 & 1.48 & 25.73 & 8.99 & -0.04 & 1.37 & 0.007 & 0.017\\
  2457412.44675 & -0.34 & 1.07 & 18.53 & 5.74 & 0.58 & 0.91 & -0.004 & 0.014\\
  2457412.45384 & 1.56 & 1.16 & 19.84 & 7.39 & 3.0 & 0.98 & -0.008 & 0.014\\
\hline
\end{longtable}

\clearpage
\begin{longtable}{l c c c c c c c c}
\caption{CARMENES VIS RVs \Stephc{and some activity indicators} of HD~238090.}\\
%\label{Tab: RV Data HD238090}\\
\hline
\hline
  \multicolumn{1}{l}{BJD } &
  \multicolumn{1}{c}{RV\,[m\,s$^{-1}$]} &
  \multicolumn{1}{c}{$\sigma_{\tx{RV}}$\,[m\,s$^{-1}$]} &
    \multicolumn{1}{c}{CRX\,[m\,s$^{-1}$]} &
  \multicolumn{1}{c}{$\sigma_{\tx{CRX}}$\,[m\,s$^{-1}$]} &
    \multicolumn{1}{c}{dLW\,[]} &
  \multicolumn{1}{c}{$\sigma_{\tx{dlW}}$\,[]} &
    \multicolumn{1}{c}{H$\alpha$\,[\r{A}]} &
  \multicolumn{1}{c}{$\sigma_{\tx{H\alpha}}$\,[\r{A}]} \\
  \hline
  \endhead
   \hline
  \endfoot
\hline
  2457395.77165 & -0.33 & 2.08 & -20.26 & 19.68 & -46.95 & 3.22 & 0.017 & 0.02\\
  2457400.72878 & -1.91 & 1.48 & 31.97 & 12.52 & -29.18 & 2.14 & -0.017 & 0.016\\
  2457401.72977 & -0.68 & 1.75 & 5.26 & 16.27 & -43.01 & 3.82 & 0.003 & 0.017\\
  2457414.64092 & -1.82 & 1.65 & 27.71 & 14.44 & -16.41 & 2.3 & -0.012 & 0.016\\
  2457415.64256 & 1.02 & 3.06 & -4.62 & 30.17 & -32.62 & 4.07 & 0.032 & 0.019\\
  2457418.59670 & 0.79 & 1.64 & 32.05 & 14.06 & -15.35 & 1.97 & -0.024 & 0.017\\
  2457441.69153 & -6.72 & 1.6 & -5.75 & 11.48 & -9.3 & 1.47 & 0.003 & 0.015\\
  2457473.55778 & -1.02 & 1.93 & 14.38 & 12.04 & -30.97 & 2.14 & 0.021 & 0.013\\
\hline
\end{longtable}

\clearpage
\begin{longtable}{l c c c c c c c c}
\caption{CARMENES VIS RVs \Stephc{and some activity indicators} of Lalande~21185.}\\
%\label{Tab: RV Data Lalande21185}\\
\hline
\hline
  \multicolumn{1}{l}{BJD } &
  \multicolumn{1}{c}{RV\,[m\,s$^{-1}$]} &
  \multicolumn{1}{c}{$\sigma_{\tx{RV}}$\,[m\,s$^{-1}$]} &
    \multicolumn{1}{c}{CRX\,[m\,s$^{-1}$]} &
  \multicolumn{1}{c}{$\sigma_{\tx{CRX}}$\,[m\,s$^{-1}$]} &
    \multicolumn{1}{c}{dLW\,[]} &
  \multicolumn{1}{c}{$\sigma_{\tx{dlW}}$\,[]} &
    \multicolumn{1}{c}{H$\alpha$\,[\r{A}]} &
  \multicolumn{1}{c}{$\sigma_{\tx{H\alpha}}$\,[\r{A}]} \\
  \hline
  \endhead
   \hline
  \endfoot
\hline
  2457395.67686 & -3.41 & 0.77 & -7.16 & 5.45 & 13.23 & 0.64 & -0.042 & 0.017\\
  2457398.68170 & -1.05 & 1.54 & 8.99 & 14.23 & -13.03 & 1.86 & -0.065 & 0.02\\
  2457400.76874 & 2.01 & 0.78 & 14.79 & 4.93 & 12.44 & 0.88 & -0.057 & 0.017\\
  2457408.63475 & 0.49 & 5.75 & 27.24 & 13.73 & -11.09 & 1.81 & -0.06 & 0.018\\
  2457408.65109 & -55.17 & 11.69 & 9.26 & 124.35 & -75.85 & 17.63 & -0.216 & 0.144\\
  2457413.51085 & -2.2 & 1.1 & 17.42 & 6.29 & 9.45 & 1.14 & -0.056 & 0.022\\
  2457413.53853 & -1.17 & 1.26 & 8.74 & 9.68 & 5.92 & 1.04 & -0.061 & 0.021\\
  2457413.54289 & -2.67 & 1.2 & 11.07 & 8.7 & 4.2 & 1.18 & -0.057 & 0.021\\
\hline
\end{longtable}

\end{appendix}

\end{document}